\documentclass[twocolumn,aps,prl,a4paper]{revtex4}

\usepackage{graphicx}

\begin{document}
\title {HBT Interferometry: Historical Perspective}
%\date{\today}
%\vspace{-1.35cm}\hfill
%\begin{flushright}
%{\small IFT-P.092/2002}\\
%\end{flushright}
%}
%
\author{Sandra S. Padula}

\affiliation{Instituto de F\'\i sica Te\'orica - UNESP, Rua
Pamplona 145, 01405-900 S\~ao Paulo, Brazil}
\begin{abstract}
I review the history of HBT interferometry, since its discovery in
the mid 50's, up to the recent developments and results from
BNL/RHIC experiments. I focus the discussion on the contributions
to the subject
%field
given by members of our Brazilian group.
\end{abstract}
%\pacs{25.75.-q, 25.75.Gz, 25.70.Pq}
%\vskip -1.35cm

\maketitle

\bigskip

\section{I. Introduction}

I will discuss here the fascinating method invented decades ago,
which turned into a very active field of investigation up to the
present. This year, we are celebrating the $50^{th}$ anniversary
of the first publication of the phenomenon observed through this
method. In this section, I will briefly tell the story about the
phenomenon in radio-astronomy, the subsequent observation of a
similar one outside its original realm, and many a posteriori
developments in the field, up to the present.

\subsection{I.1~ HBT}

\begin{figure}[!htb1]
\begin{center}
\includegraphics*[angle=0, width=8cm]{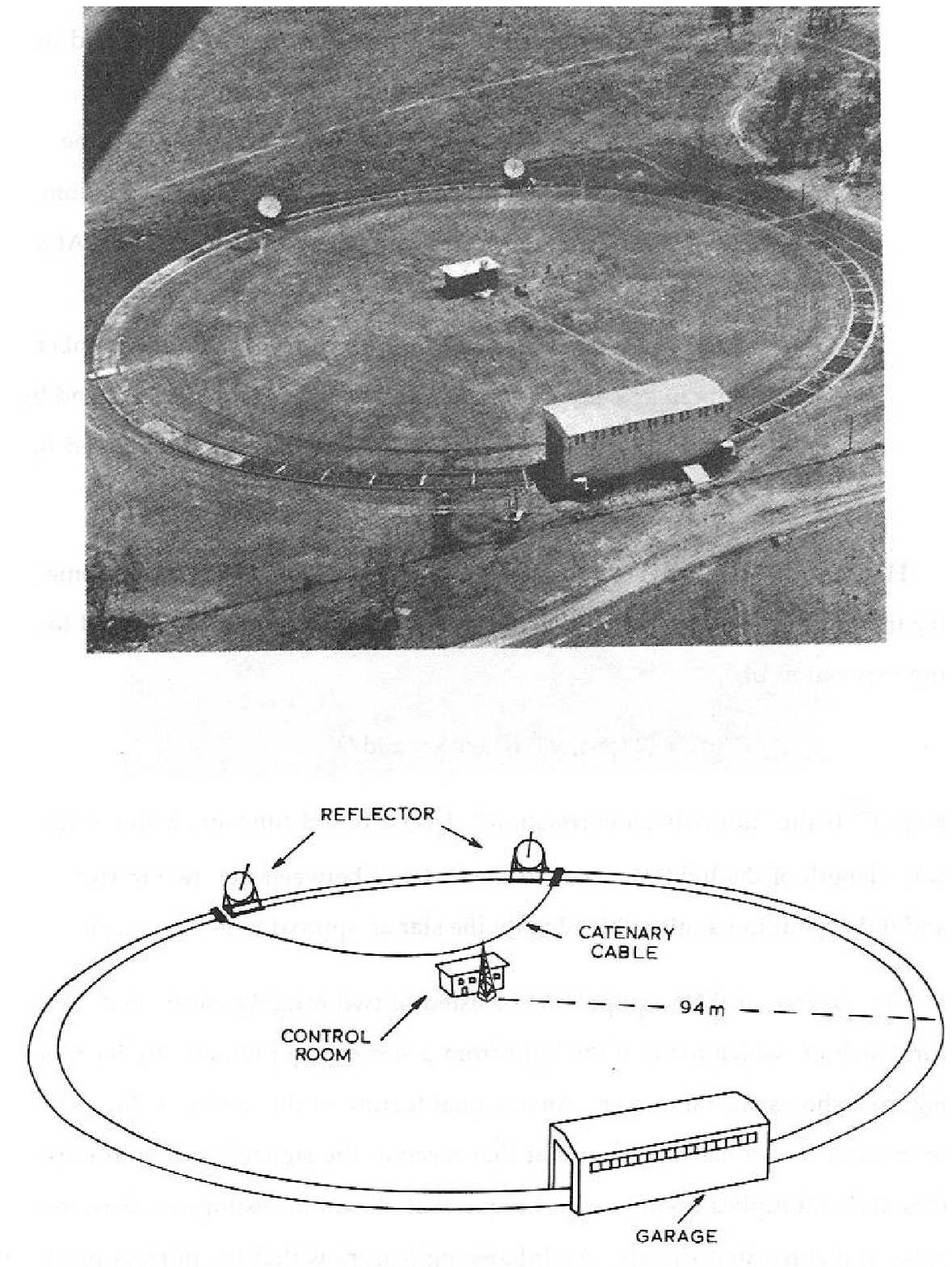}\\
\end{center}
%\caption{\emph{\small Photos and illustration of the original
%HBT apparatus. The upper part shows an aerial view of whole set and
%the lower part is a close picture of the two telescopes.
%These figures have been extracted from Ref.\cite{gold}.}}
\caption{\emph{\small Aerial photo and illustration of the original
HBT apparatus.
They have been extracted from Ref.\cite{gold}.}}
\end{figure}
    HBT interferometry, also known as two-identical-particle correlation,
was idealized in the 1950´s by Robert Hanbury-Brown, as a means to
measuring stellar radii through the angle subtended by nearby
stars, as seen from the Earth's surface.
\begin{figure}[!htb2]
\begin{center}
\includegraphics*[angle=0, width=9cm]{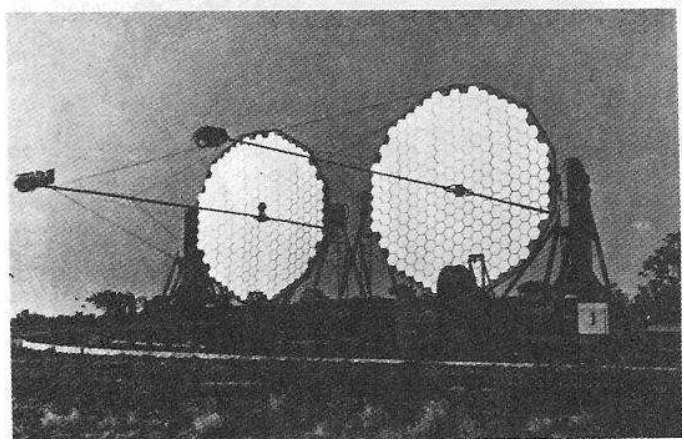}
\end{center}
\caption{\emph{\small Picture of the two telescopes used in the
HBT experiments.
The figure was extracted from Ref.\cite{gold}.}}
\end{figure}
Before actually performing the experiment, Hanbury-Brown invited
Richard Q. Twiss to develop the mathematical theory of intensity
interference (second-order interference)\cite{HBT}. A very
interesting aspect of this experiment is that it was conceived by
both physicists, who also built the apparatus themselves, made the
experiment in Narrabri, Australia, and finally, analyzed the data.
Nowadays, the experiments doing HBT at the RHIC/BNL accelerator
have hundreds of participants. We could briefly summarize the
experiment by informing that it consisted of two mirrors, each one
focusing the light from a star onto a photo-multiplier tube. An
essential ingredient of the device was the {\it correlator}, i.e.,
an electronic circuit that received the signals from both mirrors
and multiplied them. As Hanbury-Brown himself described it, they
{\it \bf `` ... collected light as rain in a bucket ... ''}, there
was no need to form a conventional image: the (paraboloidal)
telescopes used for radio-astronomy would be enough, but with
light-reflecting surfaces. The necessary precision of the surfaces
was governed by maximum permissible field of view. The draw-back
they had to face in the first years was the skepticism of the
community about the correctness of the results. Some scientists
considered that the observation could not be real because it would
violate Quantum Mechanics. In reality, in 1956, helped by Purcell
\cite{purcell}, they managed to show that it was the other way
round: not only the phenomenon existed, but it also followed from
the fact that photons tended to arrive in pairs at the two
correlators, as a consequence of Bose-Einstein statistics. A very
interesting review about these early years was written by Gerson
Goldhaber\cite{gold}, one of the experimentalists responsible for
discovering the identical particle correlation in the opposite
realm of HBT: the microcosmos of high energy collisions.

\subsection{I.2~ GGLP}
In 1959, Goldhaber, Goldhaber, Lee and Pais performed an
experiment at the Bevalac/LBL, in Berkeley, CA, USA, aiming at the
discovery of the $\rho^0$ resonance\cite{GGLP}. In the experiment,
they considered $\bar{p} p$ collisions, at 1.05 GeV/c. They were
searching for the resonance by means of the decay $\rho^0
\rightarrow \pi^+ \pi^-$, by measuring the unlike pair, $\pi^+
\pi^-$, mass-distribution and comparing it with the ones for like
pairs, $\pi^\pm \pi^\pm$. Afterwards, they concluded that there
was not enough statistics for establishing the existence of
$\rho^0$. Nevertheless, they observed an unexpected angular
correlation among identical pions! Later, in 1960, they
successfully reproduced the empirical angular distribution by a
detailed multi-$\pi$ phase-\-space calculation using symmetrized
wave functions for LIKE particles. Being so, they concluded the
effect was a consequence of the Bose-Einstein nature of $\pi^+
\pi^+$ and $\pi^- \pi^-$. They were not aware of the experiment
Hanbury-Brown and Twiss had performed previously. Thus, they had
discovered, by chance, the counterpart of the HBT effect in high
energy collisions. They parameterized the observed correlation as:
\begin{eqnarray}
C(Q^2) &=& 1 + e^{-Q^2 r^2}
= 1 + e^{(q^2_0 - {\bf q}^2) r^2} \\
Q^2 &=& -q^2=-(k_1-k_2)^2=M^2_{12}-(m_1+m_2)^2
%\; ; \; Q^2=-q^2=-(k_1-k_2)^2 \nonumber
\;. \nonumber\label{HBT} \end{eqnarray}
The Gaussian form in the above equation, and several of its
variant options, would be widely used in the years to come, mainly
by the experimentalists, due to the simplicity of the emission
source and analytical results allowed by this profile. We will see
which are the parameters and interpretations derived from it in a
while.

\subsection{I.3~ SIMPLE PICTURE}
At this point, it is natural to ask the question: How to
understand interferometry, or two-particle correlation, in a
simple way? First of all, we should anticipate that it follows
from considering two essential points: the adequate quantum
statistics and chaotically emitting sources, which was already
emphasized by Bartknik and Rz\c a$\dot{z}$ewski\cite{bart}. Let me
illustrate it by a simple example of only two point sources, as
shown in Fig.3:

\bigskip

\thicklines
\begin{figure}[!htb3]
\begin{picture}(360,140)(0,0)
\setlength{\unitlength}{1pt}

%\put(160,-30){\bf Figure 1}

%\put(100,40){\circle*{4}}
{ \bf \put(42,40){\circle*{10}}}
{ \bf \put(39,80){\circle*{10}}}
\put(207,-10){\line(0,1){20}}
\put(207,110){\line(0,1){20}}

{\bf  \put(43,80){\line(4,1){160}}}
{\bf  \put(40,40){\line(4,-1){160}}}
{\bf  \multiput(40,78)(20,-10){8}{\line(2,-1){10}}}
{\bf  \multiput(40,40)(20,10){8}{\line(2,1){10}}}
\put(45,28){\bf $x_2^{\mu}$}
\put(45,90){\bf $x_1^{\mu}$}
\put(25,40){\bf II}
\put(27,77){\bf I}
\put(203,-20){\bf ($x_B^{\mu}$)}
\put(203,135){\bf ($x_A^{\mu}$)}
\put(210,0){\bf B}
\put(210,120){\bf A}
\put(190,10){\bf $k_2^{\mu}$}
\put(190,105){\bf $k_1^{\mu}$}
\put(202,-10){\line(1,0){5}}
\put(202,10){\line(1,0){5}}
\put(202,110){\line(1,0){5}}
\put(202,130){\line(1,0){5}}
{\bf \put(29,60){\oval(30,70)}}
\end{picture}
\vskip1cm
%\begin{center}
\caption{{\it \small  Simplified picture: two point sources, I and
II, emit quanta considered as plane waves, which are observed in
detectors A and B, respectively, with momenta $k_1^\mu$ and
$k_2^\mu$. Since the quanta are indistinguishable, there are two
possible combinations for this observation, illustrated by the two
continuous and the two dashed lines.}}
%\end{center}
\end{figure}
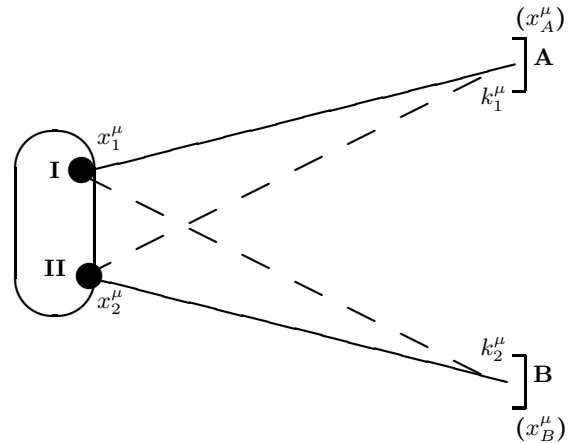
\bigskip

The amplitude for the process can be written as

%{\large
\begin{eqnarray}
A(k_1,k_2) &=& \frac{1}{\sqrt{2}} [ e^{-i k_1 . (x_A-x_1)} e^{i { \phi_1}}
 e^{-i k_2 . (x_B-x_2)} e^{i { \phi_2}}
\nonumber
\\&&
{ \bf \pm}
\; e^{-i k_1 . (x_A-x_2)} e^{i { \phi'_2}} e^{-i k_2 . (x_B-x_1)}
 e^{i { \phi'_1}} ]
,\; \label{A12}\end{eqnarray} \\
where the ($+$) sign refers to bosons and the ($-$) one, to
fermions. In the above equation, $\phi_i$ corresponds to an
aleatory phase associated to each independent emission (completely
chaotic sources), i.e., one phase at random in each emission.
These phases are also considered to be independent on the momenta
$k$ of the emitted quanta.

The probability for a joint observation of the two quanta with
momenta $k_1$ and $k_2$ is given by \noindent
\begin{eqnarray}
&&P_2(k_1,k_2) =\langle|A(k_1,k_2)|^2\rangle =
\nonumber
\\
&=& \frac{1}{2}[ 2\pm
(e^{i(k_1-k_2) . (x_1-x_2)}
{ \langle} e^{\pm i({ \phi_1}+{ \phi_2}-
{ \phi'_1}-{ \phi'_2})} { \rangle} + c.c.) ]
\nonumber
\\
&=& 1 \pm \cos[(k_1-k_2) . (x_1-x_2)]
%,\nonumber\label{P12simpl}\end{eqnarray}
.\label{P12simpl}\end{eqnarray}

The emission being chaotic, we have to consider an average over
random phases, i.e.,
\bigskip
{\large
\begin{equation}
{ \langle} e^{\pm i({ \phi_1}+{ \phi_2}-{ \phi'_1}- { \phi'_2)}} {
\rangle} = \delta_{{ \phi_1 \phi'_1}} \delta_{{ \phi_2 \phi'_2}} +
\delta_{{ \phi_1 \phi'_2}} \delta_{{ \phi_2 \phi'_1}} \;
.\label{phases}\end{equation} }
\bigskip
The two-particle correlation function can be written as
\begin{equation}
C(k_{1},k_{2}) = {{P_2(k_{1},k_{2}) } \over {P_1(k_{1})P_1(k_{2})}}
= 1 \pm \cos[(k_1-k_2) . (x_1-x_2)] \; , \label{c12}\end{equation}
where $P_i(k_i)$ is the single-inclusive distribution. It is
estimated in a similar way as in the simultaneous detection
discussed above, i.e.,
\begin{eqnarray}
A(k_i)\!\!&=&\!\!\frac{1}{\sqrt{2}} [ e^{-i k_1 . (x_A-x_1)} e^{i {
\phi_1}} { \bf \pm} \; e^{-i k_1 . (x_A-x_2)} e^{i { \phi_2}}]
\nonumber \\
\nonumber P_1(k_i)\!\!&=&\!\!\langle|A(k_i)|^2\rangle\!=\!\frac{1}{2}[ 2\pm
e^{ik_i . (x_1-x_2)} { \langle} e^{\pm i({
\phi_1}-{ \phi_2})} { \rangle}\!+\!c.c.]\!\!\\
\label{A1}\end{eqnarray}
In the above case, we would have ${ \langle} e^{\pm i({ \phi_1}-{
\phi_2})} { \rangle}=\delta_{\phi_1\phi_2}$. Since the source is
supposed to be chaotic, the two aleatory phases of emission would
be equal only if they were emitted at the same space-time point.
However, since we are considering here that the probability of two
simultaneous emission by the same source is negligible, we would
be forced to conclude that only possible solution to this problem
that would satisfy this criterium is that the average over phases
is null, in the case of observation by a single detector. We see
then that $P_i(k_i)=1$ in this case and then the result on
Eq.(\ref{c12}) follows.

Already from the very simple example discussed above, se can see
that, in the case of two identical bosons (fermions),  we expect
to see that $C(q=k_1-k_2=0)=2\; (0)$ for completely chaotic
sources. On the contrary, in the case of total coherence
$C(q=k_1-k_2)=1$ for all values of the momentum difference. For
large values of their relative momenta, however, the correlation
function should tend to one, which is clearly not the case in Eq.
(\ref{c12}). But this is merely the consequence of considering an
oversimplified example of only two point sources.

\subsection{I.4~ EXTENDED SOURCES}
 More generally, for extended sources in space and time, if $
\rho(x)$ is the normalized space-time distribution, we have

\noindent
\begin{eqnarray}
&&P_2(k_1,k_2) =
\nonumber
\\
&& = P_1(k_1) P_1(k_2) \int d^4 x_1 \int d^4 x_2 \; |A(k_1,k_2)|^2
\rho(x_1) \rho(x_2)
\nonumber
\\
&&=  P_1(k_1) P_1(k_2) [ 1 \pm | \tilde{\rho}(q) |^2]
%,\nonumber\label{P12simpl}\end{eqnarray}
,\label{P12simpl}\end{eqnarray}
where
\begin{equation}
 \tilde{\rho}({ q}) = \int
d^4x\; e^{i q^\mu x_\mu} \rho(x)
\end{equation}
is the  Fourier transform of $ \rho(x) $. Conventionally, we
denote the 4-momentum difference of the pair by {\bf $q^\mu =
(k^\mu_1-k^\mu_2)$}, and its average by {\bf   $K^\mu =
\frac{1}{2}(k^\mu_1+k^\mu_2)$}.

Then, the two-particle correlation function can be written as
\begin{equation}
C(k_{1},k_{2}) = {{P_2(k_{1},k_{2}) } \over {P_1(k_{1})P_1(k_{2})}}
= 1 \pm { \bf \lambda} \; | \tilde{\rho}({ q}) |^2.
\label{idealc12}\end{equation}
%
%\noindent

In Eq.(\ref{idealc12}) we added, as historically done, the
parameter {\bf $\lambda$}, later called {\bf incoherence} or {\bf
chaoticity parameter}. This was introduced by Deutschmann et
al.\cite{deutsch}, in 1978, as a means for reducing systematic
errors in the experimental fits of the correlation function. The
origin of the large systematic errors was the Gaussian fit. The
reason was that the experimentalists tried to fit the data points
with Gaussian functions whose maxima in $q=0$ were 2, although the
data never reached that maximum value. This led to discrepancies
and to large systematic errors. The easiest way out of this
apparent inconsistency was to add a fit parameter, {\bf
$\lambda$}, thus reducing the systematic errors by the
introduction of this extra degree of freedom.

\begin{figure}[!htb4]
\begin{center}
\includegraphics*[angle=0, width=8cm]{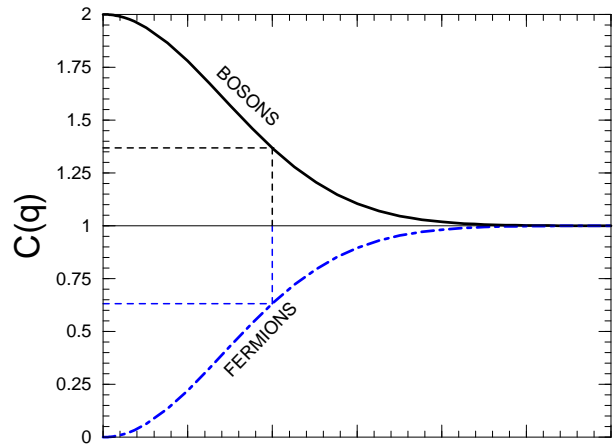}
\end{center}
\caption{\emph{ \small Simple illustration corresponding to the
ideal Gaussian source. The upper curve represents to the bosonic
case, while the lower curve, the fermionic one. The parameter $R$
is the r.m.s. radius of the emitting region.}}
\end{figure}

To illustrate the correlation function as written in
Eq.(\ref{idealc12}) with a simple analytical example, let us
consider the Gaussian profile, i.e.,
\begin{equation}
\rho(x) = e^{- x^\mu x_\mu/(2R)^2} \; \longrightarrow \; \rho(q) =
e^{-q^2 R^2/2} \;. \label{gaus}\end{equation} Consequently, in this
very simple example, a typical correlation function is written as
\begin{equation}
C(k_{1},k_{2}) = 1 \pm { \bf \lambda} \; e^{-q^2 R^2}.
\label{c12gen}\end{equation}

In equation (\ref{c12gen}), as we denoted before, the plus sign
refers to the bosonic case, and the minus sign to the fermionic
one. We easily see that, in this simple example, we would expect
experimental ideal HBT data to behave as sketched in Fig. 4, where
the upper part refers to bosons and the lower one, to fermions. We
see that, in the two-boson (two-fermion) case, there is an
enhancement (depletion) of the correlation function in the region
where the relative momenta of the pair are small. In both cases of
this simple example, the typical size of the emission region
corresponds to the inverse width of the $C(k_{1},k_{2})$ curve,
plotted as a function of $q=k_{1}-k_{2}$.

Returning to the discussion of the fit parameter $\lambda$, I
would like to point out that there is a very simple explanation to
reconcile this apparent inconsistency, without the need to
introduce this extra degree of freedom. Limited statistics is
behind it, since it is virtually impossible to measure two
identical particles with exactly the same momenta. This led the
experimentalists to split the momenta of the particles in small
bins. In more recent times, these bins can be projected in two or
more dimensions. %%
For instance, along the income beam direction in fixed target
heavy ion collisions ($q_L$), and in the direction transverse to
it ($q_T$). Good quality data allow the experimentalists to
consider very small bin sizes. Nevertheless, their range is
finite. Being so, when the correlation function is projected
along, say, the $q_T$ direction, the smallest value of $q_L$ is
not zero, but within the first (smaller) bin size, in case of high
enough statistics. Consequently, we immediately see that the
correlation function plotted as a function of $q_T$, will not
reach the maximum (minimum) value of 2 (0) for bosons (fermions)
at $q_T=0$. Naturally, the larger the first bin size, the bigger
is the deviation from the maximum (minimum) expected value for the
correlation function at $q=0$. This can be better seen with the
help of Fig. 5, where the two upper curves represent the pure
correlation function and the two lower ones, the theoretical
correlation functions corrected by the Gamow factor,
$\Upsilon(q)$, a multiplicative factor taking into account 2-body
Coulomb final state interaction. This factor distorts the pattern,
mainly at small values of the momentum difference of the pair, and
is written as
\begin{equation}
\Upsilon(q)=\frac{q_c/q}{\exp(q_c/q) - 1} \; \; ; \; \; q_c=2\pi\alpha m
%; \; q=\sqrt{-(k_1-k_2)^2}
\;,\label{gamow}\end{equation} where $q$ is the four-momentum
difference and $\alpha$ is the fine structure constant. The Gamow
factor simply multiplies the entire expression in
Eq.(\ref{idealc12}), (\ref{c12gen}),
%(\ref{qKPG}), (\ref{c12nr}), (\ref{bertsch}), (\ref{cce}),
and all other forms of correlation function for two charged,
identical particles. Fig. 5 was generated by the code CERES, whose
hypothesis and formulation will be discussed later in this
manuscript, in Section II.2.
\begin{figure}[!htb5]
\begin{center}
\includegraphics*[angle=0, width=8cm]{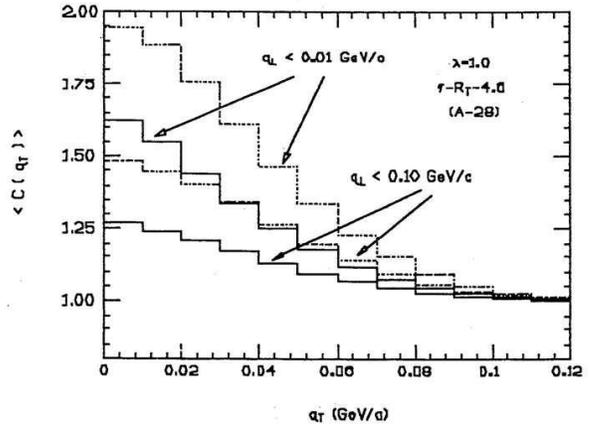}
\end{center}\vskip-5mm
\caption{\emph{ \small The plot illustrates the fact that we can
have the maximum of the bosonic correlation curve below unity,
even when the intercept parameter is fixed to be $\lambda=1$. This
normally happens in realistic cases of finite statistics. In the
plot, we see the two-pion correlation as a function of $q_T$, for
two possible bin sizes, i.e., for $q_L\le 0.01$ GeV/c and $q_L\le
0.1$ GeV/c. This plot was extracted form Ref.\cite{NPA525}}}
\end{figure}

We should emphasize, however, that the simple relation between the
two-particle correlation function and the Fourier transform of the
space-time distribution, as written in Eq. (\ref{c12gen}) is not
straightforward, in general. It is observed only when we can
consider the phase-space as decoupled, i.e., $f({\bf x}, {\bf p})
= \rho({\bf x}) g({\bf p})$, where $\rho({\bf x})$ is the source
space-time distribution, and $ g({\bf p})$ is the energy-momentum
distribution. In general, it cannot be decoupled in this way. This
is, for example, the case in relativistic heavy ion collisions,
where the source expands during its life-time. The reason is that
HBT is not only sensitive to the source geometry at particle
emission but is also sensitive to the underlying dynamics. This
makes the analysis model-dependent and more powerful formalisms
(like the one proposed by Wigner, the Covariant Current Ensemble,
etc.) must be adopted. We will discuss more about this limitation
later on.

\subsection{I.5~ FURTHER APPLICATIONS}
In the 1970's, Kopylov, Podgoretski\u\i, and Grishin\cite{kpg}
used second-order interferometry to study several interesting
problems. For example, they modelled the nucleus as a static
sphere with radius R, emitting pions from its surface and got the
following correlation function
\begin{equation}
 C(k_1,k_2) = 1 \pm [\frac{2 J_1(q_T R)}{q_T
R}]^2 [1+(q_{_0} \tau)^2]^{-1}
 \;, \label{KPG1}\end{equation} where
\begin{eqnarray}
 q_\parallel &=& {\bf q} . \frac{\bf K}{|\bf K|} \nonumber\\
{\bf q_T} &=& {\bf q} - {\bf  q_\parallel} \nonumber\\
q_{_0} &=& E_1 - E_2 \approx \frac{1}{2m}({\bf k}^2_1 - {\bf k}^2_2)
\nonumber\\
 &\approx& \frac{1}{2m} ({\bf k}_1 - {\bf k}_2) . ({\bf k}_1 + {\bf k}_2)
\propto q_\parallel \;.\label{qKPG}\end{eqnarray} \\
The two variables in Eq.(\ref{qKPG}) are nowadays known as Kopylov
variables. With respect to the same parametrization,
Cocconi\cite{cocconi} re-interpreted in 1974 the quantity
$c\tau=\delta(<R)$ as the thickness of the pion emission layer. They
used similar forms for studying: i) the lifetime of excited nuclei
through the interferometry of evaporated neutrons, ii) shape and
size of multiple production region with $\pi^\pm\pi^\pm$
correlations, etc. They also applied to CERN/ISR data on $pp,
\bar{p} p$.

Many other scientists contributed to the field during that decade.
Just to mention a few names, I would quote Shuryak;  Biswas;
Fowler \& Weiner; Giovannini \& Veneziano; Grassberger; Yano \&
Koonin; Gyulassy, Kauffmann \& Wilson, etc.  Many of these
contributions were organized in a collection of reprints, edited
by R. M. Weiner\cite{WeinerBook},  in the late 1990's, which is a
very good source of these reference papers. Among them, the papers
by Gyulassy, Kauffmann \& Wilson\cite{GKW}, as well as those by
Fowler and Weiner\cite{weiner}, represented important steps in the
field, for introducing more powerful formalisms for studying the
cases of coherent, chaotic, and partially coherent sources. On the
other hand, Grassberger\cite{grass} called the attention to the
fact that resonances could play an important role in
interferometry, since the long lived ones could distort the
correlation function in the region where it was more significant,
i.e., at small values of ${\bf q}$, thus changing the {\it
chaoticity parameter} considerably. This is due to the fact that a
resonance of 4-momentum $k$, mass $M$, and width $\Gamma$, would
travel a distance $\sim k/(M\Gamma)$ before decaying, causing
interference effects whenever $k.q \le M\Gamma$. The first attempt
to analyze the effect of resonances on interferometry in detail
was made more than a decade afterwards. I will discuss it later in
Sec. II.2.

Despite the comment made earlier in the text, regarding the
limitations of the static Gaussian fit, this has always been the
preferred source model, due to its simplicity. Along this line, it
is instructive to observe that there is a Gaussian limit of
Eq.(\ref{KPG1}), corresponding to

\begin{equation}
[\frac{2 J_1(q_T R)}{q_T R}]^2 \approx e^{-q^2_{_T} R^2_T/4}
\; ; \; \frac{1}{[1+(q_{_0} \tau)^2]}  \approx e^{- q^2_{_0}\tau^2}
\;,
\label{approx}\end{equation} \\
a variation of which suggests a non-relativistic parametrization of
the correlation function, i.e.,
\begin{equation}
C(k_1,k_2) = 1 \pm {\lambda} \; e^{-q^2_{_0} \tau^2/2} e^{-q^2_T
R^2_T/2} e^{-q^2_L R^2_L/2} \;.\label{c12nr}\end{equation}

The expression in Eq. (\ref{c12nr}) has been widely employed since
the beginning of high energy heavy ion collisions, becoming the
standard form to analyze two-particle interferometry, particularly
among the experimentalists. In Eq.(\ref{c12nr}), ${\bf q_L}$ is
the momentum difference along the direction of the incident beam,
${\bf q_T}$ is the component transverse to beam direction, and
$q_{_0}$ is the time component. In the late 80's, the most popular
form changed slightly, according to the suggestion by
Bertsch\cite{bertsch}, becoming
\begin{equation}
C(k_1,k_2) = 1 \pm {\lambda} \; e^{-q^2_{\small S} R^2_S}
e^{-q^2_{\small O} R^2_O} e^{-q^2_L R^2_L/2}
\;,\label{bertsch}\end{equation} similarly to the previous
definition. However, in Bertsch's suggestion, there was a
decomposition of the transverse component, partially incorporating
the definition introduced by Kopylov and Podgoretski\u\i , i.e.,
${\bf q_{O}}$ and ${\bf q_{S}}$ are both perpendicular to beam
direction but ${\bf q_{O} \parallel K_T}$ [$= \frac{1}{2}({\bf
k_{1_T}}+{\bf k_{2_T}})$], and ${\bf q_{S} \perp K_T}$. As before,
${\bf q_{L}}$ represents the component of the pair momentum
difference along the beam direction. Latter, Heinz et al.,
suggested to include a {\sl out-longitudinal} cross term in
Gaussian fits to the data, i.e., the correlation in this case
would be written as\cite{heinzchsc}
\begin{equation}
C(k_1,k_2) = 1 \pm {\lambda} \; e^{-q^2_{\small S} R^2_S}
e^{-q^2_{\small O} R^2_O} e^{-q^2_L R^2_L} e^{-2q^2_{OL} R^2_{OL}}
\;.\label{heinz}\end{equation}

Ever since, this field has been under constant development and
expansion, both in the theoretical and in the experimental grounds.
I will briefly highlight only some of the
theoretical contributions to the field, mainly focusing at the ones
from the Brazilian group and some collaborators from abroad, since a
complete discussion of the contribution along theses 50 years is
beyond the scope of the present review. For a complete survey of the
subject, as well as of the theoretical and experimental progress in
the field, I would strongly encourage the reader to look into Refs.
\cite{WeinerBook,zajc,boal,wong,heinz1,csorgo,rhic}.

\section{II. CONTRIBUTIONS FROM GROUP MEMBERS}

The first contact of the group, whose contributions we are
discussing here, with HBT interferometry started in the mid- to
late eighties, and was the subject of my PhD
Thesis\cite{SandraPhD}. In fact, it happened a few years before
the group itself began performing as a group. Nevertheless, this
topic consistently appeared during the group meetings along these
years and, since this is a historical perspective, it is worthy to
insert the subject in this context. Around the beginning of the
decade of 1980, there was already an emerging subject that was
attracting the attention of the high energy community: the
possible existence of a new state of matter, the Quark-Gluon
Plasma (QGP), expected to be produced at high enough temperatures
and/or densities. The QGP is a state in which quarks and gluons,
the constituents of the hadrons, would be free to wander around a
volume much bigger than the usual hadronic size. This state was
expected to exist for a brief period of time, since only usual
hadrons, with quarks and gluons confined in their interior, have
been observed empirically. This imposed the need to look for
probes of its existence. Among them, Interferometry was suggested,
as a means to estimate the dimensions of the system formed in high
energy collisions, thus testing if it was produced in such a new
state of matter. In fact, James D. Bjorken was the person who
suggested pion interferometry as the subject of my Ph.D.
thesis\cite{Bj83}.

\subsection{II.1~ EXPANSION EFFECTS IN HBT}
In the first paper on the subject, we started by making the
hypothesis that the Quark-Gluon Plasma was already being
produced in~ $p p$ and $\bar{p} p$
collisions at the CERN/ISR. We considered\cite{sandra} that the
system produced in such collisions expanded before emitting the
final particles (hadrons), according to the one-dimensional
Landau Hydrodynamical Model \cite{Landau}.
%, in two stages.
In the initial stage, the system was formed in the QGP phase at a
certain temperature, $T_0$, started expanding and cooling down,
until it reached the critical temperature, $T_c$, which we assumed
to be of order of pion mass. It could be imagined that, once $T_c$
was reached, the hadronization occurred instantaneously, followed
by the particle emission. This simplifying hypothesis was actually
adopted in the general study of the effects on the correlation
function caused by the system expansion. On the other hand, the
energy density of the ideal QGP fluid once $T_c$ is reached, is
much higher that the correspondent one for a hadronic system, due
to the statistical degeneracy factors. More explicitly,
$\epsilon_{QGP} (T_c)= \frac{\pi}{30}(g_g+\frac{7}{8}g_q)T_c^4+B$,
where $g_g=8(color)\times 2(spin)$, $g_q=2(\bar{q}q)\times
3(color)\times 2(spin)\times N_f(flavors)$ are the gluon and quark
degeneracy factors. The constant $B$ is the vacuum pressure in the
MIT Bag model. And, the hadronic correspondent for an ideal gas of
pions and kaons, at $T_c$, is $\epsilon_{\pi}(T_c) =
\frac{1}{2\pi^2}\left[(g_\pi)\phi(m_\pi/T_c) + g_K
\phi(m_K/T_c)\right]$, being $g_\pi=3$ and $g_K=4$, respectively
the statistical factor for pions and for kaons (The function
$\phi(z)$ is a combination of Bessel modified functions of second
class, $K_i(z)$, $ \phi(z)=z^2\sum_{m=0}^\infty(\pm 1)^m
\left\{\frac{3K_2[z(1+m)]}{(1+m)^2} + z
\frac{K_1[z(1+m)]}{(1+m)}\right\} $). Nevertheless, the large
ratio of the QGP to the hadronic statistical degeneracy factors,
together with the entropy conservation during the phase
transition, make the duration of the mixed phase very long. And
mesons would be emitted during all that period. This was a more
realistic hypothesis that was adopted when comparing our
predictions with experimental data. However, for the sake of
simplicity, we considered in the calculation that the emission
occurred at a typical average freeze-out time, $<\tau_f>$.

In our calculation, we neglected the
transverse expansion and used the asymptotic Khalatnikov solution,
i.e.
\begin{eqnarray}
\xi &\equiv& \ln \left(\frac{T}{T_0} \right) \simeq -c_0^2 \ln \left(
\frac{\tau}{\Delta}\right)\nonumber \\
\alpha &\simeq& \frac{1}{2}\ln\left(\frac{t+x}{t-x}\right)
\end{eqnarray}
where $\alpha$ is the system rapidity, $c_0\approx 1/\sqrt{3}$ is
the sound velocity $\sqrt{(1-c_0^2)/\pi l}$, being $l$ the initial
thickness of the fireball (solution valid whenever $\xi \gg
|\alpha|$). This is essentially the Bjorken picture of
hydrodynamics but with different initial conditions. In the
version we adopted of the hydrodynamics, the initial temperature
$T_0$ depends on the value of the fireball, with mass $M$,
initially formed in high energy hadronic collisions.
%, i.e., the amount of energy  initially deposited in the central collision.
The hypothesis we used was that a large Lorentz-contracted
fireball was formed around one of the incident
particles\cite{hama}, constituted of quarks and gluons, with
initial radius $R \simeq R_{proton}$. The fireball mass was
estimated as the {\sl missing mass}, i.e., by discounting the
fraction of the energy available in the center of mass of the
$\bar{p}(p) p$ collision that was dragged by the dominant particle
after the collision happened. For relating this initial
temperature with the fireball mass, we equated the number of
produced hadrons at $T_c$ to the (conserved) entropy, which can
easily be estimated for a QGP as $S(T_0)=s(T_0)V_{QGP}$. The
initial volume of the QGP can be related to the initial energy
density, $\epsilon_0$, which can be simply written as
$\epsilon_0=M/V_{QGP}$. The initial entropy density, $s(T_0)$, can
be estimated through statistical relations, leaving the
calculation of the initial QGP volume to be made. Since we
considered $\bar{p}(p)p$ collisions, we assumed that $V_{QGP} =
\frac{4}{3}\pi R_0^3 \frac{2 m_p}{M}$, where $R_0$ and $m_p$ are,
respectively, the proton radius and mass. At the end, the final
proportionality coefficients were estimated with the help of the
experimental data on charged multiplicity versus the missing mass.
Finally, the initial temperature was related to the fireball mass
by a numerical factor, $T_0 = 0.0989 \sqrt{M}$. From that, we
estimated $\tau_c$, instant corresponding to the beginning of the
phase transition, $ \tau_c = \Delta \left(\frac{T_0}{T_c}\right)
$, and also the instant it ended, $\tau_f$, as well as the typical
(average) duration, $<\tau_f>$ (see Ref. \cite{sandra} for
details).

We had adopted the Kopylov variables described before, in Eq.
(\ref{qKPG}) and sketched in Fig. 6, as the relevant momentum
difference of the pair of pions.
%\vskip-.5cm
\begin{figure}[!htb6]
\begin{center}
\includegraphics*[angle=0, width=7cm]{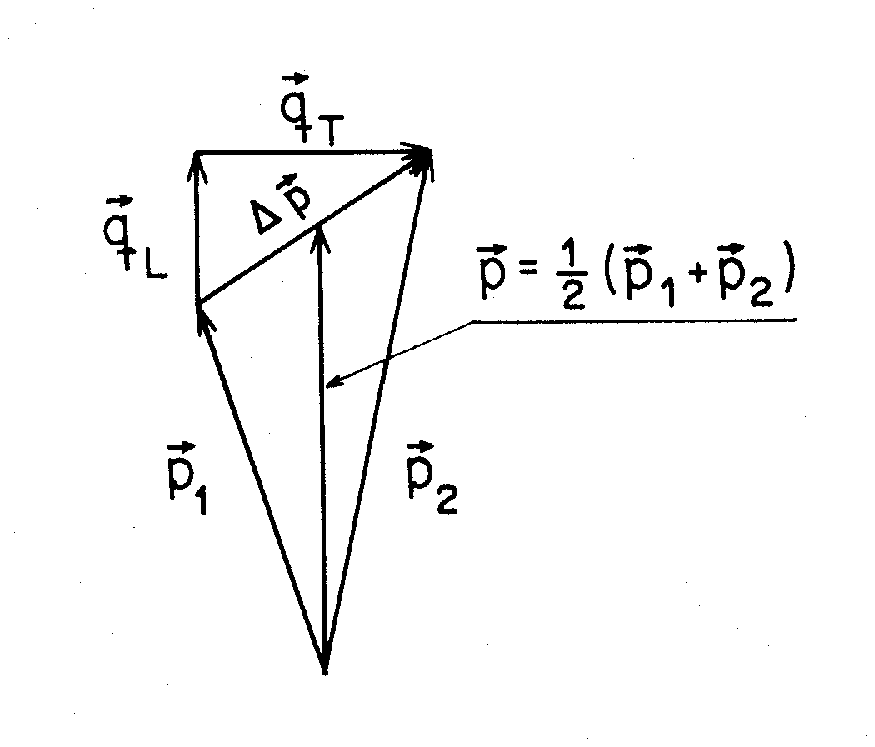}
\end{center}\vskip-3mm
\caption{\emph{ \small Illustration of the Kopylov variables:
we see that $\vec{q_L} \equiv \vec{q_\parallel} \parallel \vec{K}$ and
$\vec{q_T}\equiv \vec{q_\perp} \perp \vec{K}$. The figure also shows
other notations commonly used: $\vec{q}\equiv \vec{\Delta p}$
and $\vec{K}\equiv \vec{p}$.}}
\end{figure}

For studying the general behavior of the correlation function
under the influence of the expansion effects, we assumed that each
point on the surface $\tau=\tau_f$ of the QGP, where $T=T_c\simeq
m_\pi$, was an independent chaotic source with momentum spectrum
given by
\begin{equation}
f(p) \simeq \frac{1}{(2\pi)^3}\frac{u^\mu p_\mu}{E}
\exp{(-\frac{u^\mu p_\mu}{T_c})} \;,\label{spectrum}\end{equation}
where $u^\mu=(\cosh\alpha, \sinh\alpha,0,0)$ is the 4-velocity of
the fluid, and $p^\mu=(E,p_x,p_y,p_z)$ is the 4-momentum of the
emitted particle. The amplitude for a particle emitted at $x'$ to
be observed at $x$ is written as
\begin{equation}
A(x,x')=\int d{\vec p} \sqrt{f(p)} \exp{[-ip_\mu(x^\mu-x'^\mu)]}
e^{i\phi(x')} \;, \label{ampl}\end{equation} where $\phi(x')$ is a
random phase. We followed the formulation and notation of
Ref.\cite{shuryak} for writing the probability of detecting two
quanta of momenta $p_1$ and $p_2$ in an event, as
\begin{equation}
W(p_1,p_2)=\tilde{I}(0,p_1)\tilde{I}(0,p_2)+
|\tilde{I}[(p_1-p_2),\frac{1}{2}(p_1+p_2)]|^2
\;, \label{2part}\end{equation}
where
\begin{equation}
\tilde{I}[(\Delta p),p]\!=\!\int dx \; d(\Delta x) e^{i(x \;\Delta
p + \Delta x \;p)} \int dx' I(x,\Delta x,x') \;,
\label{shuwig}\end{equation} and
\begin{equation}
\langle A^*[x-\frac{\Delta x}{2},x'] A^*[x+\frac{\Delta
x}{2},x']\rangle =\delta(x'-x) I(x,\Delta x,x') \;.
\label{aver}\end{equation} The average indicated in the above
equation is taken over the random phases $\phi(x')$ and
$\phi(x'')$. In particular, we see that the single-inclusive
distribution is written as
\begin{equation}
W(p_i)=\langle |\int dx' \int dx e^{ip_i.x} A(x,x') dx|^2 \rangle = \tilde{I}[0,p_i]
\;, \label{1part}\end{equation}

%\vskip 6cm
\begin{figure}[!htb7]
\begin{center}
\includegraphics*[angle=0, width=8cm]{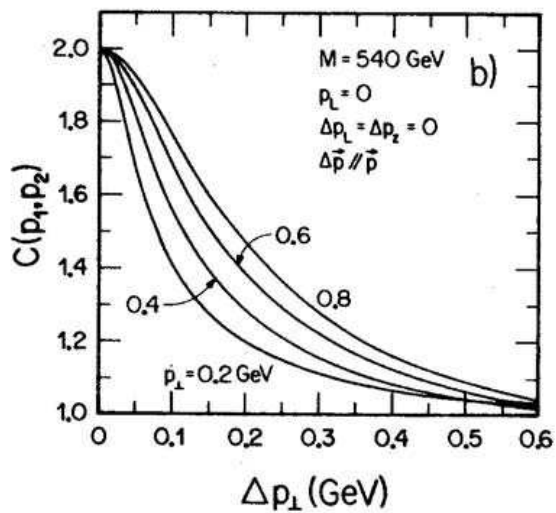}
\end{center}
\caption{\emph{ \small $C(p_1,p_2)$ as a function of the momentum
difference $\Delta p_T$ ($\equiv q_T$, in the text), for several
values of the average momentum $p_\perp$ ($\equiv K_T$ in the
text). A clear dependence is seen: the curves broaden with
increasing average momentum, showing a progressively smaller
effective emission region. This plot is a reproduction of the one
originally published in Ref.\cite{sandra}.}}
\end{figure}

From this formulation we obtained the expression for the
two-particle correlation function. With that, we studied several
different kinematical zones, trying to apprehend the lessons that
idealized theoretical cuts could teach us. Many interesting and
important results came out of that study (see Ref. \cite{sandra}
for details).

The first, unexpected effect, was showing a clear influence of
different emission times at which the pions were emitted from the
source. I will write it in a simple analytical form, later in the
text. This effect showed itself into the correlation function when
plotted as a function of $q_T$, the Kopylov variable parallel to
the average momentum of the pair, ${\bf K}=\frac{1}{2}({\bf
k_1}+{\bf k_2})$. It is illustrated in Fig. 7. Independently of
our knowledge, S. Pratt had also suggested that the time would
influence the correlation function, so that large short-lived
sources could result into a similar correlation function as a
short long-lived one \cite{pratt}.

The dependence on the average momentum of the pair, ${\bf K_T}$,
shown above, was a symptom of effects coming from the underlying
dynamics, and reflected the break-up of the naive picture, in which
the correlation function depended exclusively on the variable ${\bf
q_T}$, as in the Gaussian example discussed above.

%\vskip 6cm
\begin{figure}[!htb8]
\begin{center}
\includegraphics*[angle=0, width=9cm]{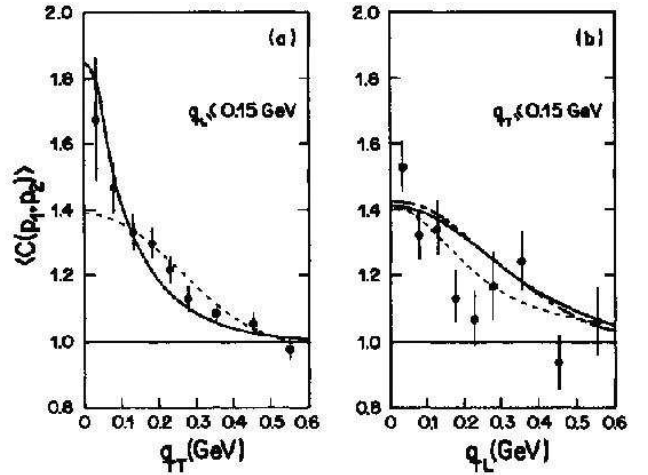}
\end{center}
\caption{\emph{ \small In (a), the two-pion correlation function
is shown in as a function of the Kopylov variable $q_T$, averaged
over the interval $q_L \le 0.15$ GeV (we adopted natural units, in
which $\hbar=c=1$). In (b), it is shown as a function of the other
component, ${\bf q_L}\parallel {\bf K}$, for $q_T \le 0.15$ GeV.
The data points are from Ref. \cite{israfs}. These plots were
originally published in Ref.\cite{sandra}.}}
\end{figure}

We also compared our results with data of $p p$ and $\bar{p} p$
collisions at CERN/ISR ($\sqrt{s}=53$ GeV) and could efficiently
describe the trend of data. This was maybe the only successful
description of that particular experimental result, reflecting the
need for more powerful formalisms when describing HBT
interferometry at high energies. In this case, in an effort to
make the estimate more realistic, we considered that the emission
occurred later, at a typical instant of time $<\tau_f>$, averaged
over the long period that lasted the first order phase transition.
We should notice that the curves in Fig. 8 are not fits, but
predictions from the model, obtained without the need for
introducing the $\lambda$ parameter (which is equivalent to fixing
$\lambda=1$). Similar treatment within the same model was also
given to two-kaon interferometry data from the same
experiment\cite{sandra}, successfully describing data.

Fig. 7 and 8 above also show another important result from that
study: the observation of strong distortions in the correlation
function, definitely departing from the Gaussian shape, due to the
dynamical effects related to the expansion of the system.

\subsection{II.2~ NON-IDEAL EFFECTS}

More than 10 years after Grassberger pointed out the important
role resonances could have in interferometry we investigated it in
detail, in collaboration with M. Gyulassy\cite{gyupad89}. In
particular, we analyzed the effect of resonances decaying into
pions, following the predictions of resonance fractions from the
ATTILA version of the Lund model\cite{att}. Very briefly, it can
be understood as follows: long lived resonances, such as $\omega$,
$\eta$, $\eta'$, can mimic sources with longer life-times, even if
they freeze-out simultaneously as the direct $\pi$'s.
%\vskip 6cm
\begin{figure}[!htb9]
\includegraphics*[angle=0, width=9cm]{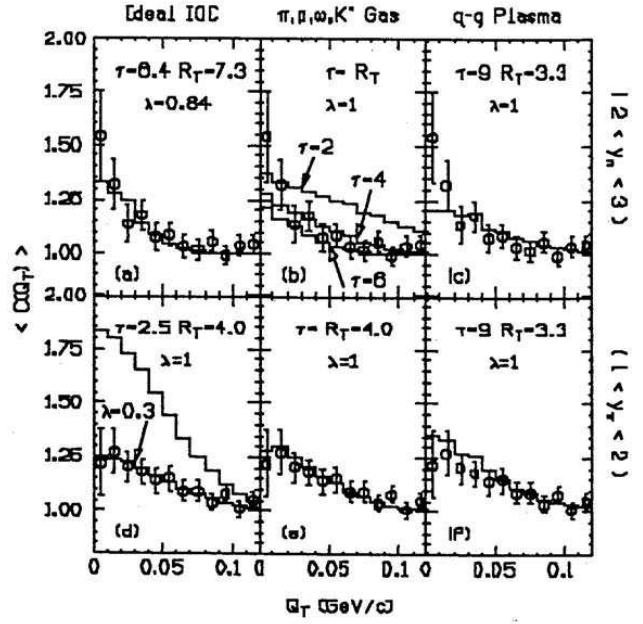}
\caption{\emph{\small The two-pion correlation function is shown
in three different scenarios, as a function of $q_T$. The first,
in parts (a) and (d) were calculated assuming an ideal
inside-outside cascade source (IOC) as in the 1-D Bjorken
picture\cite{Bj}. In parts (b) and (e), a non-ideal resonance gas
source is considered with parameters suggested by the ATTILA
version of the Lund model. Parts (c) and (f) correspond to the QGP
model of Ref.\cite{gb}. Plots in the upper panel were calculated
in the central rapidity region ($2<y_{\pi}<3$), whereas the ones
in the lower panel refer to $1<y{_\pi}<2$. This figure is a
reproduction of the one originally published in
Ref.\cite{gyupad89}.}}
\end{figure}

I call the attention to the fact that, although denoted by the
same letters, the variables $q_L$ and $q_T$ of momentum
difference, appearing in Fig. 9, are defined as the components
parallel and perpendicular to the incident beam direction,
respectively, as became a convention in high energy heavy ion
collisions.

It was very important to verify if we could explain the
preliminary results obtained by the NA35 collaboration
(CERN/SPS)\cite{NA35} colliding O+Au at 200 GeV/nuclear in a more
conventional way, without assuming QGP formation. This last
hypothesis had been suggested by G. Bertsch\cite{gb} at that time
as the only possible explanation. Actually, by introducing
resonances decaying into pions explicitly into interferometry, we
managed to demonstrate that our result, assuming the formation of
a regular hadronic system and no QGP, led to an equally good
agreement with data, as can be seen from Fig. 9. In fact, we
showed in Ref. \cite{gyupad89} (and presented in the Quark matter
'88 Conference \cite{qm88}) that the NA35 preliminary data of that
time was consistent with a wider range of pion source parameters
when additional non-ideal dynamical and geometrical degrees of
freedom were incorporated into the analysis by extending the
Covariant Current Formalism\cite{GKW,gyupad89,KG}.

\subsection{II.2.1~ Ideal Bjorken IOC Picture}

In the Covariant Current Ensemble formalism\cite{gyupad89,GKW,KG}, the
correlation function for identical bosons (since mainly
bosonic HBT will be discussed in this review) can be expressed as
\begin{equation}
C(k_1,k_2)=C(q,K)=1+ \lambda \frac{|G(q,K)|^2}{G(k_1,k_1)G(k_2,k_2)}\;\;,  \label{cce}
\end{equation}
where $q^\mu =k_1^\mu -k_2^\mu $ and $K^\mu =\frac 12(k_1^\mu +k_2^\mu
)$.

The complex amplitude, $G(k_1,k_2)$, can be written as
\begin{equation}
G(k_1,k_2)=\int d^4xd^4p\;e^{iq^\mu x_\mu }D(x,p)j_0^{*}(u_f^\mu k_{1\mu
})j_0(u_f^\mu k_{2\mu })\;\;,  \label{g12}
\end{equation}
where $D(x,p)$ is the break-up phase-space distribution\cite{gyupad89,pg:nioc}
and the currents, $j_0(u_f.k_i)$, contain information about the production
dynamics. The one-particle spectrum is obtained from Eq. (\ref{g12})
by imposing $k_1=k_2$, which leads to
\begin{equation}
G(k_i,k_i)=\int d^4xd^4p\;D(x,p)|j_0(u_f^\mu k_{i\mu })|^2\;\;.  \label{gii}
\end{equation}

The currents $j_0(u_f.k_i)$ in Eqs. (\ref {g12},\ref{gii}) can be
associated to thermal models, and can be written in a covariant
way as
\begin{equation}
j_0(k)\propto \sqrt{u^\mu k_\mu }\exp {\{-\frac{u^\mu k_\mu }{2T}\}}
\;.\label{j0thermal}\end{equation}
However, to make the computation easier, we adopted
a more convenient parametrization
\begin{equation}
j_0(u.k)=\exp {\{-\frac{u^\mu k_\mu }{2T_{ps}}\}}\;\;,  \label{j0pseudo}
\end{equation}
where the so-called pseudo-temperature $T_{ps}$ is related with
the true temperature $T$ according to\cite{KG}
\begin{equation}
T_{ps}(x)=1.42T(x)-12.7\text{ MeV}\;.  \label{tps}
\end{equation}
This mapping between $T(x)$ and $T_{ps}(x)$ was later shown to be a good
approximation also in the case of kaon interferometry\cite{padrol}.

With the covariant pseudo-thermal parametrization as in Eq.
(\ref{j0pseudo}), the complex amplitude can be rewritten in a
simpler form,
\begin{equation}
G(k_1,k_2) = \langle e^{i q x_{af}}
e^{- K p_{af}/(mT)}
\rangle
\; \; ,\label{cegij}\end{equation}
where the bracket $<...>$ denote an average over the pion freeze-out
phase-space coordinates.

In the case of Bjorken ideal Inside-Outside Cascade (IOC) picture, the
phase-space distribution involves a fixed freeze-out proper time $\tau_f$ and a
perfect correlation between $\eta$ and $y$.
The correspondent phase-space distribution is written as
\begin{equation}
D(x,p) = \frac{1}{E_f} \rho \frac{1}{\tau_f} \delta(\tau_f - \tau)
\delta(\eta - y) \delta(p_0 - E_{\bf p}) g({\bf p}_{T})
\frac{e^{\frac{-x_{T}^2}{R_T^2}}}{\pi R_T^2} \;,
\label{ioc}\end{equation} where $E_{\bf p}={\sqrt {{\bf
p}^2+m^2}}$ is the energy and $g({\bf p}_T)$ is the transverse
momentum distribution; the rapidity distribution is considered to
be uniform, i.e., $\frac{dN}{dy} = \rho$. In the ideal IOC
picture, there is a perfect correlation in phase-space between the
space-time rapidity
\begin{equation}
\eta=\frac{1}{2}\ln[\frac{t+z}{t-z}]
\end{equation}
and the energy-momentum rapidity,
\begin{equation}
y=\frac{1}{2}\ln[\frac{E+p_z}{E-p_z}] \;,\end{equation} i.e., they
are indistinguishable.

To obtain simple analytical equations, we assume a very narrow
distribution of ${\bf p}_T$ around small momenta, i.e.,
$g({\bf p}_T)=\delta^2({\bf p}_T)$. The finite pion wave-packets
generate the finite $p_T$ distribution in our case.
    By substituting $D(x,p)$ from (\ref{ioc}) into (\ref{g12})  and
considering the pseudo-thermal parametrization (\ref{j0pseudo}) for the
currents, the function $G(k_1,k_2)$ was found to be\cite{KG}

\begin{equation}
G(k_1,k_2)=2<\frac{dN}{dy}>\{\frac 2{q_TR_T}J_1(q_TR_T)\}K_0(\xi )\;\;,
\label{gk1k2}
\end{equation}
where %{\small
\begin{eqnarray}
\xi ^2 &=&[\frac 1{2T_{ps}}(m_{1T}+m_{2T})-i\tau (m_{1T}-m_{2T})]^2+
\nonumber \\
&&2(\frac 1{4T_{ps}^2}+\tau ^2)m_{1T}m_{2T}[\cosh (\Delta y)-1]\;\;,
\label{xi}
\end{eqnarray}
%}
and $\Delta y=y_1-y_2$.

The single-inclusive distribution is then written as
\begin{equation}
G(k_i,k_i)=E\frac{d^3N}{dk_i^3}=2<\frac{dN}{dy_i}>K_0(\frac{m_{i_T}}{T_{ps}}%
)\;.  \label{gkk}
\end{equation}

To compare theoretical correlation functions with data projected
onto two of the six dimensions, we computed the projected correlation
function trying to mimic what is done in the experiment, i.e.,
\begin{equation}
C_{\!proj}(q_T,q_L)\!=\!\!\!\frac{ \int d^3 k_1 d^3 k_2 P_2({\vec
k}_1,{\vec k}_2) A_2(q_T,q_L;{\vec k}_1,{\vec k}_2)} {\int d^3 k_1
d^3 k_2 P_1({\vec k}_1)P_1({\vec k}_2) A_2(q_T,q_L;{\vec
k}_1,{\vec k}_2)} \;,\label{c35}\end{equation} where $P_1$ and
$P_2$ are, respectively, the single- and two-pion inclusive
distributions. It is essential to have perfect correspondence
between the experimental information and the theoretical estimates
concerning the number of dimensions into which the correlation
data is projected, the cuts in momentum and rapidity, the sizes of
the bins, etc. $A_2$ is the experimental two-particle binning and
acceptance function, through which we approximate the theoretical
estimate to the empirical cuts.

We should notice that, experimentally, the two-particle correlation
function in high energy collisions is obtained by measuring the
following ratio
\begin{eqnarray}
C_{exp}(i,j) &=& {\cal N}_{exp} \times \frac{A(i,j)}{B(i,j)}
\nonumber
\\
\Delta C_{exp}(i,j) &=& C_{exp}(i,j) \sqrt{ \left[\frac{\Delta A(i,j)}{A(i,j)}\right]^2 +
\left[\frac{\Delta B(i,j)}{B(i,j)}\right]^2 }\nonumber\\
\; \; .\label{cexp}\end{eqnarray} The numerator $A(i,j)\pm \Delta
A(i,j)$ corresponds to the combination of pairs of identical
particles from {\sl the same} event, and the denominator,
$B(i,j)\pm \Delta B(i,j)$, represents the background; ${\cal
N}_{exp}$ is an experimental normalization. Historically, the
background for identically charged pions have been the combination
of unequally charged ones. However, later it was realized that
$\pi^+ \pi^-$ could frequently come from the decay of resonances,
which would distort the background and cause strange pattern for
the correlation function built in this way. They soon realized
that a better way to construct the background was to combine
identically charged particles, but from different events. Another
possibility used sometimes is a Monte Carlo simulated background,
taking into account the experimental cuts and acceptance.

\subsection{II.2.2~ Non-ideal IOC}

The non-ideal picture mentioned before referred to the underlying
effects that would be important to incorporated into the
interferometric analysis, even restricting the attention to
completely chaotic sources. For instance, the $\pi^-$ rapidity
distribution at 200 AGeV was clearly not uniform, as assumed in
the asymptotic Bjorken picture, but would be better described by a
Gaussian with width $Y_c \approx 1.4$\cite{NA35}. On the other
hand, a large fraction of the pions could arise from the decay of
long lived resonances, such as $\omega$, $K^*$, $\eta$, etc, as
was suggested by Grassberger\cite{grass}. In coordinate space, the
finite nuclear thickness, together with resonance effects, could
lead to a large spread ($\Delta \tau$) of the freeze-out proper
times, and to a wide distribution of transverse decoupling radii
($R_T$). In phase-space, there is not the perfect correlation
between the space-time and energy-momentum rapidity variables
present in the ideal Bjorken picture. Instead, as suggested by the
ATTILA version of the Lund model, they would be better related by
a Gaussian with finite width $\Delta \eta$. Besides, other
correlations may have to be considered if collective hydrodynamic
flow occurs, for instance, between transverse coordinates (${\vec
x}_\perp$) and transverse momentum component (${\vec p}_\perp$).
All these effects together were generically called as the
non-ideal picture, which is equivalent to considering a more
realistic picture than the one idealized by the Bjorken in his
version of the 1-D hydrodynamics. The phase-space distribution
representing these effects together can be obtained from Eq.
(\ref{ioc}) by replacing
%\begin{eqnarray}
%D(x,p)\!&\propto&\!\tau
% \exp\left\{-\frac{\tau^2}{ \Delta \tau^2}-\frac{(y-y^*)^2}{2 Y_c^2}
%-\frac{(\eta - y)^2}{2 \Delta \eta^2}
%-\frac{x_T^2}{R_T^2}\right\} \nonumber\\
%&&  \delta(E - E_{\vec p})
%\delta^2({\vec p}_T)
%\\;.\label{non-ideal}\end{eqnarray}
\begin{eqnarray}
&&\rho \frac{1}{\tau_f} \delta(\tau_f -
\tau) \delta(\eta - y) \rightarrow
\frac{2}{\Delta \tau^2} \exp\left(\frac{-\tau_f^2}{\Delta
\tau^2}\right) \times \nonumber \\
&&\exp\left[-\frac{(y-y^*)^2}{2 Y_c^2}\right]
\frac{1}{\sqrt{2\pi}\Delta \eta}\exp\left[-\frac{(\eta-y)^2}{2 \Delta\eta^2}\right]
\;.\label{non-ideal}\end{eqnarray}
Besides the modification in Eq.(\ref{non-ideal}), there is a major
correction to be added, i.e., the effect of  long-lived resonances
decaying into pions.  This can be included in the semiclassical
approximation\cite{qm88,gyupad89}.
The pion freeze-out coordinates, $x_a^\mu$, can be
related to the parent resonance
production coordinates, $x_r^\mu$, through
\begin{equation}
 x_a^\mu =x_r^\mu + u_r^\mu \tau\; \; ,
\label{clas} \end{equation}
where $u_r^\mu$  is the resonance four velocity and
$\tau$ is the proper time of its decay.
Summing over resonances $r$ of widths $\Gamma_r$,
and averaging over their decay proper times,
we obtain, instead of Eq.(\ref{cegij}) the final
expression\cite{qm88,gyupad89}
\begin{equation}
G(k_1,k_2) \approx \langle \; \sum_{r} f_{\pi^-/r} \frac{\exp(i q
x_r -K u_r/ T_r)}{\left(1-iq u_r/ \Gamma_r \right)} \;\rangle \;
\; , \label{ceresgij}\end{equation} where $f_{(\pi^-/r)}$ is the
fraction of the observed $\pi^-$'s arising from the decay of a
resonance of type $r$, and the temperature $T_r$ characterizes the
decay distribution of that resonance. According to the Lund model,
the main resonance contributing to the negative pion yield at
CERN/SPS energies are $f_{(\pi^-/\omega)}=0.16$,
$f_{(\pi^-/K^*)}=f_{[\pi^-/(\eta+\eta^\prime)]}=0.09$,
$f_{(\pi^-/\rho)}=0.40$, $f_{(\pi^-/direct)}=0.19$. Although we
included direct pions and the ones coming from $\rho$ decay
independently, they are hardly distinguishable, since $\rho$'s
decay very fast.

All these effects combined were simulated in a Monte Carlo code,
named CERES, which was also able to include simplified subroutines
which mimicked the experimental cuts and
acceptance\cite{gyupad89,qm88,pg:nioc}.

\begin{figure}[!htb10]
\begin{center}
\includegraphics*[angle=0, width=9cm]{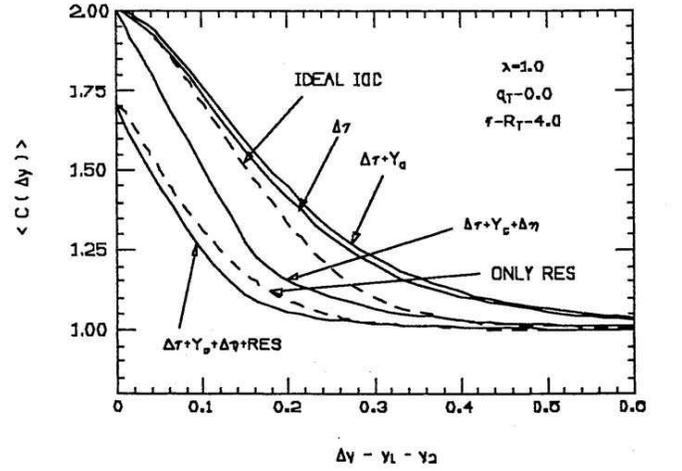}
\end{center}
\caption{\emph{\small Numerical results for the pion correlation
function versus $\Delta y$ are shown in the central rapidity
region, when the non-ideal effects are introduced, one by one.
This plot is a reproduction of the one originally published in
Ref.\cite{pg:nioc}.}}
\end{figure}

For explicitly demonstrating how the above results would show
themselves into the correlation function, we derived in Ref.
\cite{pg:nioc} several analytical relations where the non-ideal
effects were progressively introduced. We started with the ideal
Bjorken 1-D picture. Then, instead of the instant freeze-out
assumed there, we considered a spread ($\Delta \tau$) and,
subsequently, all the other effects mentioned above, one by one.
For fixed $q_T=0$, the results as a function of the rapidity
difference $\Delta y$ ($\approx q_L/m_T$ for $\Delta y \ll 1$),
can be seen in Fig. 10. For generating these plots, we estimated
the correlation function using the code CERES and adopting the
following values for the parameters: $R_T \approx 4$fm, $\tau_f
\approx 4$fm/c, fixed the intercept parameter $\lambda=1$, $\Delta
\tau =4$ fm/c; $Y_c \approx 1.4$, $\Delta \eta \approx 0.8$, plus
the above fractions $f_{\pi/r}$ from the Lund model, when adding
resonances.
%$\Delta p^2/m = T = 0.17$ GeV/c (from the pion inclusive distribution)

In the curves shown in Fig. 9, nevertheless, we had three sets of
parameters, corresponding to the different models compared there:
in the Lund resonance case, besides the fractions $f_{\pi/r}$, we
used $Y_c \approx 1.4$, $\Delta \eta \approx 0.7$, $R_T \approx
3$fm, $\tau_f \approx 3$fm/c, $y^*=2.5$ and fixed $\lambda=1$; for
mimicking the QGP model of Ref.\cite{gb}, we used no resonance,
$\tau_f=9.0$ fm/c, $R_T=3.3$ fm, $\Delta \eta \approx 0.76$,
assuming $Y_c \rightarrow \infty$ and $\lambda=1$.
%The other parameters are given in Fig. 9.

\bigskip
The results shown previously in  Fig. 9 clearly posed
another problem to the interferometric probes of high energy heavy
ion collisions: several very distinct dynamical scenarios could
lead to approximately the same final correlation function and similar
experimental HBT results.
%\vskip 6cm
\begin{figure}[!htb11]
\begin{center}
\includegraphics*[angle=0, width=8cm]{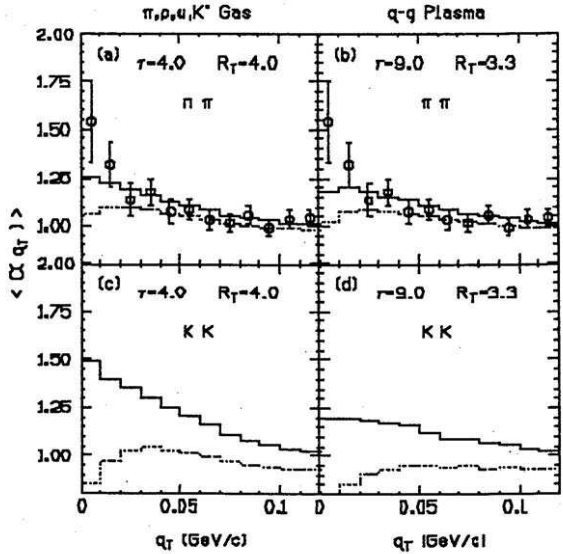}
\end{center}
\caption{\emph{ \small Comparison of two-pion, in parts (a) and
(b), and two-kaon projected correlation functions, in parts (c)
and (d) is considered as a function of the transverse momentum
difference, $q_T$. The plots are calculated in the central
rapidity region and with $q_L \le 0.1$ GeV. Solid (dashed) lines
indicate correlations without (with) Coulomb distortions. Parts
(a) and (c) correspond to predictions based on the Lund
model\cite{att}, and parts (b) and (d), to the plasma
hydrodynamical model of Ref.\cite{gb}. The pion data is from
Ref.\cite{NA35}. This figure was extracted from
Ref.\cite{gyupad89a}.}}
\end{figure}
One possibility to discriminate among
very different scenarios, suggested in Ref. \cite{gyupad89a},
was to explore distinct freeze-out geometries by
comparing pion and kaon interferometry.
This suggestion was motivated
by the fact that an entirely different set of hadronic resonances decay
into pions than into kaons.
In the first case, according to the ATTILA
version of the Lund fragmentation model, long lived resonances such as
$\omega$, $\eta$ and $\eta'$, contribute to the final pion yield,
whereas, in the second case, half of the kaons are produced by direct
string decay, and the other half by the decay of $K^*$. On the other
hand, in the QGP model considered previously in Ref.\cite{gyupad89}
for comparison, the freeze-out geometry of all hadrons was expected
to be about the same. In the case of pions, we saw from Fig. 9 that
both cases led to equally good results as compared to the experimental
points. In the case of kaons, then, an entirely different behavior
would be expected. Indeed, we see from Fig. 11 a more significant difference
between those two models, helping to separate long-lived scenarios
from those where HBT results were generated by the effect of
resonances.

The important role of resonances stressed above led us to
investigate how would be their influence for much smaller systems
at higher energies. For this, we looked into $\bar{p}p$ and $pp$
data from CERN/ISR\cite{lorstad}. For simplicity, in this
calculation, we assumed the ideal Bjorken picture to describe the
freeze-out distribution. Surprisingly, the result\cite{npa544}
turned out to be neither compatible with the absence of resonance
nor with the full resonance fractions predicted by the Lund
fragmentation model mentioned before. Instead, the data seemed to
be best described by the scenario with about half the resonance
fractions predicted by the Lund model, as can be seen in Fig.12.
\begin{figure}[!htb12]
\begin{center}
\includegraphics*[angle=0, width=8cm]{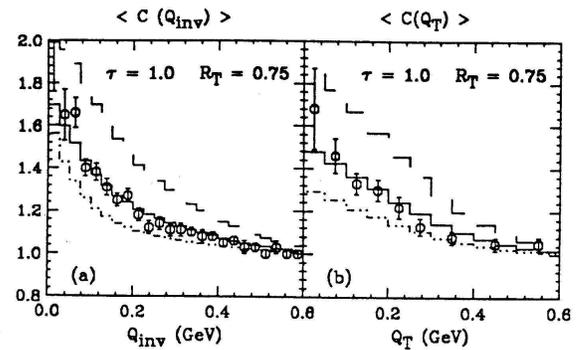}
\end{center}
\caption{\emph{ \small Negative pion correlation in $\bar{p}p$ and
$pp$ reactions at CERN/ISR energies are shown as a function of
$Q_{inv}$ on the left plot, and as function of $q_T$ (with $q_L
\le 0.15$ GeV/c) on the right plot. Dashed, solid and dot-dashed
histograms indicate the calculated correlation functions with,
respectively, no, half and full Lund resonance abundances. The
data points are from Ref.\cite{lorstad} on the left, and from Ref.
\cite{israfs}, on the right hand side. Both plots were extracted
from Ref.\cite{NPA525}.}}
\end{figure}

Meanwhile, we also derived a general and powerful
formulation\cite{pgg}, based on the Wigner density formalism,
which allows to treat complex system, by including an arbitrary
phase-space distribution (i.e., the momentum distribution, $g({\bf
p})$, and the space-time one, $\rho({\bf x}) $, could be
entangled) and multi-particle correlations. This formalism
corresponds to a semiclassical generalization of the n-particle
phase-space distribution, in which it is allowed for a Gaussian
spread of the coordinates around the classical trajectories, in
order to incorporate minimal effects due to the uncertainty
principle. Also, correction terms due to pion cascading before
freeze-out were derived using this semi-classical hadronic
transport model. Such terms, however, can be neglected if the mean
free path of pions is small compared to the source size, or if the
momentum transfers are small compared to the pion momenta.
%momentum positions, reconciling the classical phase-space picture
%with the uncertainly principle, in a semi-classical approach. Then,
%the initial and the final (decoupling) phase-space configuration
%were related by transport model.
The main result of that investigation can be summarized by the
following formula for the Bose-Einstein symmetrized n-pion invariant
distribution
\begin{eqnarray}
P_n ({\bf k_1},...,{\bf k_n}) &\propto& \;\langle \sum_\sigma
\prod_{j=1}^n \exp \left[ i(k_j-k_{\sigma_j}).x_j\right]\nonumber\\
&& \; \; \; \; \; \tilde{\delta}_\Delta(k_j-k_{\sigma_j},p_j)\; \rangle
\;,\label{wignerpgg1}\end{eqnarray}
with the smoothed delta function given by
\begin{eqnarray}
\tilde{\delta}_\Delta(k-k',p-K) &=&
\frac{\!\exp\{[p-\frac{1}{2}(k+k')]^2/(2 \Delta p^2)\}}{{(2\pi\Delta
p^2)}^{\!-3/2}}\nonumber \\
&& \times  \; \; \exp[\frac{\Delta x^2}{2}(k-k')^2] \;.
\label{deltapgg1}\end{eqnarray} The brackets $<...>$ denote an
average over the $8n$ pion freeze-out coordinates
$\{x_1,p_1,...,x_n,p_n\}$, as obtained form the output of a
specific transport model, such as a cascade\cite{gb} or the Lund
hadronization model\cite{att}. In the form written in Eq.
(\ref{wignerpgg1}) and (\ref{deltapgg1}), this formulation is
ideally suited for Monte Carlo computation of pion interference
effects. The smoothed delta function results from the use of
Gaussian wave packets with widths $\Delta x$ and $\Delta p$, which
depend on details of the pion production mechanism. The sum runs
over $n!$ permutations $\sigma=(\sigma_1,...,\sigma_n)$ of the
indices; $x, k, p, ...$ denote 4-vectors and all momenta are
on-shell. This is a generalization of the Wigner type of
formulation proposed in Ref.\cite{shuryak} and used in the pioneer
work of HBT in the group, discussed in Section II.1 and in Ref.
\cite{SandraPhD,sandra}. As a special limit, we found out that,
for minimum Gaussian packets, i.e., $\Delta x \Delta p =1/2$,
having $\Delta p \simeq m T_{ps}$, where $m$ is the particle mass
and $T_{ps}$ is the pseudo temperature of Eq. (\ref{tps}) we
recovered the interferometric relations derived within the
Covariant Current Ensemble formalism\cite{GKW}, which, for the
sake of simplicity, was adopted as a first approach to the study
of non-ideal effects on Interferometry described above.

An equivalent alternative way of expressing Eq.(\ref{wignerpgg1})
is \cite{pgg}
\begin{equation}
P_n ({\bf k_1},...,{\bf k_n}) \approx \;\langle C_{N,n}\rangle
\sum_\sigma \prod_{j=1}^n D_\Delta[q(j,\sigma_j),{\vec K}(j,\sigma_j)]
\;,\label{wignerpgg2}\end{equation}
with
\begin{equation}
D_\Delta(q,{\vec K})=\int d^4 x \int d^3p \; e^{iqx} D(x,{\vec p})
\tilde{\delta}_\Delta(q,p-K)
,\label{deltapgg2}\end{equation}
where $D(x,{\vec p})$ can be given by, for example, Eq.(\ref{ioc}),
and $C_{N,n}\equiv N!/(N-n)!$, where $n=2$ in case of two-pion
correlations, and $N$ is the multiplicity of the event.

\bigskip
An interesting simple point explicitly demonstrated in Ref.
\cite{pgg}, is the dependence of the effective transverse radius
on the average momentum of the pair, which was already shown in
Fig.7, as one of the results of Ref.\cite{sandra}, and also
suggested in \cite{pratt}. This dependence on $K_T$  appears
through the time dependence of the emission process. The
demonstration was done by means of a simple Gaussian example, as
in Eq. (\ref{c12nr}). For better understanding it, we should
recall the definition of the average 4-momentum of the pair,
$K^\mu$, and their difference, $q^\mu$, i.e.,
\begin{equation}
q^\mu = (k^\mu_1-k^\mu_2)\; ; \;  K^\mu =
\frac{1}{2}(k^\mu_1+k^\mu_2) \;. \label{qK1}\end{equation}
From the
above relations, it immediately follows that
\begin{equation}
q^\mu K_\mu = q^0 K_0 - {\bf q . K} \equiv 0 \rightarrow q^0 =
\frac{{\bf q.K}}{E_K}=
\frac{{\bf q_T.K_T}+q_L K_L}{E_K}
\;, \label{q)}\end{equation}
where $E_K = \sqrt{{\bf K}^2+m^2}$ was written for the sake of simplicity.

Propagating the above result into the Gaussian correlation function,
we get
\begin{eqnarray}
R^2_{T_{\small eff}} &\approx& 2 R_T^2 + (\Delta \tau)^2 ({\bf
K}^2_T/E_K^2)
\nonumber\\
R^2_{L_{\small eff}} &\approx& R_L^2 + (\Delta \tau)^2
(K^2_L/E_K^2)
\;. \label{RTRL}\end{eqnarray}

The results on Eq. (\ref{RTRL}) show that the time spread of the source freeze-out
generally enhances the effective size measured by interferometry.

I should remark that several contributions and invited talks
presented in international conferences in the period are being
omitted here, due to the lack of space. I would address to the
Quark Matter Conference proceedings for that, as well as the
proceedings of the RANP Conference and of Hadron Physics.

\subsection{II.3~ DISCRIMINATING DIFFERENT DYNAMICAL SCENARIOS}

The coincidental agreement with data of two opposite scenarios,
such as the resonance gas and the QGP  discussed before, in Fig.
9, stressed the necessity of finding other means to more clearly
discriminate among different decoupling geometries. Although the
comparison of kaon with pion interferometry was shown to be
helpful, as seen in Fig. 11, it still lacked from more
quantitative information. Then, how to disentangle different
models in a more precise way?
\begin{figure}[!htb13]
\begin{center}
\includegraphics*[angle=0, width=9cm]{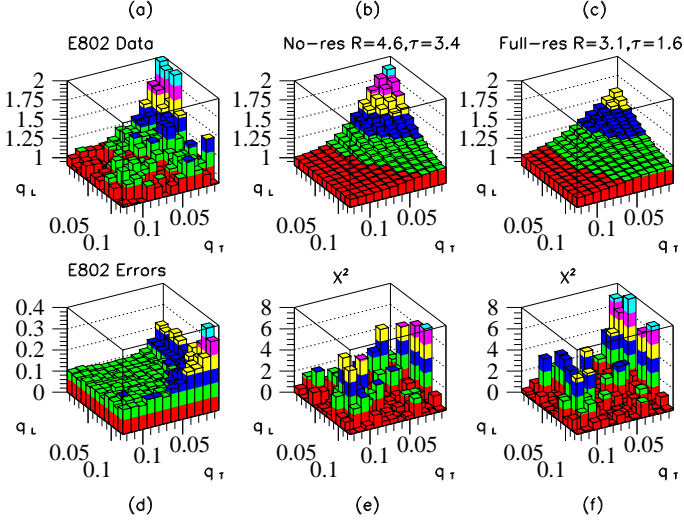}
\end{center}
\caption{\emph{ \small $\pi^-\pi^-$ correlations in central Si+Au
collisions is shown as a function of $q_T$ and $q_L$. The
preliminary E802 data\cite{e802} corrected for acceptance and
Coulomb effects are shown in part (a). Parts (b) and (c) show
theoretical correlation functions filtered with the E802
acceptance. They correspond, respectively, to cases without and
with resonance production. This figure was extracted from
Ref.\cite{PLB348}.}}
\end{figure}

In order to answer this question, M. Gyulassy and myself developed a
method, in which a 2-D $\chi^2$ analysis was performed, comparing
two-dimensional theoretical and experimental pion interferometry
results\cite{PLB348}. For illustrating the method we performed the
calculations using the code CERES mentioned above, for two very
distinct scenarios. The first one considered the effects of
resonances decaying into pions, including Lund resonance fractions.
The other one ignored the contribution of resonances. The data
points were kindly sent to us by Richard Morse, from the BNL/E802
Collaboration, and corresponded to $Si+Au$ collisions at 14.6
GeV/c\cite{e802}, as measured at the BNL/AGS. I will summarize the
method by recalling Eq. (\ref{c35}) and Eq. (\ref{cexp}). The E802
experimental acceptance functions for two particles was approximated
by
\begin{eqnarray}
&&\!\!A_2(q_T,q_L;{\vec k}_1,{\vec k}_2) = A_1({\vec k}_1) A_1({\vec k}_2) \Theta(20-|\phi_1-\phi_2|)
\nonumber \\  &\;&   \; \; \; \; \; \; \; \; \; \delta(q_L-|k_{z1}-k_{z2}|)
\delta(q_T-|{\vec k}_{T_1}-{\vec k}_{T_2}|)
\;\; .\label{A2}\end{eqnarray}
The angles are measured in degrees and the momenta in GeV/c.
The single-inclusive distribution  cuts were specified by
\begin{eqnarray}
A_1({\vec k})&=& \Theta(14<\theta_{lab}<28)\nonumber \\
& \;& \Theta(p_{lab}<2.2\;{\rm GeV/c})  \Theta(y_{min}>1.5)
\; \; .\label{a1}\end{eqnarray}
%                         & & \Omega (\theta_{lab},\phi,p_{lab},y_{min},
%y_{max},\Delta_{\phi},p_{T_{lab}},p_z) \; \; ,\eeqar{a2}
% \; \; ,\eeq{a2}
The input temperature matching the
experimentally observed pion spectrum was $T\approx170$ MeV.

For the purpose of performing a quantitative analysis of the compatibility of
different scenarios with data, we computed
the $\chi^2$ goodness of fit on a two-dimensional
grid in the $(q_T,q_L)$ plane, binned with $\delta q_T=
\delta q_L=0.01$ GeV/c. The $\chi^2$ variable was computed (as
suggested by W. A. Zajc)
through the following relation\cite{PLB348}
\begin{equation}
\chi^2(i,j) =\frac{ [A(i,j) - {\cal N_\chi}^{-1} C_{th}(i,j)  B(i,j)]^2 }
{ \{ [\Delta A(i,j)]^2 + [ {\cal N_\chi}^{-1} C_{th}(i,j)
\Delta B(i,j)]^2 \} }
\; \; ,\label{chi2}\end{equation}
where ${\cal N}_\chi$ is a normalization factor
estimated as to minimize the
average $\chi^2$, which depends on the range in the
$q_T,q_L$ plane under analysis.
The minimization of the average $\chi^2$ was performed by exploring
the parameter space of the transverse radius $R_T$ and the time
$\tau$, and computing the $<\chi^2>$, averaging over a 30x30 grid in
the relative region $q_T,q_L < 0.3$ GeV/c of relevant HBT signal. In
the vicinity of the minimum we determined the parameters of the
quadratic surface
\begin{equation}
\langle \chi^2(R_T,\tau)\rangle =\chi^2_{min} +  \alpha (R_T-R_{T_0})^2
 + \beta (\tau-\tau_0)^2 \; \; . \label{parab}\end{equation}

\begin{center}
{\bf TABLE 1: 2D-$\chi^2$ Analysis of Pion Decoupling Geometry} \\

%\vskip 2.9cm

%\begin{tabular}{||c||c|c||}
\begin{tabular}{|c|c|c|}
%\hline\hline
\hline
%  &  &  \\
$\chi^2(R_{T},\Delta \tau)$  & No Resonances & LUND Resonances \\
%  &  &  \\
\hline\hline
\multicolumn{3}{|c|}{ E802 Data  Gamow Corrected}\\
\hline\hline
$|\chi^{2}_{min}-1|/\sigma$ & 2.1 & 2.2 \\
\hline
$R_{0_T}$ & 4.6$\pm$ 0.9 & 3.1$\pm$ 1.3 \\
\hline
$\Delta\tau_0$ & 3.4 $\pm$ 0.7 & 1.6$\pm$ 1.0\\
\hline
$\alpha$ & 0.027 & 0.014  \\
\hline
$\beta$ & 0.042 & 0.023 \\
%\hline\hline
\hline
\end{tabular}
\end{center}

The quantitative differences could be seen in a 3-D plot of the
correlation function, projected in terms of $q_T$ and $q_L$, shown
in Fig. 13.  The main results in the case of Gamow corrected data
(i.e., where the HBT signal has recovered at small values of $q$ by
multiplying by the inverse of the Gamow factor, $\Upsilon(q)^{-1}$,
defined in Eq. ({\ref{gamow})), are shown in Table 1. For a more
complete discussion of the method, I would address the full article,
in Ref.\cite{PLB348}.
%
%\vskip 6cm
\begin{figure}[!htb14]
\begin{center}
\includegraphics*[angle=-90, width=7cm]{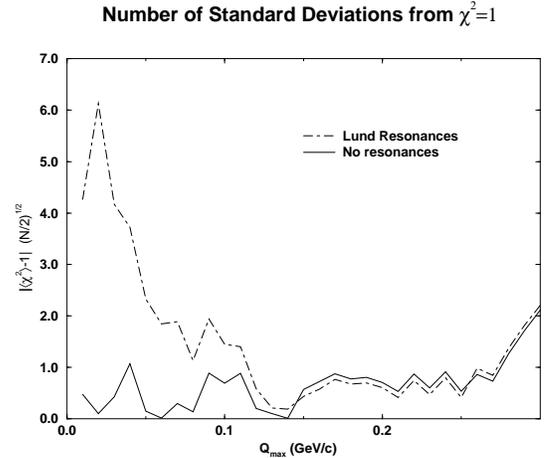}
\end{center}
\caption{\emph{ \small The plot shows the number of standard
deviations from unity, of the average $<\chi^2>$ per degree of
freedom, as a function of the range $Q_{max}$ of the analysis in the
$(q_T,q_L)$ plane. This plot was extracted from Ref.\cite{PLB348}.}}
\end{figure}
We see from the upper plots in Fig. 13 that the 2-D correlation
function for the non-resonance case is clearly different from the
resonance scenario but only at very small values of the momentum
difference $q_T$ and $q_L$. Nevertheless, if we look into the
lower panel where we plotted the $\chi^2$ distribution of the
theoretical curves compared to the experimental points, we see
that the distinction is blurred by the large fluctuations of data,
mainly at the edge of the acceptance. The most efficient measure
of the goodness of the fit in this case was obtained by studying
the variation of the average $<\chi^2>$ per degrees of freedom in
the ($q_T,q_L$) plane with respect to the unity, i.e.,
$|<\chi^2_{\small min}>-1|/\sigma$. In this way, we found out that
the resolving power of the distinction between different scenarios
was magnified, as shown in Fig. 14.

\bigskip

The method was further tested later, in a more challenging
situation\cite{padrol}, by comparing the interferometric results
of $K^+ K^+$ of two distinct scenarios, i.e., Lund predicted
resonances (only $K^*$'s and direct kaons contribute significantly
to this particle yield) with the non-resonance case. This was done
by Cristiane G. Rold\~ao in her Master Dissertation, under my
supervision. The data on $K^+K^+$ interferometry from $Si+Au$
collisions at $14.6$ GeV/c was sent to us by Vince Cianciolo, from
E859 Collaboration (an upgrade of the previous E802 experiment at
BNL/AGS).

\vskip 0.7cm
\begin{center}
{\bf TABLE 2: 2D-$\chi^2$ Analysis of Kaon Decoupling Geometry} \\
\begin{small}

\begin{tabular}{|c|c|c|}
%\hline\hline
\hline
$\chi^2(R_{T},\Delta \tau)$  & No Resonance  &
LUND Res.  \\
  & $(f_{K_{dir}}=1)$ & $(f_{K_{dir}}=f_{K/K^*}=0.5)$ \\
\hline\hline
%\multicolumn{3}{||c||}{Optimized $R_T$ and $\Delta \tau$}\\
\multicolumn{3}{|c|}{Optimized $R_T$ and $\Delta \tau$}\\
\hline\hline
$\langle \chi^{2}_{min} \rangle _{30 \times 30}$ & 1.03 & 1.02 \\
\hline
$\langle \chi^{2}_{min} \rangle _{10 \times 10}$ & 1.17 & 1.30 \\
\hline
$R_{T0}$  & 2.19$\pm$ 0.76 & 1.95 $\pm$ 0.89 \\
\hline
$\Delta \tau_0 $ & 4.4$\pm$ 2.0 & 4.4$\pm$ 2.6 \\
\hline
$\alpha$ & 0.0410 & 0.0299  \\
\hline
$\beta$ & 0.0058 & 0.0034 \\
%\hline
%\multicolumn{3}{|c|}{Optimized $R_T$ ($\Delta \tau=0)$}\\
%\hline
%$\langle \chi^{2}_{min} \rangle_{30 \times 30}$ & 1.29 & 1.33 \\
%$\langle \chi^{2}_{min} \rangle_{10 \times 10}$ & 4.04 & 2.92 \\
%%$R_{0_T}$ &  10.6 $\pm$ 8.9 & 4.8 $\pm$ 0.9 \\
%$R_{0_T}$ & $\sim$ 10.6 & $\sim$ 4.8 \\
%$\alpha$ & 0.0003 & 0.0280  \\
\hline
\end{tabular}

\end{small}
\end{center}

\bigskip
The acceptance function for the E859 experiment was
approximated\cite{e859} by
\begin{eqnarray}
&&A_2(q_T,q_L;{\boldmath k}_1,{\boldmath k}_2) =
A_1({\boldmath k}_1) A_1({\boldmath k}_2)
\Theta(22-|\phi_1-\phi_2|)\nonumber\\
 && \; \; \; \; \; \; \; \; \; \; \; \;\delta(q_L-|k_{z1}-k_{z2}|)
  \delta(q_T-|{\boldmath k}_{T_1}-{\boldmath k}_{T_2}|)
\; \; .\label{a2exp}\end{eqnarray}
The angles were measured in degrees and the momenta in GeV/c.
The single inclusive distribution  cuts are specified by
\begin{eqnarray}
A_1({\boldmath k}) &=& \Theta(14 <\theta_{lab}< 28)
\Theta(p_{lab}< 2.9\;{\rm GeV/c}) \nonumber\\
&& \; \; \;  \; \; \; \; \; \; \Theta(y_{min}> 0.75) \; \;
.\label{a1exp}\end{eqnarray} In the case of kaons, the input
temperature matching the experimentally observed kaon spectrum was
$T\approx180$ MeV.

%\vskip 6cm
\begin{figure}[!htb15]
\begin{center}
\includegraphics*[angle=0, width=8.5cm]{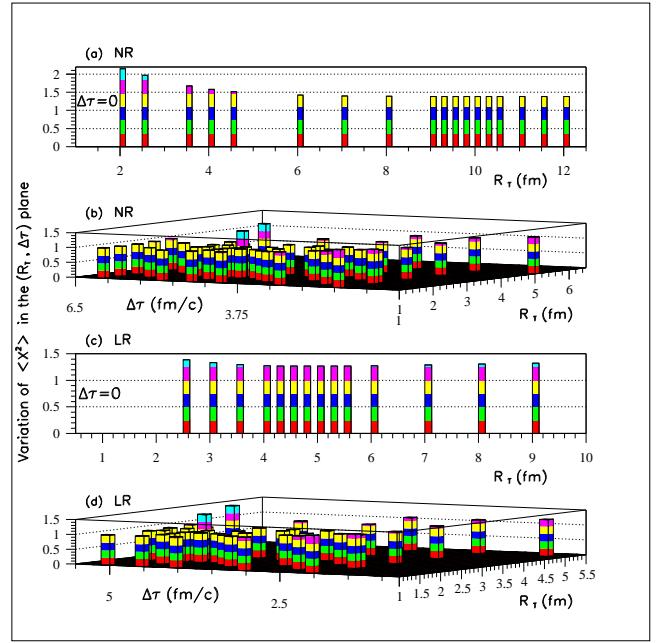}
\end{center}
\caption{\emph{ \small Zone in the ($R_{T},\Delta\tau$) plane
investigated, leading to the determination of the most probable
region where the minimum $\langle \chi^2 \rangle$, associated to
($R_{T},\Delta\tau$), could be located. Parts (a) and (b)
correspond to cases where the contribution of $K^\star$ were
ignored. Parts (c) and (d) were estimated including their
contribution to the kaon yield. In parts (a) and (c) we fixed
$\Delta\tau=0$, and optimized only $R_{T}$. Figure extracted from
Ref.\cite{padrol}.}}
\end{figure}

It was expected to be harder to differentiate both scenarios due
to the lack of contribution from long-lived resonances. It was
found that they could still be separated, with data favoring the
non-resonance scenario at the 14.6 $Si+Au$ collisions (BNL/AGS).
As in the two- pion interferometric analysis, the variation of the
average $<\chi^2>$ per degrees of freedom in the ($q_T,q_L$) plane
was the significant quantity to look at, as can be seen n Fig. 16.
The main fit results found in this analysis are summarized in
Table 2. For more details, see Ref.\cite{padrol}.

%\vskip 6cm
\begin{figure}[!htb16]
\begin{center}
%\hskip-3mm
\includegraphics*[angle=0, width=8.7cm]{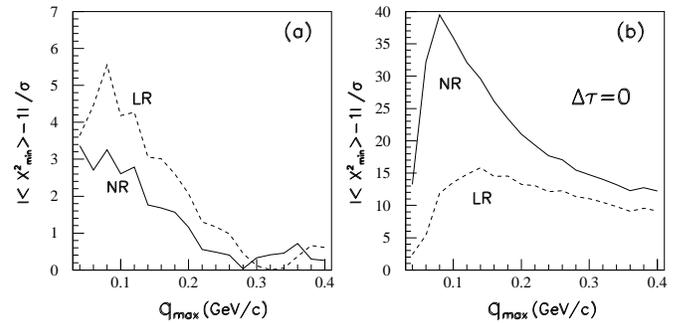}
\end{center}
\caption{\emph{ \small The number of standard deviations from unity
of the average $<\chi^2>$ per degree of freedom  is shown for increasing
number of bins. In part (a), the transverse radius parameter, $R_T$,
and the freeze-out spread, $\Delta \tau$,
used for generating the theoretical plots, were optimized. In (b),
only $R_T$ was optimized, whereas $\Delta \tau$ was fixed to be zero.
Figure extracted from Ref.\cite{padrol}.}}
\end{figure}

The tests also ruled out the possibility of a zero decoupling
proper-time conjectured by that experimental results of AGS/E859
Collaboration\cite{e859}. This can be seen from part (b) of Fig.
16: by fixing $\Delta \tau$ to be zero the numerical deviations of
the average $<\chi^2>$ per degrees of freedom are completely
meaningless.

\subsection{II.4~ SONOLUMINESCENCE BUBBLE}

In the beginning of this review, we saw that the HBT
interferometry was originally proposed for measuring the large
sizes (of order $R \sim 10^{10}$ m) of stelar sources in
radio-astronomy . On the other hand, in the totality of the cases
discussed here so far, the dimensions went down to the order of
the hadronic or to the nuclear size (roughly, $R \sim 10^{-15}$
m).
\begin{figure}[!htb17]
%%\begin{center}
%\includegraphics*[angle=0, width=8cm]{sonolumifig1a}
%%\includegraphics*[angle=0, width=7.5cm]{sonolumifig1}\\
%%\includegraphics*[angle=0, width=7.5cm]{sonolumifig2}\\
%\includegraphics*[angle=0, width=8cm]{sonolumiFigs9612046}
%%%\end{center}
  \resizebox{14pc}{!}{\includegraphics{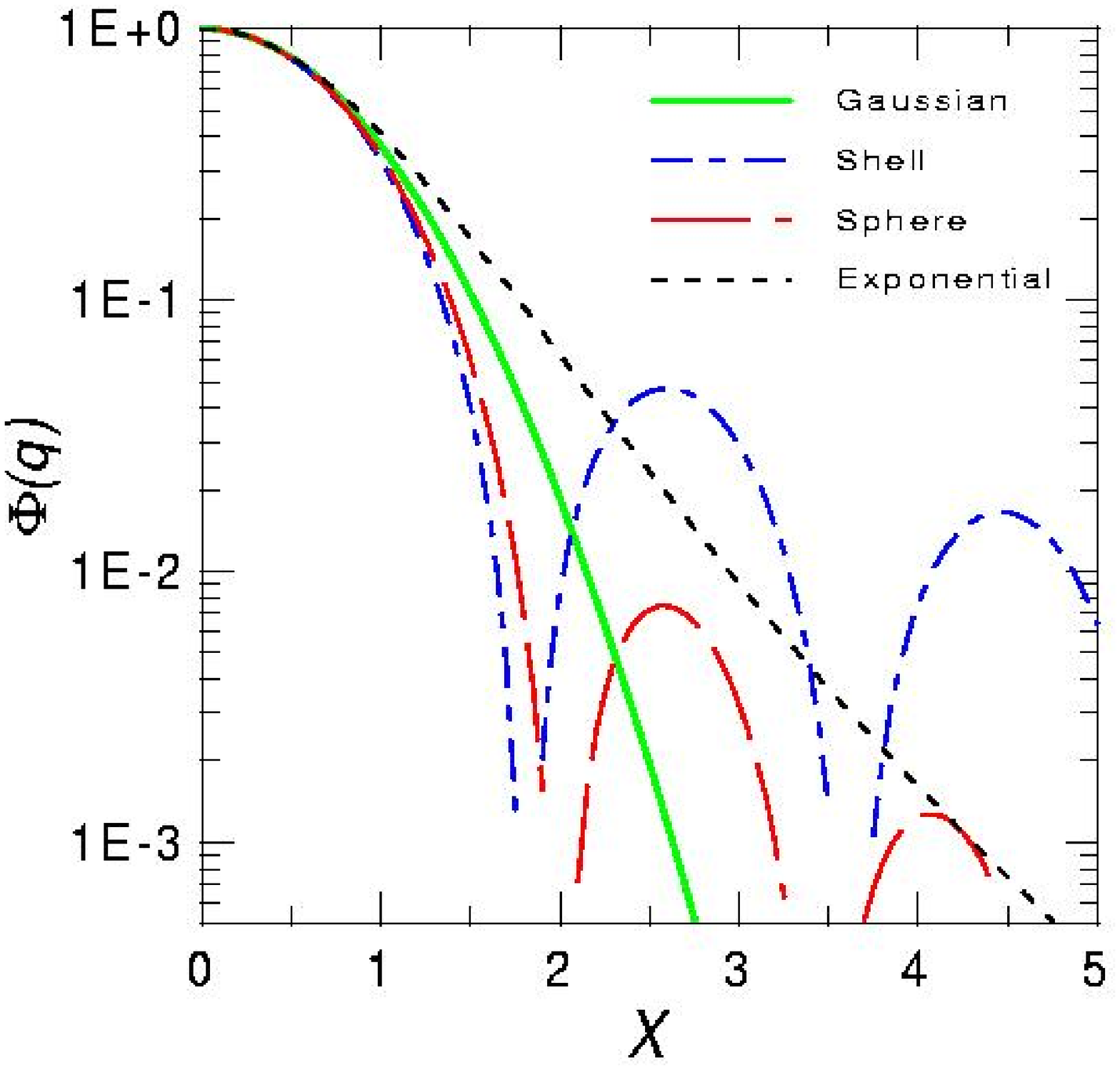}}\\
  \resizebox{15pc}{!}{\includegraphics{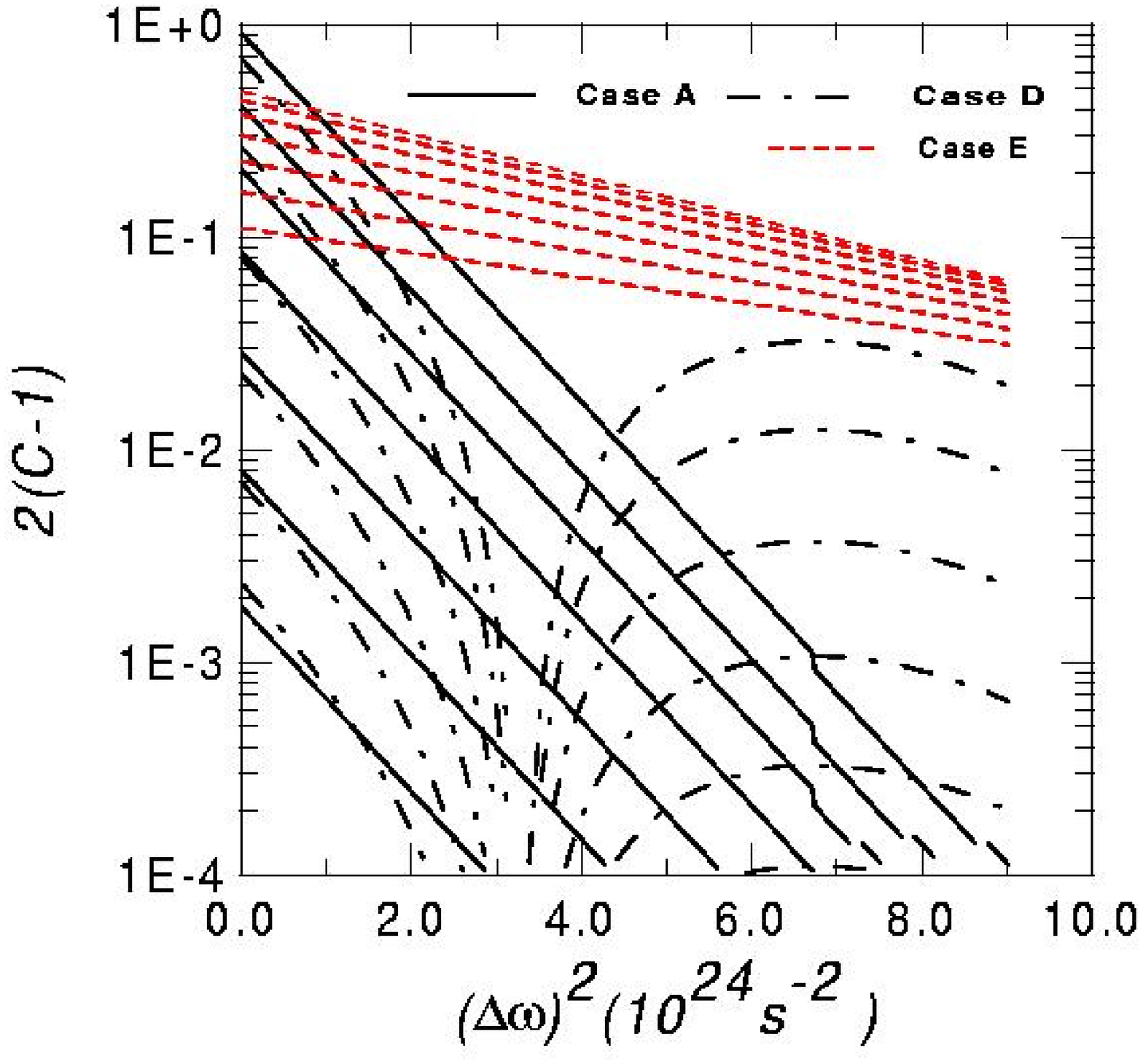}}
\caption{\emph{\small The upper part of the figure shows
correlation functions plotted as function of $\Delta \omega =
\omega_1-\omega_2 = c(k_1-k_2)$. Cases A to E correspond,
respectively, to the first up to the fifth examples in Table 2.
The lower part, shows the geometrical form factor $\Phi(q)$
plotted as a function of $X=\sqrt{-(1/2)[d^2\Phi(0)/dq^2]} q$.
Figures extracted from Ref.\cite{hkp}.}}
\end{figure}
In between these two very different scales, Yogiro Hama, Takashi
Kodama, and myself \cite{hkp} discussed in 1995 an interesting
approach to a beautiful problem, in a distinct environment. The
focus was in a small sonoluminescent bubble, whose radius would
lie in the range $R \sim 10^{-5}$m.
The phenomenon had been discovered long time ago, in 1934, at the
Univ. of Cologne, but its single bubble version was found out by
Gaitan et al.\cite{sono1} only in 1988. In this last case, a
single bubble of gas (usually filled with air) formed in water, is
trapped by standing acoustic waves, contracting every 10-12
pico-sec approximately, and simultaneously emitting light. In
other words, the sonoluminescence process converts the acoustic
energy in a fluid medium into a short light pulse emitted from the
interior of a collapsing small cavitating bubble\cite{sono1}.
%Since the discovery of
%the technique for trapping a single cavitating bubble by standing acoustic
%wave\cite{sono1}, many remarkable properties have been revealed
%\cite{sonoprop}
%(see Ref.\cite{hkp} for an extensive list of references).
The spectrum
of emitted light is very wide, extending from the visible to the ultraviolet
regions.
%The acoustic-luminous conversion process requires some mechanism of
%extraordinary concentration of energy and is apparently related to keeping
%the sphericity of the collapsing bubble while its radius shrinks more than
%one order of magnitude.
An important fact is that the light emission takes place within a
very short period of time. The small size of the emission region
and the short time scale of the emission process make it difficult
to obtain precise geometrical and dynamical information about the
collapsing bubble. Besides, a particular aspect of the emission
process is still controversial: some authors attribute the light
emission to the quantum-electrodynamic vacuum property based on
the dynamical Casimir effect\cite{casimir}, whereas others
consider that thermal processes\cite{thermal}, such as a
black-body type of radiation, should be the natural explanation.
%Besides this two, another
%candidate explanation could be non-equilibrium atomic collision processes.
%
In Ref.\cite{hkp}, we suggested that two-photon interferometry
could shed some light to the problems raised above, by estimating
the very small size and life-time of the single bubble
sonoluminescence phenomenon. In fact, the simple existence of a
HBT type of signal from such a collapsing bubble selects the
scenario of explanation between the two classes mentioned above.
This is possible because the emission processes are opposite in
the case of thermal models and in the Casimir based ones. In the
first case, the emission is chaotic, a condition that is essential
for observing the two-identical particle HBT correlations, and an
interferometric pattern would be observable. For the Casimir type
of models, however, the emission is coherent and no HBT effect
would be observable since, in that case, the correlation function
acquires the trivial value $C(k_1,k_2)=1$ for all values of $k_i$.

In this work we neglect all the dynamical effects discussed in the
previous studies, and considered that the space and time
dependence would be decoupled as well. We adopted the notation
used in \cite{Heinz2}, applied for our case.
\begin{equation}
C(\vec{k}_1,\vec{k}_2)=1+\frac 12\frac{| \tilde{S}(\vec{q},\vec{K}%
)|^2}{\tilde{S}(0,\vec{k}_1)\,\tilde{S}(0,\vec{k}_2)}\,\;\;,
\label{C(k1k2)}
\end{equation}
where
\[
\tilde{S}(\vec{q},\vec{K})=\int d^4x\,e^{-iqx}\rho (x)\,j^{*}(\vec{k}_1)
j(\vec{k}_2) \;,
\]
$\vec{q}=\vec{k}_1-\vec{k}_2$\thinspace , $\vec{K}=(\vec{k}_1+\vec{k}_2)/2\,$%
, and $j(\vec{k}_i)$ is the amplitude for the emission of a photon
with the wave-vector $\vec{k}_i$ at the source point $x$. The
factor 1/2 in the second term of Eq. (\ref{C(k1k2)}) comes from
the spin 1 character of the photon\cite{Heinz2}.

We suggested a set of candidate
models that could try and describe the system, as summarized in
Table 3.

\begin{center}
{\bf TABLE 3}\\
Analytic expressions for some source parameterizations
and the corresponding correlation functions are shown.
\end{center}

\[
\begin{tabular}{|l|l|}
\hline
$\;\;\;\;\;\;\;\;\;\;\;\;$ Source $\rho(r,t)$ & $\,\,\;\;\;\;\;\;\;\;\;\;\;\;C(\vec{k}%
_1,\vec{k}_2)-1$ \\
%$\;\;\;\;\;\;\;\;\;\;\; \rho(r,t)$ &
\hline\hline
{\bf A:} $e^{-r^2/2R^2}\;e^{-t^2/2\tau ^2}$ & $\,e^{-(\Delta \omega )^2\tau
^2\;}e^{-q^2R^2}/2$ \\
 \hline
{\bf B:} $\delta (r-R)\;e^{-t^2/2\tau ^2}$ & $\,e^{-(\Delta \omega )^2\tau
^2\,}[{\rm \sin }(qR)/(qR)]^2/2\,$ \\
\hline
{\bf C:} $\Theta (R-r)\;e^{-t^2/2\tau ^2}$ & $\;\;\;\;\; \left( 9/2 \right)
\,e^{-(\Delta \omega
)^2\tau ^2} \{(qR)\}^{-4} $ \\
& $\{\,[{\rm \cos }(qR)-{\rm \sin }(qR)/(qR)]\}^2$ \\
 \hline
{\bf D:} $e^{-r/R}\;\Theta (3\tau ^2-t^2)$ & $\left[ \sin (\Delta \omega
\sqrt{3}\,\tau )/\Delta \omega \sqrt{3}\,\tau \right] ^2$ \\
& $\;\left( 1+q^2R^2\right) ^{-4}/2$ \\
 \hline
{\bf E:}  & $9\left|
I\right| ^2/(8\mu ^6),$ \\
$\Theta (\dot{R}\,t-r)\,e^{-t^2/\tau ^2}\Theta (t)$& $I=-i\sqrt{\pi } [ (1+\mu z^{+})W(z^{+})-$ \\
& $(1-\mu z^{-})W(z^{-}) ] -2\mu $ \\
 \hline
\end{tabular}
\]

As far as I know, no HBT experiment has been made yet with the
photons emitted by the sonoluminescence bubble. T. Kodama and
members of his group had initiated the experiment to produce
single cavitating bubbles, with intension to perform the
interferometric test I have just described above. In any case, the
simple observation of a two-photon HBT correlation would be enough
to rule out one of the two classes of models that aim at
explaining the phenomenon, i.e., the ones based on the Casimir
effect, in which the emission is coherent. It would automatically
enforce the other class of thermal models, for which the emission
is chaotic.

\subsection{II.5~ CONTINUOUS EMISSION}

In Section II.2.1, we discussed the Bjorken Inside-Outside Cascade
(IOC)\cite{Bj} picture. It considers that, after high energy
collisions, the system formed at the initial  time $\tau =
\tau_0$, thermalizes with an initial temperature $T_{0}$, evolving
afterwards according to the ideal 1-D hydrodynamics, essentially
the same as Landau's version discussed in Section II.1, but with
different initial conditions. During the expansion, the system
gradually cools down and later decouples, when the temperature
reaches a certain freeze-out value, $T_{f}$. In this model there
is a simple relation between the temperature and the proper-time,
i.e., $ \tau \propto \tau_0 \left( \frac{T_0}{T_{f}} \right)^3 . $
%
%The momentum distribution of the produced hadrons in the idealized picture
%is computed by using the Cooper-Frye integral\cite{CF}
%\begin{equation}
%E\frac{d^3 N}{dp^3} = \int_{T_f} d\sigma_\mu p^\mu f(x,p)
%\;,\label{CooperFrey}\end{equation}
%over the hyper-surface $T=T_f$, where $d\sigma _\mu$ is the vector
%normal to the freeze-out surface, $p^\mu$ is the particle momentum,
%and $f(x,p)$ is its distribution function.

A very interesting alternative picture of the particle emission
was proposed by  Grassi, Hama and Kodama\cite{ghk}: instead of
emitting particles only when these crossed the freeze-out surface,
they considered that the process could occur continuously during
the whole history of the expanding volume, at different
temperatures. In this model, due to the finite size and lifetime
of the thermalized matter, at any space-time point $x^\mu $ of the
system, each particle would have a certain probability of not
colliding anymore. So, the distribution function $f(x,p)$ of the
expanding system would have two components, one representing the
portion already free and another corresponding to the part still
interacting, i.e., $ f(x,p)=f_{free}(x,p)+f_{int}(x,p) . $

In the Continuous Emission (CE) model, the portion of free
particles is considered to be a fraction of the total distribution
function, i.e., $ f_{free}(x,p)={\cal P}f(x,p)\Longrightarrow
f_{int}(x,p)=(1-{\cal P})f(x,p) $ or, equivalently,
\begin{equation}
f_{free}(x,p)=\frac{{\cal P}}{1-{\cal P}}f_{int}(x,p)\;\;.  \label{ffree}
\label{ffreeint}\end{equation}

The interacting part is assumed to be well represented by
a thermal distribution function
\begin{equation}
f_{int}(x,p)\approx f_{th}(x,p)=\frac g{(2\pi )^3}\frac 1{\exp {\ \left[
p.u(x)/T(x)\right] }\pm 1}\;\;,  \label{fth}
\end{equation}
which poses a constraint on the applicability of the picture,
since by continuously emitting particles the system would be too
dilute to be considered in thermal equilibrium in its late stages
of evolution. In Eq. (\ref{fth}), $u^\mu $ is the fluid velocity
at $x^\mu $ and $T$ is its temperature in that point.

The fraction of free particles ${\cal P}$ at each space-time
point, $x^\mu$, was computed by using the Glauber formula, i.e.,
$%\begin{equation}
{\cal P}=\exp {\ \left( -\int_t^{t_{out}}n(x^{\prime })\sigma
v_{rel}dt^{\prime }\right) },  %\label{P}
$%\end{equation}
 where
$%\begin{equation}
t_{out}=t+(-\rho \cos \phi +\sqrt{R_T^2-\rho ^2\sin ^2\phi })/(v\sin \theta
).
$%\label{tout}\end{equation}

The model also considered that, initially, the energy density could be
approximated by a constant (i.e., $\epsilon =\frac{\pi ^2}{10}T_0^4$ for all
the points with $\rho \le R_T$ and zero for $\rho >R_T$). Then, the
probability ${\cal P}$ may be calculated analytically, resulting in
$%\begin{equation}
{\cal P}=(\tau /\tau _{out})^a;\;\;a\sim 3\frac{1.202}{\pi ^2}T_0^3\tau_0
\sigma v_{rel},
$%\label{PGl}\end{equation}
where $v_{rel}\approx 1$. The factor ${\cal P}$ can be interpreted
as the fraction of free particles with momentum $p^\mu$ or,
alternatively, as the probability that a particle with momentum
$p^\mu $ escapes from $x^\mu $ without further collisions.

Their early results for the spectra can be seen in Ref.\cite{ghk}
and in the review  by Frederique Grassi, in this volume, which is
a good source of further details and discussions on the Continuous
Emission model.

Later, in collaboration with F. Grassi,
Y. Hama, and O. Socolowski\cite{pghs}, we developed the
formulation for applying this new freeze-out criterium
into $\pi \pi$ interferometry. Naturally, we would like to
further explore if the above model
would present striking differences when compared to the standard
freeze-out picture (FO). One expectation would be that the space-time
region from which the particles were emitted would be quite
different in both scenarios. In particular, as we saw in Section
II.1 and II.2.2, a non-instantaneous emission process strongly influences
the behavior of the correlation function. Being so, a sizable
difference was expected when comparing the instant freeze-out
hypothesis and the continuous emission version, since in this last
one, the emission process is expected to take much longer.

For treating the identical particle correlation within the
continuous emission picture, instead of using the Covariant Current
Ensemble formalism discussed in item II.2, we adopted a different
but equivalent form for expressing the amplitudes in Eq. (\ref{cce})
as in Ref.\cite{ghk}. Then, the
single-inclusive distribution, can be written as
\begin{equation}
G(k_i,k_i)=\int d^4x\;{\cal D}_\mu \left[ k_i^\mu f_{free}\right]
\;\;,
\label{giice}
\end{equation}
where  ${\cal D}_\mu $ is the generalized divergence operator. In
Ref.\cite{ghk}, it was shown that, in the limit of the usual
freeze-out, Eq. (\ref{giice}) is reduced to the Cooper-Frye
integral, $E\frac{d^3 N}{dp^3} = \int_{T_f} d\sigma_\mu p^\mu
f(x,p)$, over the freeze-out hyper-surface $T=T_f$, being $d\sigma
_\mu$ the vector normal to this surface. Or equivalently, Eq.
(\ref{giice} ) in the instant freeze-out picture is reduced to Eq.
(\ref{gii}), with the currents given by Eq. (\ref{j0thermal}), or
even by Eq. (\ref{j0pseudo}), which is a simplified parametric
form of describing thermal currents used for obtaining analytical
results in the Bjorken picture. We see more easily that it is
indeed the limit if we replace the distribution function in
Eq.(\ref{fth}) by its Boltzmann limit. We see more clearly that
Eq. (\ref{gii}), or Eq. (\ref{gkk}) in the Bjorken picture, are
the natural limits of the proposed continuous emission spectrum,
in case of instant freeze-out.

Analogously, the two-particle complex amplitude is written
%, instead of Eq. (\ref{g12}),
as
\begin{equation}
G(k_1,k_2)=\int d^4xe^{iqx}\;\{{\cal D}_\mu \left[ k_1^\mu f_{free}\right]
\}^{\frac 12}\;\{{\cal D}_\mu \left[ k_2^\mu f_{free}\right] \}^{\frac 12%
}\;\;.  \label{g12ce}
\end{equation}

We saw above that the expression for the spectrum is reduced to
the one in the Covariant Current Ensemble formalism in the limit
of the instant freeze-out. Analogously, the above expression in
this limit should yield to the result in Eq. (\ref{g12}). We see
that this is indeed the case if we replace the individual momenta
$k_i^\mu$ in E.(\ref{g12ce}) by the average momentum of the pair,
$K^\mu=\frac{1}{2}(k_1^\mu+k_2^\mu)$. This replacement is even
more natural, if we remember that the $K^\mu$ is the momentum
appearing in the Wigner formulation of
interferomtry\cite{pratt,pgg,gb,heinz1,hpb}. And, it was shown in
Ref. \cite{pgg} that this also is reduced to the Covariant Current
Ensemble for minimum packets and having the packet width equated
to the pseudo-thermal temperature, as discussed in Section II.2.2.
When this is assumed, also a substantial simplification is
obtained in Eq.(\ref{g12ce}), which could then be written as
\begin{equation}
G(q,K) = \int d^{4}x \; e^{i q^\nu x_\nu}  \;{\cal D}_\mu
\left[ K^\mu f_{free} \right]
\; \; . \label{g12Kce}\end{equation}

We compare next the results for these two very different scenarios
by means of the two-pion correlation functions, assuming the
Bjorken picture for the system, i.e., neglecting the transverse
expansion. For the instant FO case, we use Eq.(\ref{gk1k2}) and
(\ref{gkk}) for obtaining $C(q,K)$. For the CE case, we use
Eq.(\ref{giice}) and (\ref{g12Kce}) , writing the four-divergence
and the integrals in cylindrical coordinates, using the symmetry
of the problem, which leads to a simpler expression:
\begin{eqnarray}
&&\!\!\!\!G(q,K)=  \nonumber \\
&&\!\!\!\!\!\!\frac 1{(2\pi )^3}\int_0^{2\pi }d\phi \int_{-\infty }^{+\infty }d\eta
\;\{\int_0^{R_T}\rho \,d\rho \;\tau _{{\cal F}}\,M_T\,\cosh (Y-\eta )
\nonumber \\
&&\!\!\!\!\!\!\times e^{i[\tau _{{\cal F}}(q_0\cos \eta -q_L\sinh \eta )-\rho q_T\cos
(\phi -\phi _q)]}\;  \nonumber \\
&&\!\!\!\!\!\!+\int_{\tau _0}^{+\infty }\!\!\tau d\tau \rho _{{\cal F}}K_T\cos \phi
\,e^{i[\tau (q_0\cos \eta -q_L\sinh \eta )-\rho _{{\cal F}}q_T\cos (\phi
-\phi _q)]}\}  \nonumber \\
&&\!\!\!\!\!\!\times \; \frac{e^{-M_T\cosh (Y-\eta )/T_{ps}(x)}}{(1-{\bf {\cal P}}_{{\cal F}%
})}\;,  \label{g12contemis}
\end{eqnarray}
where {\small \smallskip
{$M_T=\sqrt{K_T^2+M^2};\;M^2=K^\mu K_\mu -\frac 14q^\mu q_\mu ;\;K^\mu=\frac 12%
(k_1^\mu +k_2^\mu )$};} $Y$ is the rapidity corresponding to $\vec{K}$, $%
\phi $ is the azimuthal angle with respect to the direction of $\vec{K}$,
$\phi _q$ is the angle between the directions of $\vec{q}$ and
$\vec{K}$, and
%\vspace{-2mm}
{\small
\begin{eqnarray}
\tau _{{\cal F}}=\frac{-\rho \cos \phi +\sqrt{R_T^2-\rho ^2\sin ^2\phi }}{%
(k_T/E)[\sqrt{\sinh ^2\eta +{\bf {\cal P}}_{{\cal F}}^{-2/a}}-\cosh \eta ]}%
\;\;. \nonumber% \label{tauf}
\end{eqnarray}\vspace{-5mm}
\begin{eqnarray}
\rho _{{\cal F}} &=&-\tau \;(k_T/E)\cos \phi [\sqrt{\sinh ^2\eta +{\bf {\cal %
P}}_{{\cal F}}^{-2/a}}-\cosh \eta ]\;\pm \;  \nonumber \\
&&\sqrt{R_T^2-\tau ^2(\frac{k_T}E)^2\sin ^2\phi \;[\sqrt{\sinh ^2\eta +{\bf
{\cal P}}_{{\cal F}}^{-2/a}}-\cosh \eta ]^2}.  \nonumber \\
&&  \nonumber%\label{rhof}
\end{eqnarray}
}
The spectrum is obtained from the expression (\ref{g12contemis}),
by replacing $K^\mu \rightarrow k_i^\mu$, $M \rightarrow m$, and
$q^\mu \rightarrow 0$.
In Eq. (\ref{g12contemis}), $\tau _{{\cal F}}$ and $%
\rho _{{\cal F}}$, whose expressions are written above, are the
limiting values corresponding to the escape probability ${\cal
P}_{{\cal F}}$, which we fix to be ${\cal P}_{{\cal F}}\approx
0.5$, approximate value chosen for the sake of simplicity and for
guaranteeing that the thermal assumption still holds in systems
with finite size and finite lifetime. The rest of the emission for
${\cal P}_{{\cal F}}>0.5$ is assumed to be instantaneous, as in
Eq. (\ref{gk1k2}) and (\ref{gkk}).

The complexity of the expressions for the Continuous Emission (CE)
case suggested that we should look into special kinematical zones
for investigating the differences between this scenario and the
instant freeze-out picture. For details and discussions, see
Ref.\cite{pghs}. I summarize some of them. First, we observed that
the correlation function $C(q_O,q_L)$ plotted versus the outward
momentum difference, ${\bf q_{O}}$, exhibited the well-know
dependence on the average momentum of the pair, $K_T$, discussed
earlier in this section, in both cases. However, it was enhanced
in the CE scenario, as expected, since the emission duration is
longer in this case than in the standard freeze-out picture. We
also observed a slight variation with $K_T$ of the correlation
function versus ${\bf q_{S}}$ in the CE, differently from the
instant freeze-out case, were it was absent, since we were
considering only the longitudinal expansion of the system and no
transverse flow. This result showed the tendency of the curve
versus ${\bf q_{S}}$ to become slightly narrower for increasing
$K_T$, an opposite tendency as compared to the curves plotted as
functions of ${\bf q_{O}}$.
\begin{figure}[!htb18]
%\begin{center}
%\includegraphics*[angle=0, width=6.5cm]{PRC62CE2qLOK} \\
%\includegraphics*[angle=0, width=6.5cm]{PRC62CE4qOOK} %\\
%\includegraphics*[angle=0, width=6.5cm]{PRC62CE6qSOK}\\
%%%\end{center}
  \resizebox{11pc}{!}{\includegraphics{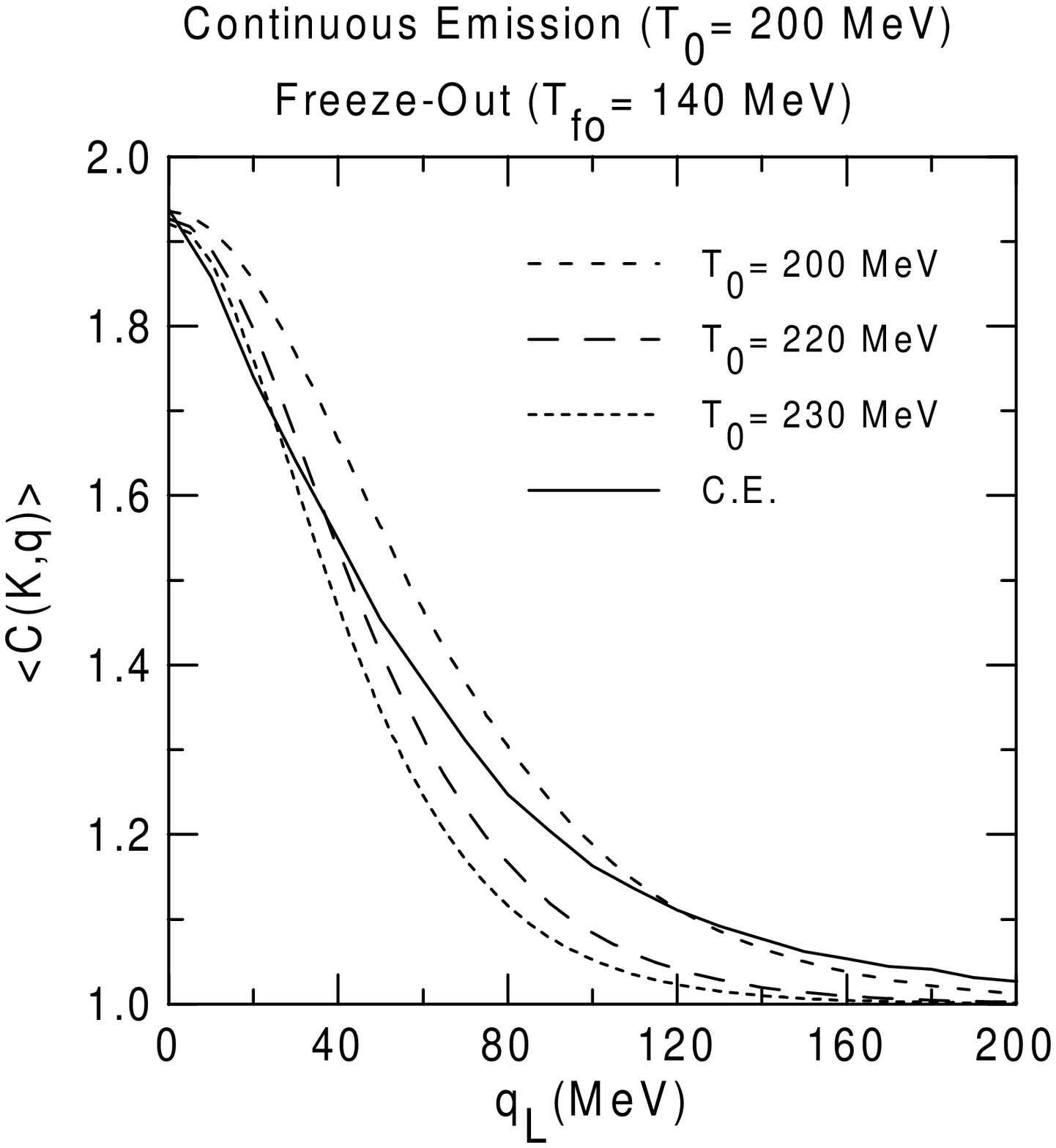}}%\\
  \resizebox{11pc}{!}{\includegraphics{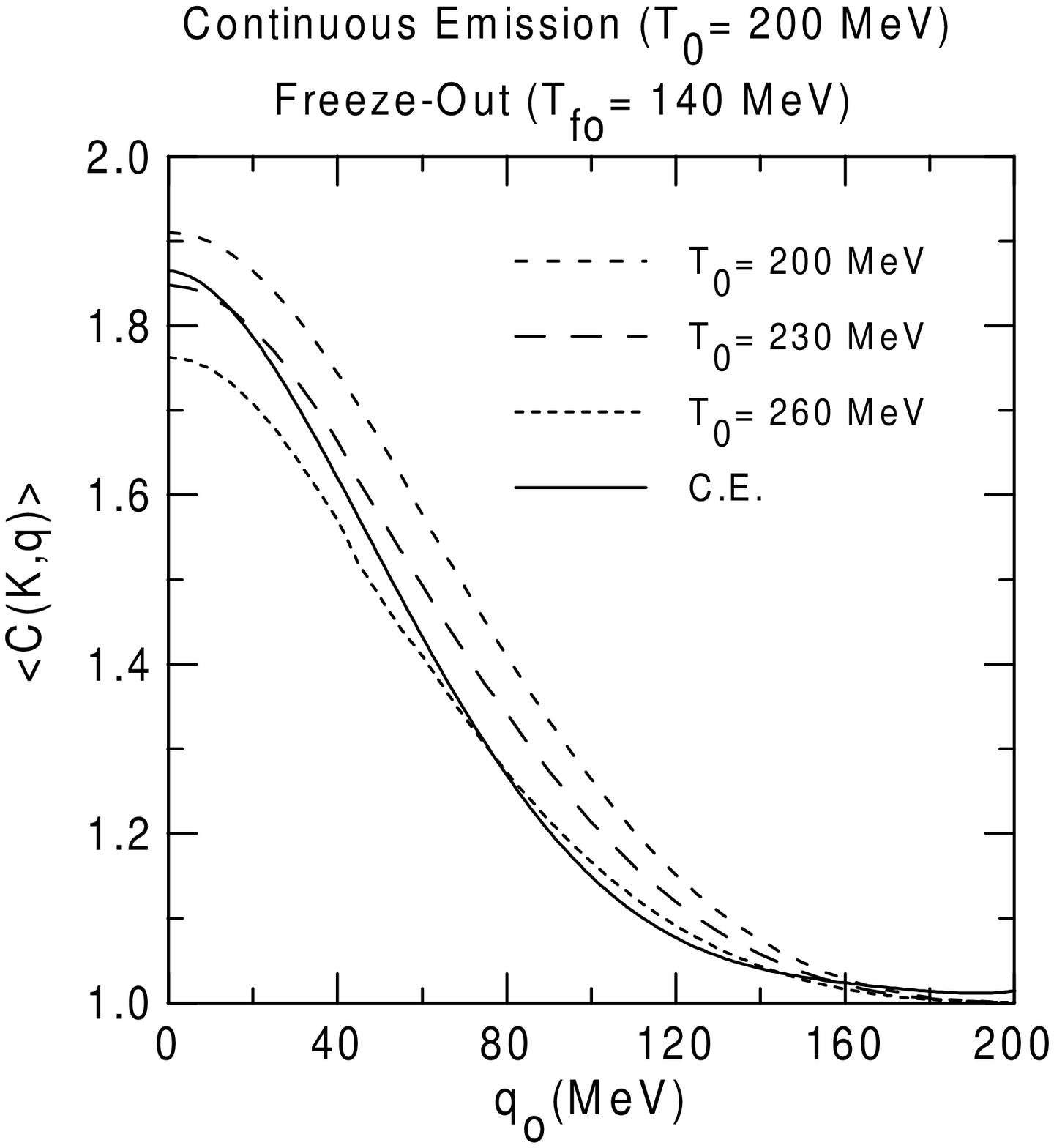}}
  \resizebox{11pc}{!}{\includegraphics{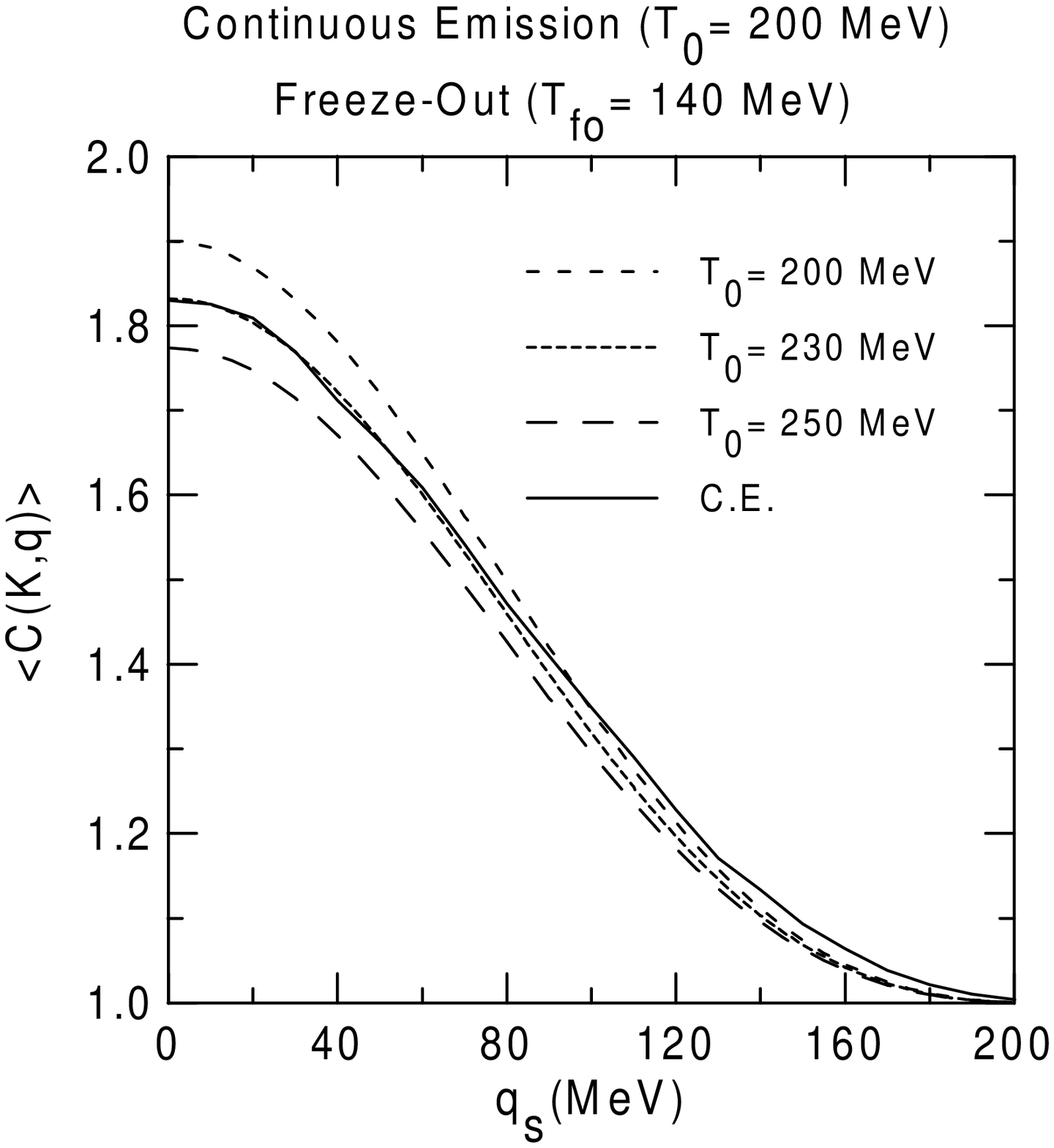}}
\caption{\emph{\small The average correlation function $\langle C
\rangle_{\pi\pi}$ is shown versus $q_L$, $q_O$, and  $q_S$,
averaging over, respectively, ($q_S$ , $q_O$), ($q_L$ , $q_S$),
and ($q_L$ , $q_O$). Comparison is made of the (solid) curve for
the CE scenario with three others corresponding to the usual
freeze-out at $T_{fo}=140$ MeV, but with different initial
temperatures, whose values are shown in the plot. Plots extracted
from Ref.\cite{pghs}. }}
\end{figure}

We also studied more realistic situations, where the correlation
function was averaged over
%angles, momentum regions, unobserved projections of the momentum
%difference, experimental kinematic cuts, etc.
kinematical zones.
Using the azimuthal symmetry of the problem, we defined the
transverse component $K_T$ along the x-axis, such that $\vec{K} =
(K_T,0,K_L)$. We then averaged over different kinematical regions,
mimicking the experimental cuts, by integrating over the components
of $\vec{q}$ and $\vec{K}$ (except over the plotting component of
$\vec{q}$). For illustration, we considered the kinematical range
of the CERN/NA35\cite{na35b} experiment on S+A collisions at 200
AGeV, as
{\small
\begin{eqnarray}
\!\!\!&&\!\!\!\langle C(q_L) \rangle = 1 +
\nonumber\\
&&\!\!\!\!\!\!\!\!\!\frac{\int^{180}_{-180} dK_L \int^{600}_{50}
dK_T \int^{30}_{0} dq_S \int^{30}_{0} dq_o C(K,q)| G(K,q) |^2 }
{\int^{180}_{-180} dK_L \int^{600}_{50} dK_T \int^{30}_{0} dq_S
\int^{30}_{0} dq_o  C(K,q) G(k_1,k_1) G(k_2,k_2)}. \nonumber
\end{eqnarray} }

With the previous equation we estimated the average theoretical
correlation functions versus $q_L$, $q_O$ and $q_S$, comparing the
results for CE and for FO scenarios. First, we considered the case
in which the initial temperature was the same in both cases
($T_0=200$ MeV) and compared the CE prediction with two outcome
curves corresponding to two FO temperatures, $T_{fo}=140$ and
$T_{fo}=170$MeV. We observed that the same initial temperature led
to entirely different results in each case. The details are shown
and discussed in Ref. \cite{pghs}. In the second situation, we
relaxed the initial temperature constraint and studied the
similarity of the curve in the CE scenario corresponding to a
certain value of $T_0$, and compared to three different curves in
the FO scenario, for which the freeze-out temperature was fixed
$T_{fo}=140$ MeV. Each one of these curves in the standard FO case
corresponded to a different initial temperature. The purpose here
was to investigate, as usually done when trying to describe the
experimental data, which initial temperatures in the FO scenario
would lead to the curve closest to the one generated under the CE
hypothesis. The results are shown in Fig. 18. Usually the shape of
the correlation curve is very different in both cases, the one
corresponding to the CE being highly non-Gaussian, mainly in the
upper left plot of Fig. 18. In this particular one, we see that
the CE correlation curve can be interpreted as showing the history
of the hot expanding matter. For instance, the tail of $\langle C
\rangle$ reflects essentially the early times, when the size of
the system is small and the temperature is high, since the tail of
the CE curve is closer to the FO one corresponding to the highest
initial temperature. On the contrary, the peak region corresponds
to the later times where the dimensions of the system are large
and the temperature low (see the compatibility with the FO curve
corresponding to the lowest initial temperature).

\subsection{II.6~ FINITE SIZE EFFECTS}

\subsection{II.6.1~ PION SYSTEM}

In 1998, a pos-doctoral fellow from China,
Qing-Hui Zhang, joined the group, spending one year with us.
Together, we investigated the
effects of finite source sizes (boundary effects) on $\pi^{\pm} \pi^{\pm}$
interferometry. We derived a general formulation for spectrum and
two-particle correlation, adopting a density matrix suitable for
treating the charged pion cases in the non-relativistic limit\cite{zp}.

A few hypotheses inspired that study. First, we considered pions,
the most abundant particles produced in relativistic heavy-ion
collisions, to be quasi-bound in the system, with the surface
tension\cite{shu,MW95,SRSAS97} acting as a reflecting boundary. In
this regard, the pion wave function could be assumed as vanishing
outside this boundary. As usual, we also considered that these
particles become free when their average separation is larger than
their interaction range and we assumed this transition to happen
very rapidly, in such a way that the momentum distribution of the
pions would be governed by their momentum distribution just before
they freeze out. We then studied the modifications on the observed
pion momentum distribution caused by the presence of this
boundary. We also investigated its effects on the correlation
function, which is known to be sensitive to the geometrical size
of the emission region as well as to the underlying dynamics.
In this formalism, the single-pion distribution can then be
written as
\begin{eqnarray}
P_1({\bf p})&=&\langle \hat{\psi}^{\dagger}({\bf p})\hat{\psi}({\bf p})\rangle
=\sum_{\lambda}\sum_{\lambda'}
\tilde{\psi}_{\lambda}^*{(\bf p)}
\tilde{\psi}_{\lambda'}^*{(\bf p)}
\langle \hat{a}^{\dagger}_{\lambda}\hat{a}_{\lambda'}\rangle
\nonumber\\
&=&\sum_{\lambda}N_{\lambda}
\tilde{\psi}^*_{\lambda}({\bf p})
\tilde{\psi}_{\lambda}({\bf p})
\; \;,\label{P1in}\end{eqnarray}
where the last equality follows from the fact that the expectation value
$ \langle \hat{a}^{\dagger}_{\lambda}\hat{a}_{\lambda'} \rangle $
is related to the occupation probability of the
single-particle state $\lambda$,
$N_\lambda$,  by
$%\begin{equation}
\langle \hat{a}^{\dagger}_{\lambda}\hat{a}_{\lambda'} \rangle
=\delta_{\lambda,\lambda'}N_{\lambda} \;;\;
$%\; \; .\label{adagaa}\end{equation}
$a_{\lambda}$ ($a^{\dagger}_{\lambda}$) is the annihilation
(creation) operator for destroying (creating) a pion in a quantum
state characterized by a quantum number $\lambda$.  In Eq.
(\ref{P1in}) $\hat{\psi}^{(\dagger)}({\bf p})=\sum_{l}
\hat{a}^{(\dagger)}_{l} \tilde{\psi}^{(*)}_{l}({\bf p}) $, where
$\hat{a}^{\dagger}$ ($\hat{a}$) is the pion creation
(annihilation) operator, and $\tilde{\psi}_{l}$ is one of the
eigenfunctions belonging to a localized complete set, satisfying
orthogonality and completeness relations.

For a bosonic
system in equilibrium at a temperature $T$ and chemical
potential $\mu$, $N_\lambda$
it is represented by the Bose-Einstein distribution
\begin{equation}
N_{\lambda}=\frac{1}{\exp{\left[\frac{1}{T} (E_{\lambda}-\mu)\right]}-1
} \; \;.\label{Nlamb}\end{equation}

The above formula coincides with the one employed in Ref.\cite{MW95}
for expressing the single-pion distribution.

   The normalized expectation value of an observable $A$ is given by
\begin{equation}
\langle \hat{A}\rangle =\frac{tr\{ \hat{\rho}
\hat{A}\}}{tr\{\hat{\rho}\}} \; \; ; \; \;
\hat{\rho}=\exp\left[-\frac{1}{T} (\hat{H}-\mu \hat{N})\right]
=\prod_{l} \hat{\rho}{_{l}}\;,
\;  \label{expect}\end{equation}
%  \item
%\begin{equation}
%\hat{\rho}=\exp\left[-\frac{1}{T} (\hat{H}-\mu \hat{N})\right]
%=\prod_{l} \hat{\rho}{_{l}},
%\; ; \label{rho}\end{equation}
where $\hat{\rho}$ is the density matrix operator for the
%$Q$-bosonic
bosonic system, and
$%\begin{equation}
\hat{\rho}_{l}=\exp\left[-\frac{1}{T} (\hat{H}_{l}-\mu \hat{N}_{l})\right]
,\; $% \label{rhol}\end{equation}
%  \item
with
\begin{equation}
\hat{H}=\sum_{l}\hat{H}_{l}; \; \; \; \;
\hat{H}_{l}=E_{l}\hat{N}_{l}; \; \; \; \;
\hat{N}=\sum_{l}\hat{N}_{l}
\; , \label{hamilt}\end{equation}
respectively, the Hamiltonian and number operators; $\mu$ is the
chemical potential, fixed to be zero in the results we present here.

Similarly, the two-pion distribution function can be written as
\begin{eqnarray}
&& \!\!\!\!\!\!\!\!\! P_2({\bf p_1,p_2})=\!\!\!\!\!\!\!\!\!
%\langle  \hat{\psi}^*({\bf p_1})\hat{\psi}^*({\bf p_2})
%\hat{\psi}({\bf p_1})\hat{\psi}({\bf p_2})\rangle
%\nonumber\\
\sum_{\lambda_1,\lambda_2,\lambda_3,\lambda_4}
\tilde{\psi}_{\lambda_1}^*({\bf p_1})
\tilde{\psi}_{\lambda_2}^*({\bf p_2})
\tilde{\psi}_{\lambda_3}({\bf p_1})
\tilde{\phi}_{\lambda_4}({\bf p_2})
\nonumber\\
&&
\; \; \; \; \; \; \; \; \; \; \; \; \; \; \; \; \; \; \; \; \; \; \;\;
\langle  \hat{a}^{\dagger}_{\lambda_1}\hat{a}^{\dagger}_{\lambda_2}
\hat{a}_{\lambda_3}\hat{a}_{\lambda_4} \rangle
\nonumber\\
&=& \sum_{\lambda_1,\lambda_2,\lambda_3,\lambda_4}
%\nonumber\\
\tilde{\psi}_{\lambda_1}^*({\bf p_1})
\tilde{\psi}_{\lambda_2}^*({\bf p_2})
\tilde{\psi}_{\lambda_3}({\bf p_1})
\tilde{\phi}_{\lambda_4}({\bf p_2})
\nonumber\\
&&
\left[\langle \hat{a}^{\dagger}_{\lambda_1}\hat{a}_{\lambda_3}\rangle
\langle \hat{a}^{\dagger}_{\lambda_2}\hat{a}_{\lambda_4}\rangle_{\lambda_1 \ne \lambda_2}
\right.
%\nonumber\\
%&&
+ \langle \hat{a}^{\dagger}_{\lambda_1}\hat{a}_{\lambda_4}\rangle
\langle \hat{a}^{\dagger}_{\lambda_2}\hat{a}_{\lambda_3}\rangle_{_{\lambda_1 \ne
\lambda_2}}
\nonumber\\
&&
\left. +\langle \hat{a}^{\dagger}_{\lambda_1}\hat{a}^{\dagger}_{\lambda_2}
\hat{a}_{\lambda_3}\hat{a}_{\lambda_4}\rangle_{\lambda_1=\lambda_2=\lambda_3=\lambda_4}
\right ]
%\nonumber\\
%&&
%=P_1({\bf p_1})P_1({\bf p_2})+ \sum_{\lambda_1}\sum_{\lambda_2}
%\nonumber\\
%&&
%N_{\lambda_1}N_{\lambda_2}
%\tilde{\psi}^*_{\lambda_1}({\bf p_1})
%\tilde{\psi}_{\lambda_1}({\bf p_2})
%\tilde{\psi}^*_{\lambda_2}({\bf p_1})
%\tilde{\psi}_{\lambda_2}({\bf p_2})
\nonumber\\
&=&P_1({\bf p_1})P_1({\bf p_2})+
|\sum_{\lambda}N_{\lambda}\tilde{\psi}_{\lambda}^*({\bf p_1})
\tilde{\psi}_{\lambda}({\bf p_2})|^2
.%\nonumber\\
\label{p2g}\end{eqnarray}

Since we are considering the case of two indistinguishable,
identically charged pions, then
$%\begin{equation}
\langle \hat{a}_{\lambda}^{\dagger}\hat{a}_{\lambda}^{\dagger}
\hat{a}_{\lambda}\hat{a}_{\lambda}\rangle = 2\langle
\hat{a}^{\dagger}_{\lambda}\hat{a}_{\lambda}\rangle^2.
$%\label{BE1}\end{equation}
\hskip2mm From the particular form proposed for the density matrix
it follows that $\langle
\hat{a}^{\dagger}_{\lambda}\hat{a}^{\dagger}_{\lambda} \rangle  =
 \langle \hat{a}_{\lambda}\hat{a}_{\lambda} \rangle = 0$,
showing that it would not be suited for describing $\pi^0 \pi^0$ and
$\pi^+ \pi^-$ cases. For this purpose, the formalism proposed in
Ref.\cite{YS94} is more adequate.

The two-particle
correlation can be written as
\begin{eqnarray}
C_2({\bf p_1,p_2})&=&\frac{P_2({\bf p_1,p_2})}{P_1({\bf p_1})
P_1({\bf p_2})}
\nonumber\\
&=&1+\frac{|\sum_{\lambda}N_{\lambda}
\psi_{\lambda}^*({\bf p_1})
\psi_{\lambda}({\bf p_2})|^2}
{\sum_{\lambda}N_{\lambda}|\psi_{\lambda}({\bf p_1})|^2
\sum_{\lambda}N_{\lambda}|\psi_{\lambda}({\bf p_2})|^2}
\; \;. \nonumber\\
\label{pi2}\end{eqnarray}

In Ref. \cite{zp} we illustrated the formalism by means of two
examples. The first considered that the produced pions were
bounded inside a confining sphere with radius $R$. In the second,
they were inside a cubic box with size $L$. Since the results were
similar in both cases, I will briefly discuss here only the first
one. We estimated Eq.(\ref{P1in}) for the confining sphere of
radius $R$ for studying the boundary effects on the spectrum. The
results can be summarized by looking directly into the top left
plot in Fig. 19, where we also show the curve corresponding to the
limit of very large system ($R \rightarrow \infty$). We see that
the finite size affects the spectrum by depleting the curve at
small values of the pion momentum and, at the same time, rising
and broadening the curve at large $p$ (momentum conservation).

\begin{figure}[!htb19]
%\begin{center}\bigskip\bigskip\bigskip
%\includegraphics*[angle=0, width=7cm]{PRC62Fig1OK} \\ \bigskip
%\includegraphics*[angle=0, width=7cm]{PRC62Fig4OK} \\ \bigskip
%\includegraphics*[angle=0, width=7cm]{PRC62Fig5OK} %\\ \bigskip
%\end{center}
\hskip-1cm  \resizebox{11pc}{!}{\includegraphics{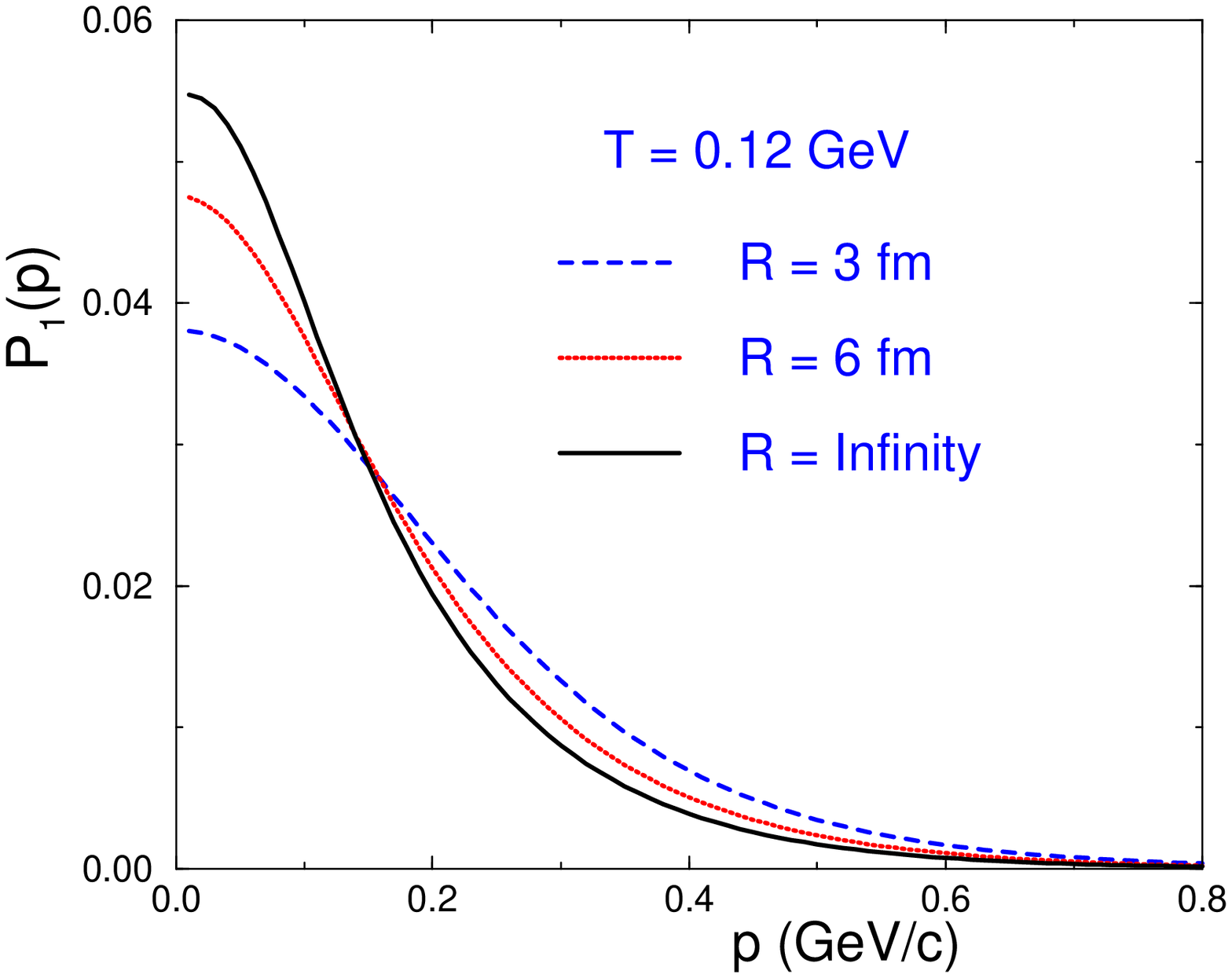}}%\\
  \resizebox{11pc}{!}{\includegraphics{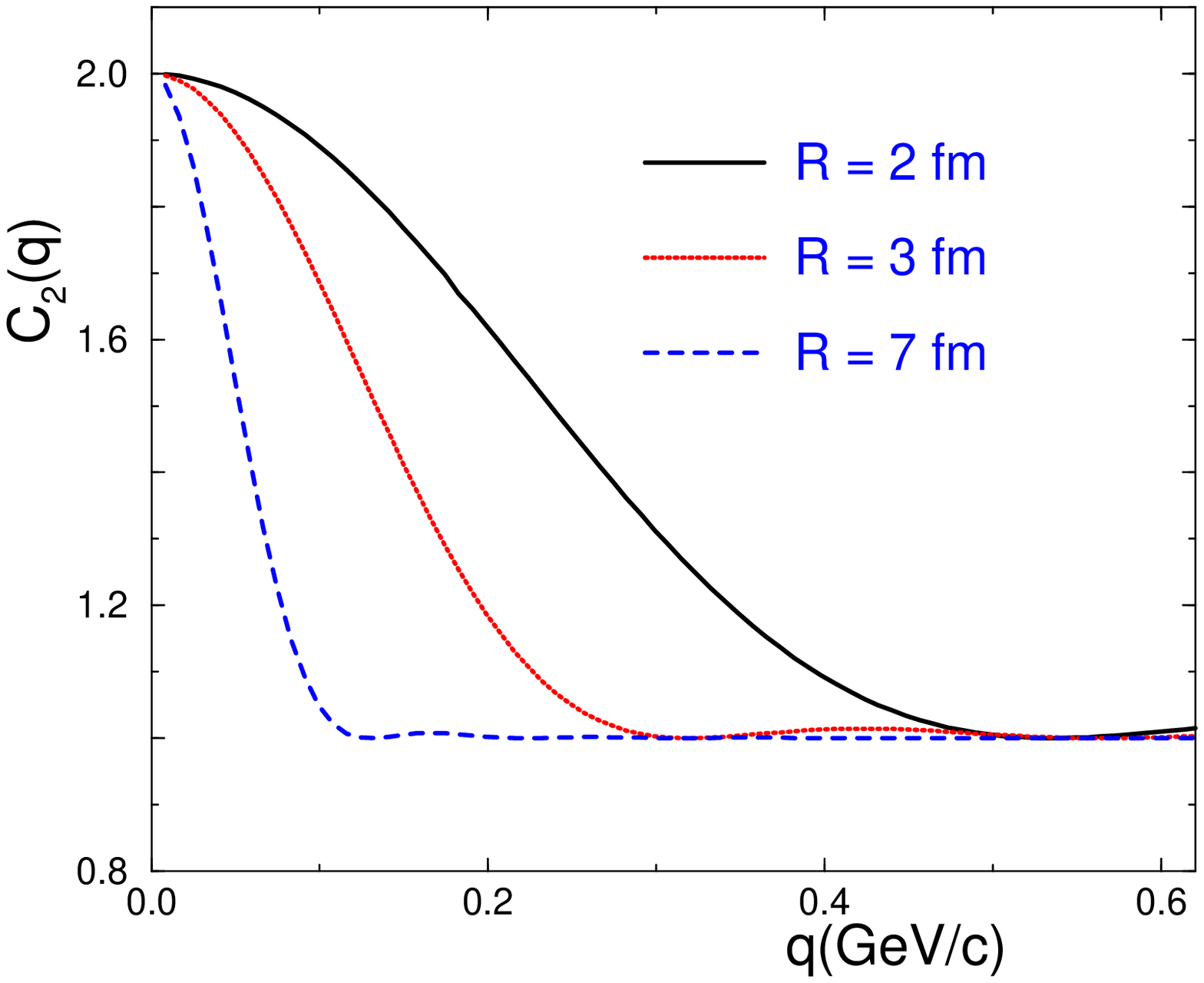}}
  \resizebox{11pc}{!}{\includegraphics{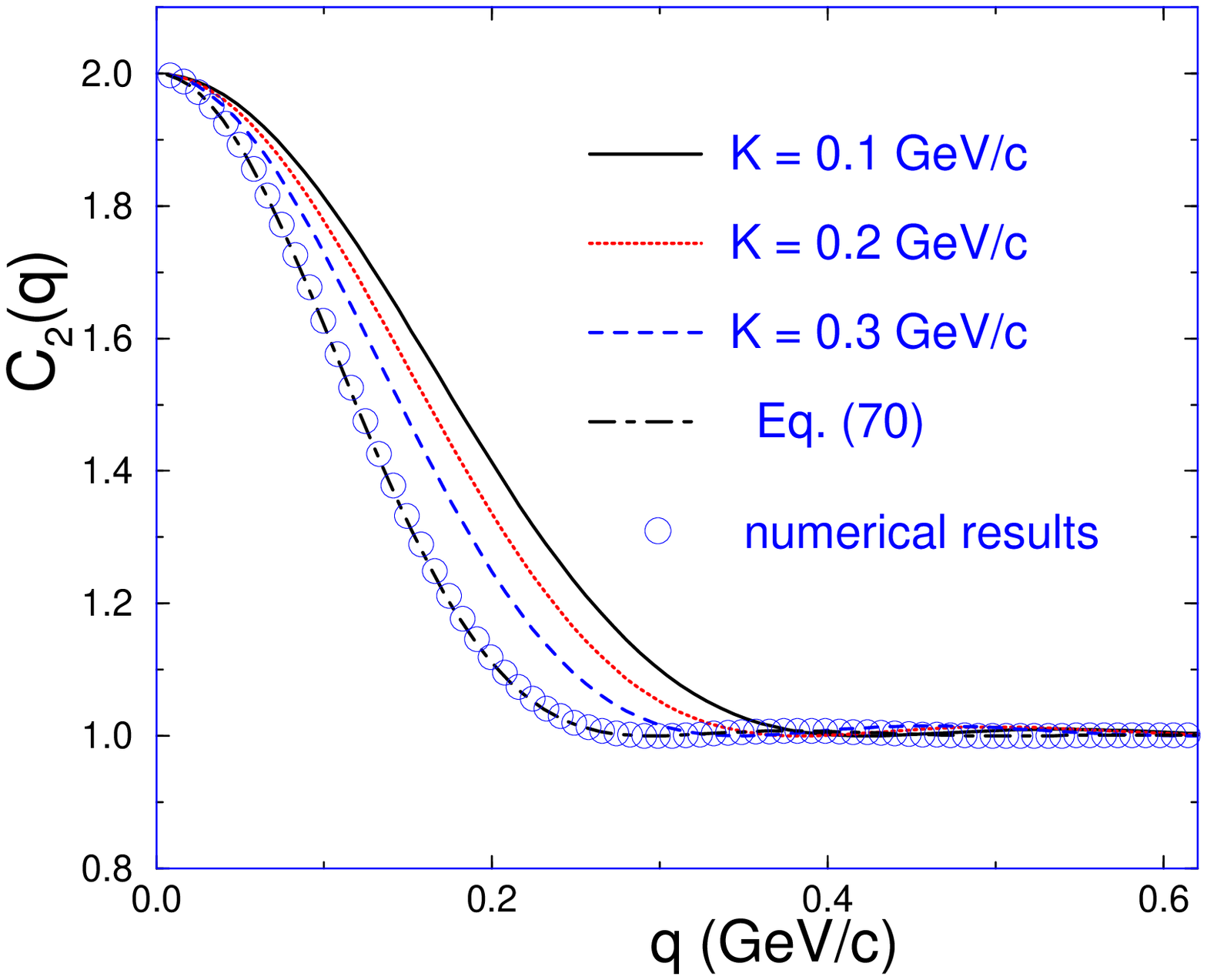}}
%  \resizebox{11pc}{!}{\includegraphics{PRC62Fig4OKold}}
\caption{\emph{\small The top plot shows the spectrum in the
example of pions confined in the sphere of radius $R$. The top
right plot shows $ C _{\pi\pi}(q)$ versus $q$, for three different
values of the source radius. The plot in the bottom shows that the
correlation function shrinks for increasing average momentum, $K$,
in opposition to the behavior seen in Fig. 7 and for expanding
systems, in general. Plots extracted from Ref.\cite{zp}. }}
\end{figure}
For studying the behavior of the correlation function we estimated
Eq. (\ref{pi2}) in the case of the bounding sphere. Some results
are shown in the top right plot of Fig. 19 and they are as
expected: the correlation shrinks (i.e., the probed source
dimension increases) with increasing values of the radius.
Nevertheless, the bottom curve, corresponding to $C(q,K)$ versus
$q$, shows an unexpected behavior for different values of the
average momentum $K$. Contrary to what is observed in expanding
system, the correlation function becomes narrower (probed region
enlarges) with increasing $K$. However, the example shown for
illustration does not take into account the expansion of the
system. The variation with $K$ merely reflects the strong
sensitivity of the results to the dynamical matrix adopted in the
formulation. It is related to the weight factor $N_\lambda$, in
Eq. (\ref{pi2}): pions with larger momenta come from higher
quantum states $\lambda$, which correspond to a smaller spread in
coordinate space. But, due to the Bose-Einstein form of the weight
factor, large quantum states give a small contribution to the
source distribution, causing the effective source radius to appear
larger. To emphasize this we include in the bottom plot the
narrowest curve, corresponding to fixing $N_\lambda=1$ (or
equivalently, by considering $T \gg 1$ in the Bose-Einstein
distribution, which makes $C(q,K)$ insensitive to $K$, due to the
large temperature).
This limit allows for an analytical expression, %i.e.,
\begin{equation} C_2(q)=1+\frac{9}{q^4 R^6}\left[ R \cos(qR) -
\frac{\sin(qR)}{q} \right]^2,\; \label{C2lim}
\end{equation} which is shown in
%%$%\begin{equation}
%%C_2(q)=1+\frac{9}{q^4 R^6}\left[ R \cos(qR) - \frac{\sin(qR)}{q}
%%\right]^2,\;
%%$%\end{equation}
the lowest curve of the bottom plot of Fig. 19. Underneath this
curve, shown by little circles in the same plot, is the
numerically generated curve for $N_\lambda=1$, confirming the
correcteness of our result.

\subsection{II.6.2~ $Q$-BOSON SYSTEM}

More recently, Qing-Hui Zhang and I extended the above formalism
for treating the interferometry of two $Q$-bosons. The concept of
{\sl quons} was suggested{\small \cite{greenberg}} as an artifact
for reducing the complexity of interacting systems, at the expense
of deforming their commutation relations by means of a {\sl
deformation parameter}, $Q$. Then, this could be seen as an
effective parameter, encapsulating the essential features of
complex dynamics.

We derived a formalism suitable for describing spectra and
two-particle correlation function of charged $Q$-bosons, which we
considered as bounded in a finite volume. We adopted the $Q$-boson
type suggested by Anchishkin et al.\cite{AGI}, according to which
the $Q$-bosons are defined by the following algebra of creation
($\hat{a}^\dag$) and annihilation operators ($\hat{a}$): $
\hat{a}_{l}\hat{a}_{l'}^{\dagger}-Q^{\delta_{l,l'}}
\hat{a}_{l'}^{\dagger}\hat{a}_{l}=\delta_{l,l'} , \;
[\hat{a}_{l},\hat{a}_{l'}]=[\hat{a}_{l}^{\dagger},\hat{a}_{l'}^
{\dagger}]=0, \;
[\hat{N}_{l},\hat{a}_{l'}]=-\delta_{l,l'}\hat{a}_{l} \;
[\hat{N}_{l},\hat{a}_{l'}^{\dagger}]=
\delta_{l,l'}\hat{a}_{l}^{\dagger}, \;
[\hat{N}_{l},\hat{N}_{l'}]=0; $ $\hat{N}_{l}$ is the number
operator,
%which can be expressed as
$
\hat{N}_{l}=\sum_{s=1}^{\infty}\frac{(1-Q)^s}{(1-Q^s)}
(\hat{a}_{l}^{\dagger})^s (\hat{a}_{l})^{s}.
$
In the limit $Q=1$, the regular
bosonic commutation relations are recovered. The deformation parameter
$Q$ is a C-number, here assumed to be within the interval $[0,1]$.

The single-inclusive, $P_1({\bf p})$, and the two-particle
distributions are derived in a similar way as in the case of
regular pions discussed above. In the $Q$-boson case,
Eq.(\ref{P1in}) continues to hold, but the weight factor,
$N_\lambda$, related to the occupation probability of a
single-particle state $l$, no longer is as written in
Eq.(\ref{Nlamb}), but is changed into a modified Bose-Einstein
distribution,
\begin{equation}
N_{l}=\frac{1}{\exp{\left[\frac{1}{T} (E_{l}-\mu)\right]}-Q }
\;.\label{NlambQ}\end{equation}

The two-particle
correlation function as also modified, and is written as
\begin{eqnarray}
&&C_2({\bf p_1,p_2})=\frac{P_2({\bf p_1,p_2})}{P_1({\bf p_1})
P_1({\bf p_2})} =
\nonumber\\
&&=1+
\left\{ \sum_{l}N_{l}|\tilde{\psi}_{l}({\bf p_1})|^2
\sum_{l}N_{l}|\tilde{\psi}_{l}({\bf p_2})|^2 \right\}^{-1}
\times
\nonumber\\
&&
\; \; \; \sum_{l,l'} \; \ N_{l} N_{l'}
\tilde{\psi}_{l}^*({\bf p_1}) \tilde{\psi}_{l'}^*({\bf p_2})
\tilde{\psi}_{l}({\bf p_2}) \tilde{\psi}_{l'}({\bf p_1})\times
\nonumber\\
&&
\left\{ 1 - \delta_{l,l'} (1-Q)\cdot
\frac{\exp(\frac{1}{T}(E_{l}-\mu))+Q}
{\exp(\frac{1}{T}(E_{l}-\mu))-Q^2} \right\}
\; \;. %\nonumber\\
\label{c21st}
\end{eqnarray}

We apply this formulation by means of similar models as in
Ref.\cite{zp}, for the confining sphere of radius $R$. We studied
the effects of different values of $Q$ on the correlation
function, under different values of the pair average momentum,
$K_T$. In Fig. 20 we show the boundary effects on the $Q$-boson
spectrum and on the two-$Q$-boson correlation function. We also
included in that plot the behavior of the of the two-$Q$ boson
intercept parameter, $\lambda$,  defined by $\lambda = C({\bf
q=0},{\bf K})-1$. The top two plots in Fig. 20 reproduce the basic
characteristics seen in Fig. 19, in particular, the correlation
function narrows as the average momentum, $K$, grows. There is,
however, a new result: the maximum of the correlation function
drops as the deformation parameter, $Q$, increases. This can also
be seen by looking into the bottom plot of the intercept
$\lambda$, as a function of $Q$.

\begin{figure}[!htb20]
\hskip-1cm\resizebox{11pc}{!}{\includegraphics{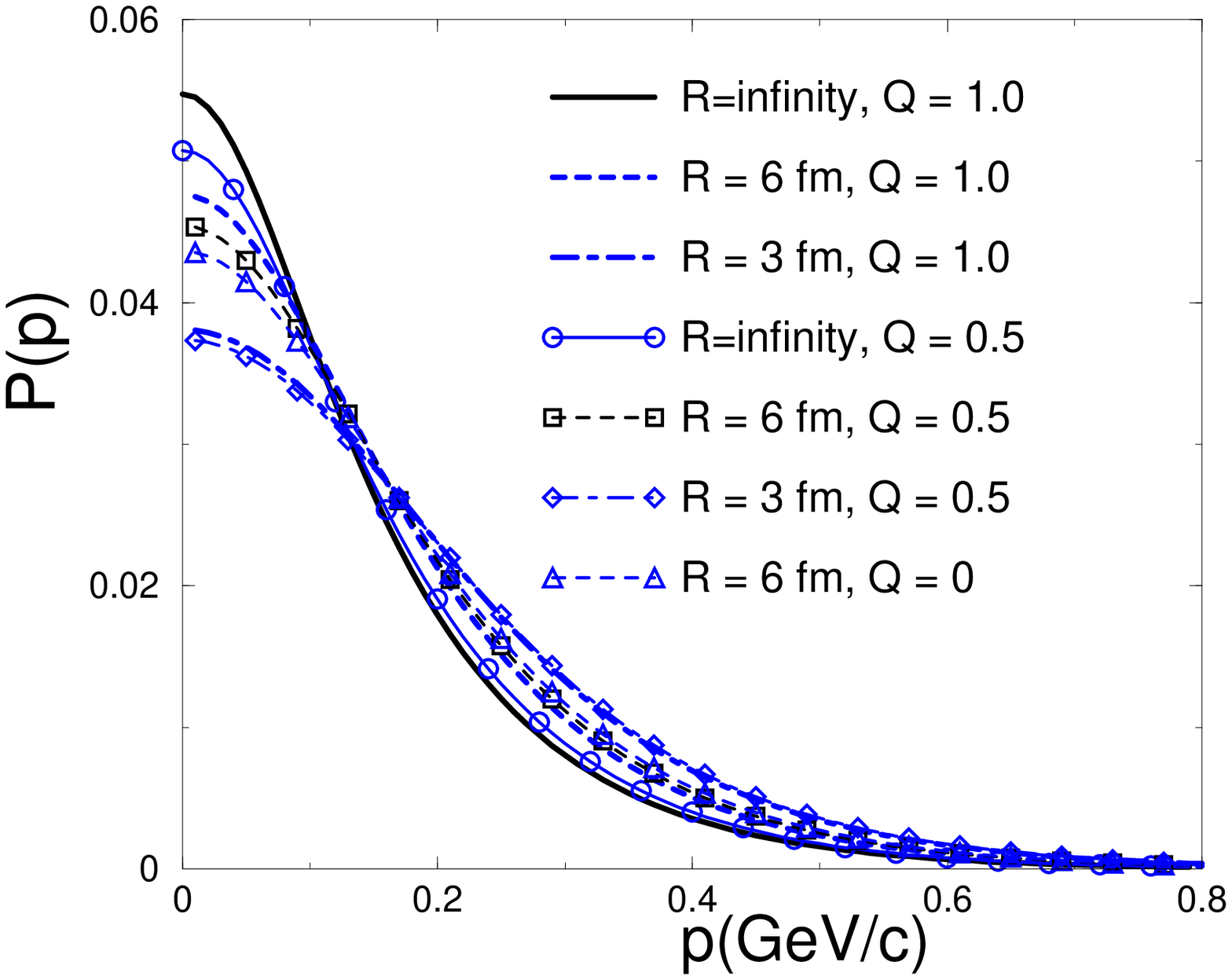}}
  \resizebox{11pc}{!}{\includegraphics{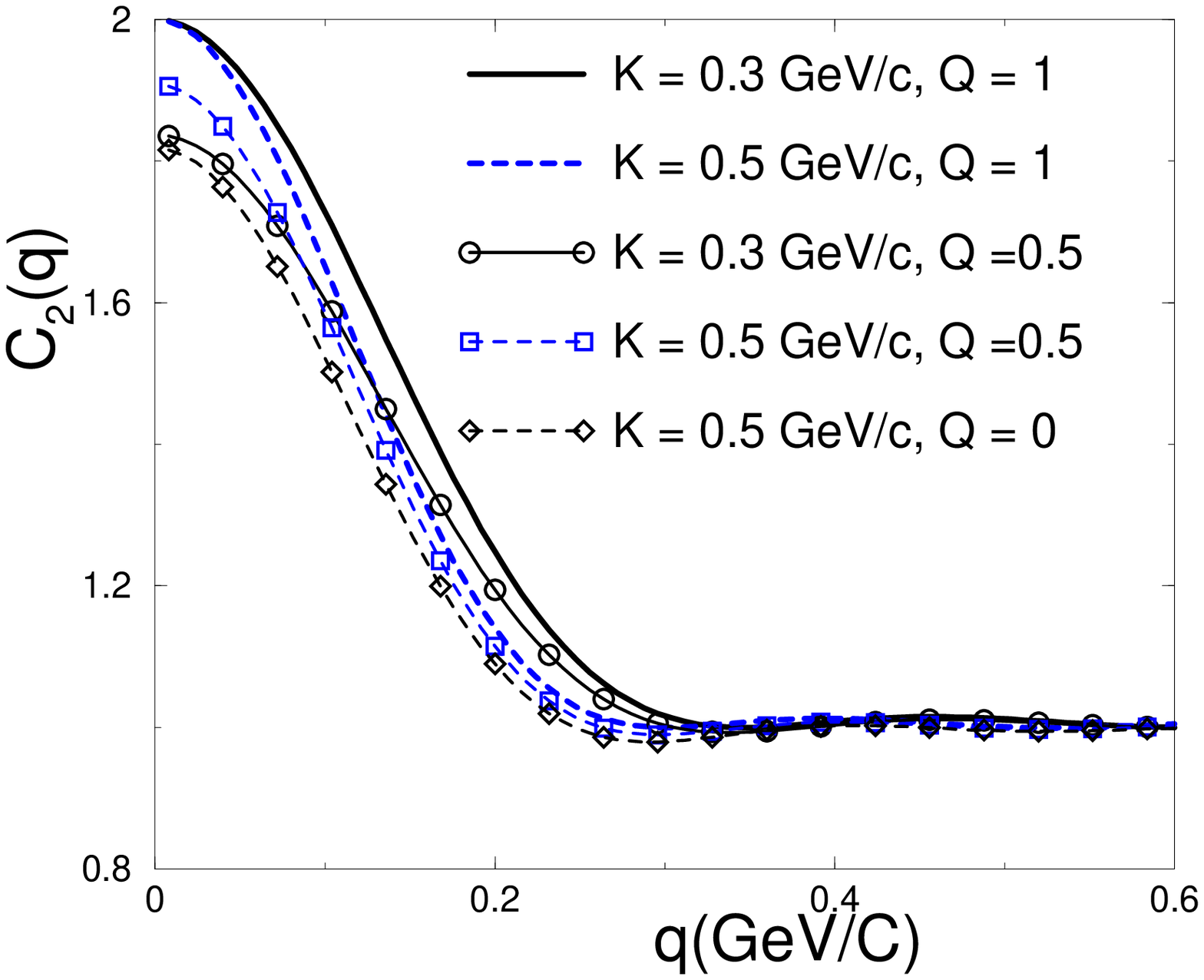}}
  \resizebox{11pc}{!}{\includegraphics{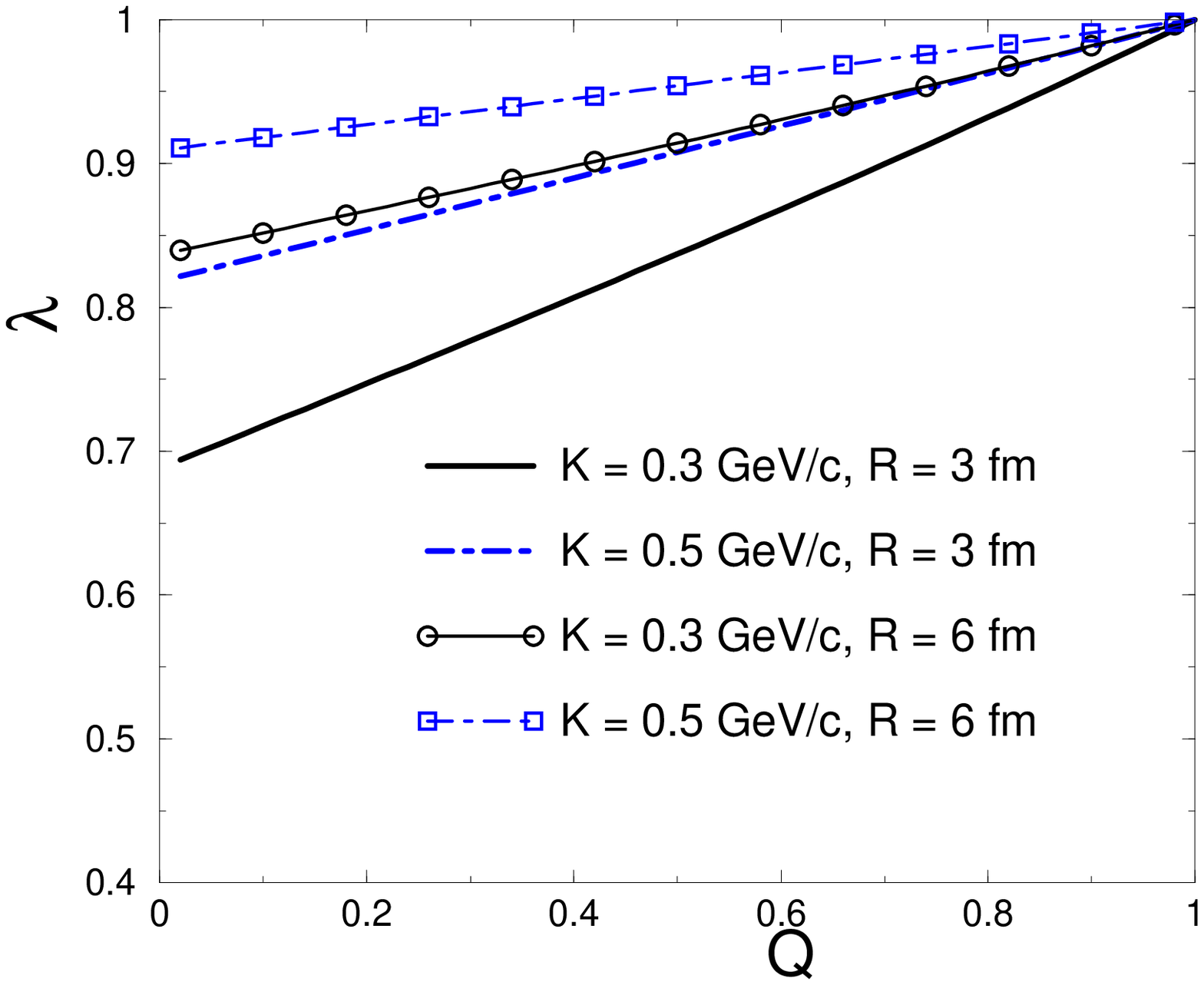}}
 \caption{
%(Color online)
The results shown here were obtained by considering the $Q$-bosons
confines inside a three-dimensional sphere of radius $R$. In the
three plots above, the temperature was fixed to $T=0.12$ GeV and
the chemical potential, $\mu=0$. The top left plot corresponds to
the normalized spectrum (in arbitrary units) vs. momentum $|${\bf
p}$|$ (in GeV/c). The solid lines correspond to the case
$R\rightarrow\infty$, the dotted ones to $R= 6$ fm, and the dashed
lines correspond to the case, $R=3$ fm/c. The bare lines refer to
$Q=1.0$ and the ones with symbols, to $Q =0.5$ ($Q =0$).
%\hspace{0.6cm}
The top right plot shows the two-pion correlation vs. momentum
difference $|{\bf q}|$ (in GeV/c), for a  $R= 3$ fm sphere. The
solid lines correspond to mean momentum $K=0.3$ GeV/c and the
dashed ones, to the case $K=0.5$ GeV/c. The bare lines refer to
the case of $Q=1.0$ and the ones with symbol, to $Q=0.5$ ($Q =0$).
%\hspace{0.6cm}
The bottom plot shows the intercept parameter, $\lambda$, vs. the
deformation $Q$. The solid lines correspond to the mean momentum
$K=0.3$ GeV/c and dashed ones to the case $K=0.5$ GeV/c. The bare
lines refer to the case $R=3$ fm and the ones with symbols, to $R
=6$ fm. Plots extracted from Ref.\cite{zp2}. }
\end{figure}

We also derived a generalized Wigner function for the $Q$-boson
interferometry, which would reduce to the regular one in the limit
of $Q \rightarrow 1$. For that, we define the Wigner function
associated to the state $l$ as
\begin{equation}
g_{l} ({\bf x},{\bf K}) =
\int \frac{d^3 \Delta x}{(2 \pi)^3} \;
e^{-i{\bf K}.{\bf \Delta x}}
\psi_{l}^*({\bf x} + \frac{\bf \Delta x}{2})
\psi_{l}({\bf x} - \frac{\bf \Delta x}{2})
\; . \end{equation}

We proceeded analogously to define the equivalent function for the
integration in ${\bf y}$ and  ${\bf \Delta y}$, remembering that
$g_{l} ({\bf x},{\bf K}) = g_{l}^*({\bf x},{\bf K})$. Then,
denoting by
\begin{equation}
g({\bf x},{\bf K}) = \sum_{l} N_l \; g_l,
\end{equation}
\vskip-0.3cm
\noindent
we finally defined the generalized Wigner
function of the problem as
\begin{eqnarray}
\hskip-.6cm&&g ({\bf x},{\bf K} ; {\bf y},{\bf K}) = g({\bf x},{\bf K}) g({\bf
y},{\bf K})
\; - \; (1-Q) \times\nonumber\\
&&\hskip-1.1cm
\sum_{l}\left\{ N_{l}^2\left[\frac{\exp(\frac{1}{T}(E_{l}-\mu))+Q}
{\exp(\frac{1}{T}(E_{l}-\mu))-Q^2}\right] g_{l} ({\bf x},{\bf K}) \;
g_{l} ({\bf y},{\bf K})\right\}.%\nonumber\\
\label{wignergen}
\end{eqnarray}
We see that, for $Q=1$, the above expression is reduced to the usual
result of the original Wigner function, i.e.,
\begin{equation}
g ({\bf x},{\bf K} ;
{\bf y},{\bf K}) = g({\bf x},{\bf K}) g({\bf y},{\bf K})
\;.\label{wigner}\end{equation}
On the other
hand, for $Q=0$, Eq. (\ref{wignergen}) is identically zero for single
modes
only, as it would be expect in the limit of Boltzmann statistics.
Nevertheless, in the multi-mode case, there seems
to be some sort of residual correlation among $Q$-bosons
even in the classical limit.

By means of this Wigner function, the two-$Q$-boson correlation
function
can be rewritten as
\begin{equation}
C_2({\bf p_1,p_2})= 1 +
\frac{\int \int e^{-i{\bf q}. ({\bf x}-{\bf y})}
g({\bf x},{\bf K} ; {\bf y},{\bf K}) d{\bf x}d{\bf y}}
{\int g({\bf x},{\bf p_1})d{\bf x} \int g({\bf y},{\bf p_2}) d{\bf y}}
%\nonumber\\
\label{c23rd}\end{equation}

Particularly interesting is the approach in Ref.\cite{gastao},
where it was shown that the composite nature of the particles
(pseudo-scalar mesons) under study could result into deformed
structures linked to the deformation parameter $Q$. In that
reference this parameter is then interpreted as a measure of
effects coming from the internal degrees of freedom of composite
particles (mesons, in our case), being the value of $Q$ dependent
on the {\sl degree of overlap} of the extended structure of the
particles in the medium. Being so, the $Q$-parameter could be
related to the power of {\sl probing lenses}, for mimicking the
effects of internal constituents of the bosons. In this case, and
for high enough {\sl magnification}, the bosonic behavior of the
$Q$-bosons could be blurred by the fermionic effect of their
internal constituents, which would result in decreasing the value
of $Q$. Our results on the two-$Q$-boson interferometry are
compatible with this interpretation, as explained in detail in
Ref.\cite{zp2}.

\subsection{II.7~ NON-EXTENSIVE STATISTICS AND HBT}

S\'ergio M. Antunes\cite{sma},
working under my supervision and in collaboration with G. Krein
during his Master Degree, studied the
effects of Tsallis\cite{tsallis} non-extensive statistics on the
two-particle correlations and spectra. In this work,
a very simple starting hypothesis was made:
that under certain circumstances, the Boltzmann limit to the pion
distribution could be replaced by an equivalent approximate expression
derived within the non-extensive statistics. This concept of
non-extensivity could be summarized very briefly by the relation:
$S_{q_{T_s}}(A + B) = S_{q_{T_s}}(A) + S_{q_{T_s}}(B) + (1 - q_{T_s})
 S_{q_{T_s}}(A) S_{q_{T_s}}(B)$, i.e., the entropy of a system
formed by two independent sub-systems $A$ and $B$, no longer is the
sum of the entropy of the two subsystems (note that by independent
it is meant that the probability of the composite system factorizes
as $p_{A+B} = p_A+p_B$). The parameter $q_{Ts}$ is a measure of the
degree of non-extensivity of the system and the Boltzmann-Gibbs
statistics is recovered in the limit $q_{T_s} \rightarrow 1$.
From the definition of the generalized entropy in the
Tsallis\cite{tsallis}
formulation, it is possible to obtain approximate analytical
expressions, for instance, for the mean occupation
number\cite{ntsall}
\begin{eqnarray}
%\begin{equation}
&& \hskip-1.2cm \exp\{[-(E(p,r)-\mu]/T\} \longrightarrow \nonumber\\
&&\{1+(q_{T_s}-1)[(E(p,r)-\mu]/T\}^{-1/(q_{T_s}-1)}
%\;.\label{tsadistr}\end{equation}
\;, \label{tsadistr}\end{eqnarray}
which is reduced to the Boltzmann distribution for
$q_{T_s}\!\!\!\rightarrow\!\!1$,
where $T$ is the temperature and $\mu$, the chemical potential.
The form given in Eq.(\ref{tsadistr}) is, however, valid for $q_{Ts}$
close to unity. Later, G. Wilk et al.\cite{wilk} showed that the above
distribution can be written in the form
\begin{equation}
G_{q_{T_s}} =
C_{q_{T_s}}\left[1-(1-q_{T_s})\frac{x}{\lambda}\right]^
{\frac{1}{1-q_{T_s}}} . \label{levy}\end{equation} Considering
$G_{q_{T_s}}$ as a probability distribution (L\'evy distribution)
in the variable $x$, with $x \;\epsilon \;(0,\infty)$, the
parameter $q_{Ts}$ must be limited to the interval $1\le q_{Ts}\le
2$. However, if the mean value of $x$ is required to be finite
($<x>=\lambda/(3-2q_{Ts})<\infty$) for $x \; \epsilon \;
(0,\infty)$, then $q_{Ts}$ has to belong to the interval $1\le
q_{Ts}\le 1.5$, which better justifies the interval of
applicability of Eq. (\ref{tsadistr}).

For investigating the above hypothesis, we
adopted a model with radial flow \cite{spratt}, leading to a non-decoupled
phase-space freeze-out distribution
\begin {equation}
D(x,\vec p) = \frac{\delta (r - R) \; e^{-t^2/\tau^2}}
{[1 + (q_{T_s} - 1)\frac{1}{T}(E'(\vec p, \hat{r}) - \mu)]^{1/(q_{T_s}-1)}},
\end{equation}
where $E'(\vec p,\hat{r}) = (E_p - \hat{r}.\vec p)(1 - v^2)^{-\frac12}$
and $\mu=0$.

With the above decoupling distribution the correlation function is
written as
\begin{eqnarray}
&&C_2(\vec K,\vec q) = 1 + \exp[-\frac12 (E_{k_1}-E_{k_2})^2 \tau^2]
\times \nonumber\\
&&\frac{\left |\int_{0}^{\pi} \frac{ \sin \theta
J_0(\mid \vec q \mid \mid \vec x \mid \sin \theta \sin \overline{\theta})
\exp \{ i \mid \vec q \mid \mid \vec x \mid \cos \theta \cos \overline{\theta} \} d\theta}
{[1 + (q_{T_s} - 1)(\frac{\gamma}{T} E_{\vec K/2} - y \cos \theta)]^{1/(q_{T_s}-1)}} \right |^2}
{\left (\int_{-1}^{+1} d\xi [1 + (q_{T_s} - 1)(\frac{\gamma}{T} E_{\vec K/2} - y
\cos \theta)]^{1/(1-q)} \right
)^2}
\nonumber\label{eq:sergio}
\end{eqnarray}
where $E_{\frac12 \vec K} = \sqrt{(\frac{\vec K}{2})^2 + m_{\pi}^2}$,
$\gamma = (1 - v^2)^{-1/2}$
and  $J_0(x)$ is the Bessel function of order $0$.
In Eq. (\ref{eq:sergio}), $\theta$ is the angle between $\vec K$ and
$\vec x$, $\overline {\theta}$ is the angle between $\vec K$ and $\vec q$,
in such a way that the angle $\alpha$ between  $\vec q$ and $\vec x$,
can be determined by $\cos \alpha = cos\theta cos \overline{\theta} +
sin \theta sin \overline{\theta} \cos(\phi - \overline{\phi})$.
But $\vec K$ and $\vec q$ define a plane, so that we can choose the
directions of these vectors in such a way that $\overline{\phi} = 0$.
Then, choosing the direction of  $\vec K$ along the $z$-axis, if we
integrate Eq. (\ref{eq:sergio}) for $\overline{\theta} = 0$ and
$\overline{\theta} = 90^0$, this corresponds, respectively, to
$\vec q \parallel \vec K$ (i.e, $q_O$) and $\vec q \perp \vec K$
(i.e., $q_S$).
\begin{figure}[!hbt21]\vskip-1.7cm
  \resizebox{11.5pc}{!}{\includegraphics{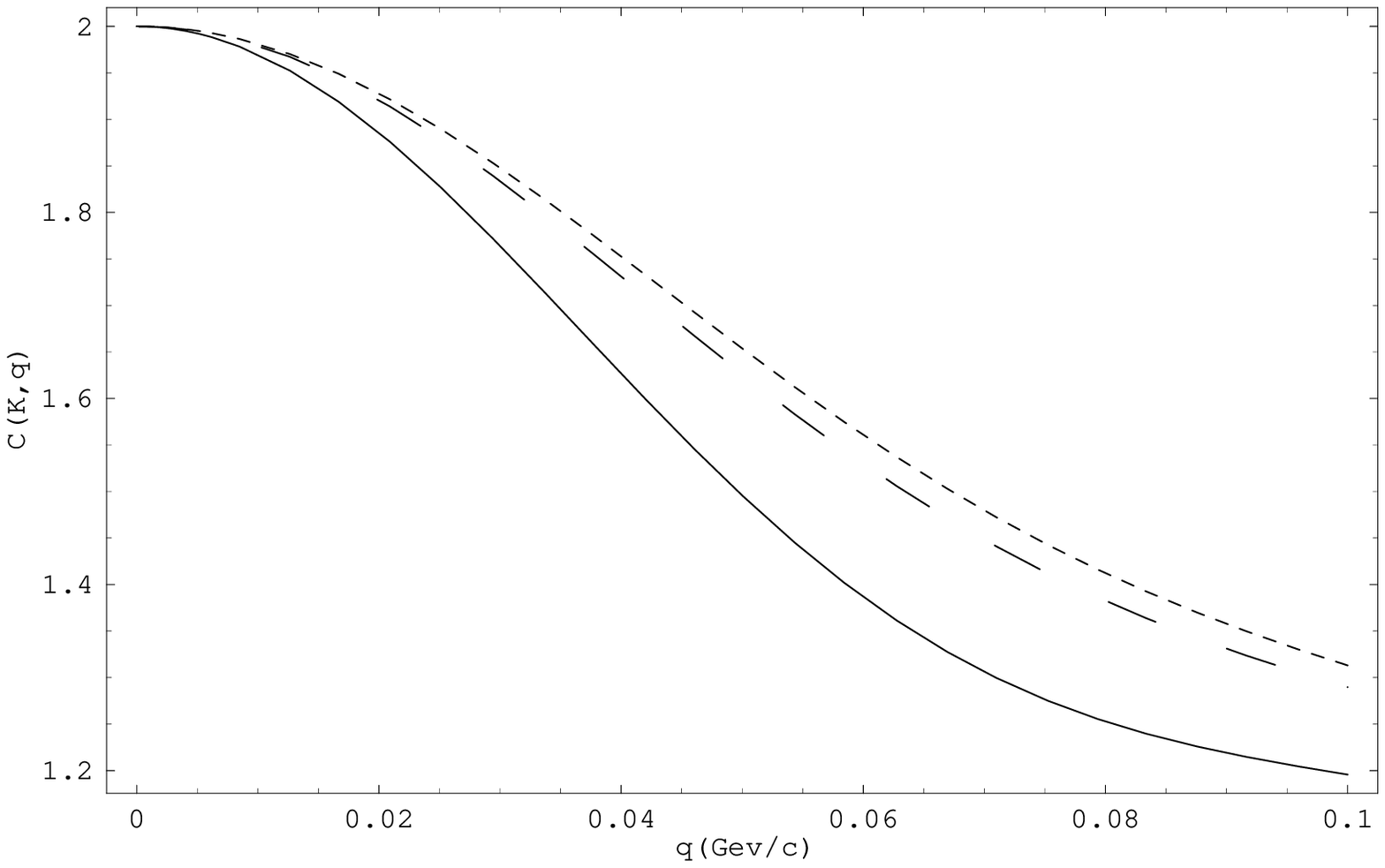}}\vskip-3.0cm
  \resizebox{11.5pc}{!}{\includegraphics{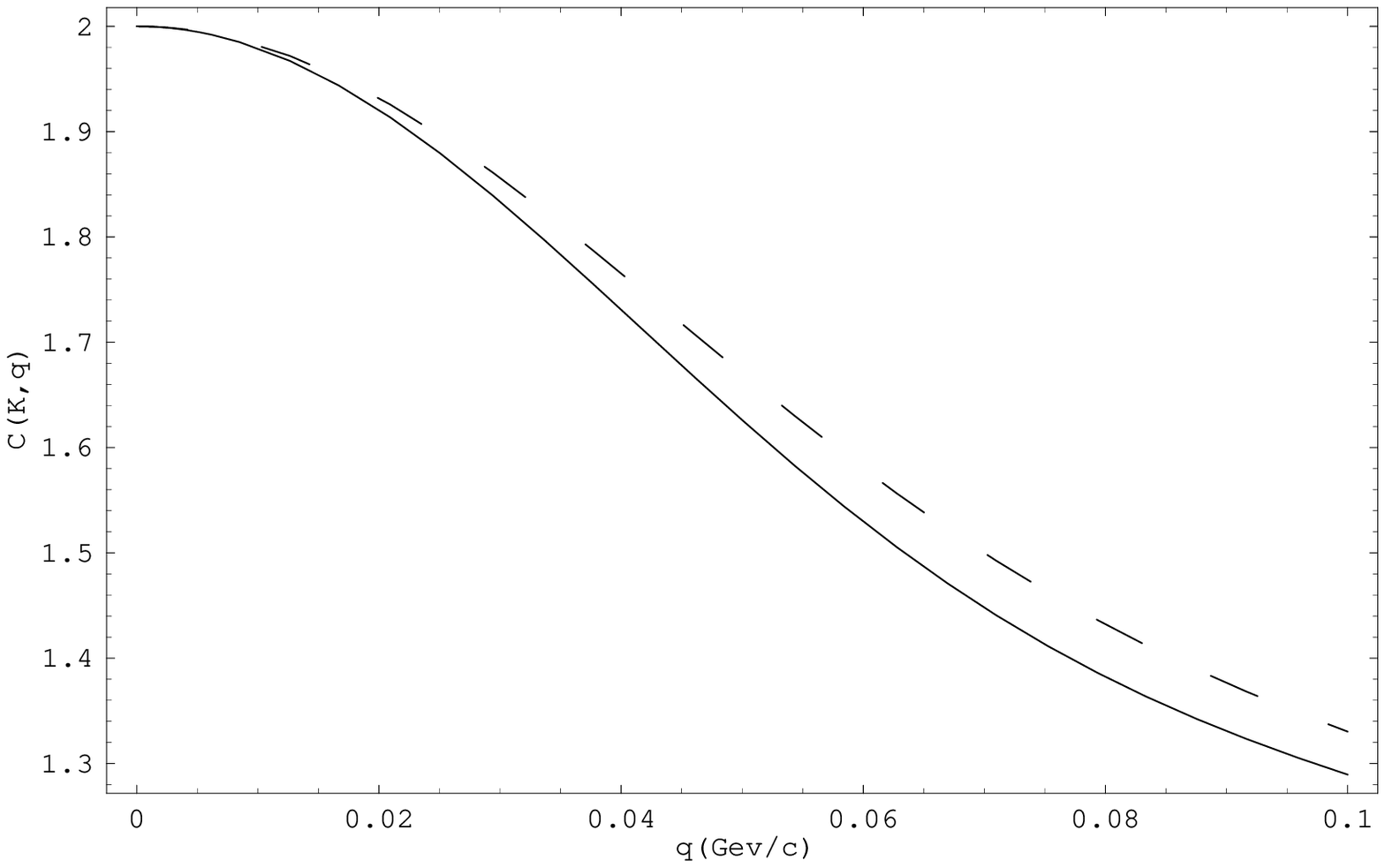}}\vskip-3.5cm
%%\includegraphics*[angle=0, width=6cm]{SMAFig5-24}
%  \resizebox{11.5pc}{!}{\includegraphics{SMAFig5-11}}\vskip-3.5cm
  \resizebox{12.5pc}{!}{\includegraphics{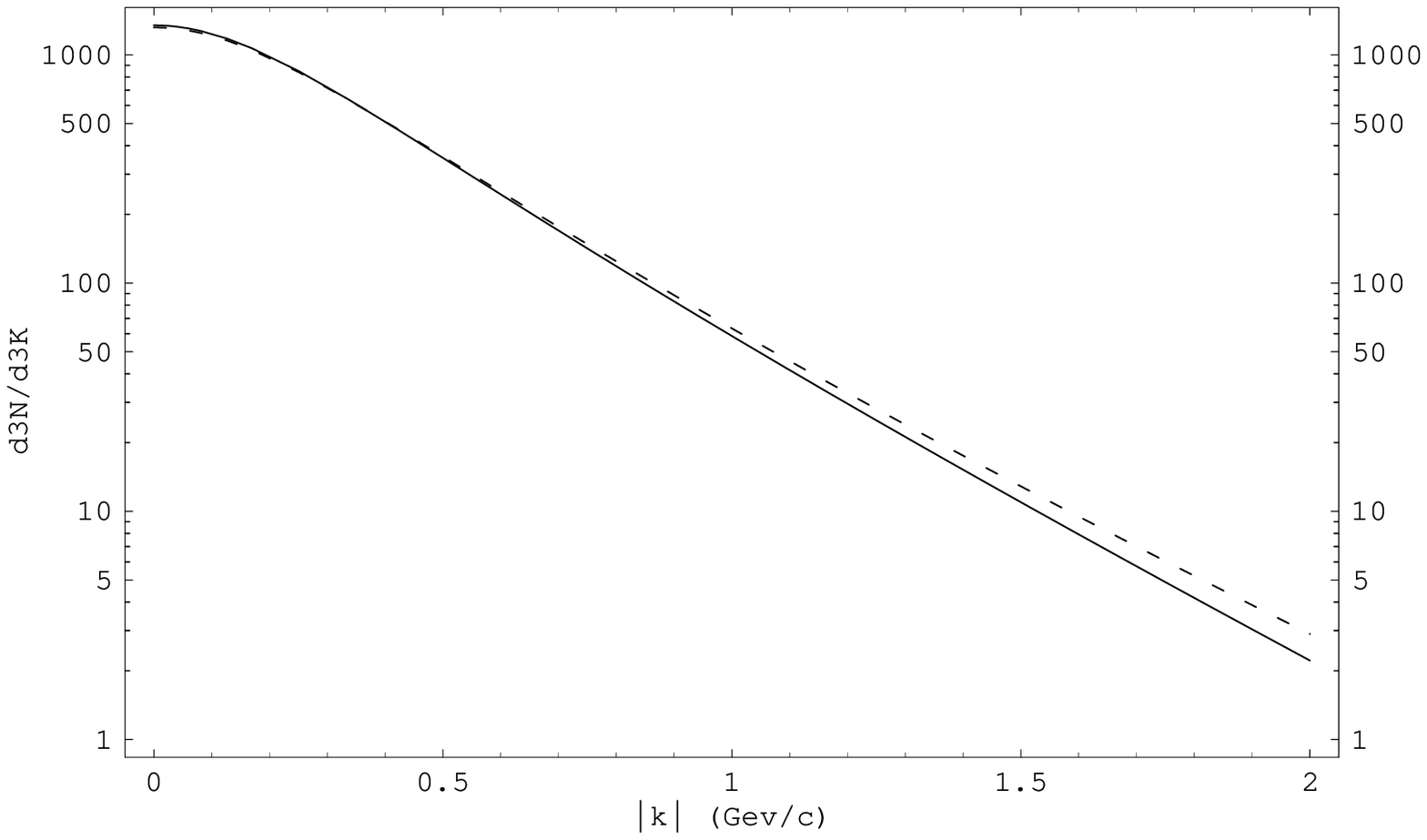}}\vskip-1.5cm
\caption{The two upper plots show results for the two-pion
correlation as function of $q_O$. The top plot has fixed values of
$|\vec K| = 0.8$ GeV/c and the factor $T/(\gamma v)=0.152$ GeV, with
$T=0.100$ MeV and $v\approx <v>=0.55$, and curves corresponding to
$q_{T_S}=1.0$ (dashed), $q_{T_S}=1.015$ (long-dashed) and
$q_{T_S}=1.10$ (continuous). The plot in the middle results from
fixing $|\vec K| = 0.6$
GeV/c, $T/(\gamma v)=0.109$ GeV,
$T=0.044$ MeV and $v\approx <v>=0.37$, with curves corresponding to
$q_{T_S}=1.0$ (long-dashed) and $q_{T_S}=1.015$ (continuous).
%The third plot used values $T/(\gamma v)=0.348$ GeV, with
%$T=0.130$ MeV and $v\approx <v>=0.55$, shows two sets of curves:
%the continuous and long-dashed ones correspond to $q_{T_S}=1.2$ and
%|\vec K| = 0.8$ GeV/c and |\vec K| = 0.5$ GeV/c, respectively. The
%broader and narrower short-dashed ones, to $q_{T_S}=1.$ and
%|\vec K| = 0.8$ GeV/c and |\vec K| = 0.5$ GeV/c, respectively.
The plot on the bottom show the spectrum corresponding to the
upper case, with corresponding curves for $q_{T_S}=1.0$
(continuous) and $q_{T_S}=1.015$ (dashed).}
\end{figure}

The correlation function versus the momentum difference,
$q=k_1-k_2$, showed a strong dependence on the combined variables
$K=k_1+k_2$ and $T/(\gamma v)$, where $v$ is the flow velocity
(with $v\sim 0.55$ as typical flow velocity at CERN/SPS, and
$v\sim 0.37$ as typical flow velocity at LBNL/Bevalac),
$\gamma=(1-v^2)^{-1}$, and $T$ is the temperature. Although not
explicitly illustrated here, the dependence on the average
momentum of the pair showed that the correlation curves became
narrower for increasing $K$, similarly to what was observed in the
results of Sect. II.6, even though the present model considers an
expanding system. Moreover, also similarly to the results on the
$Q$-boson interferometry, the correlation shrinks, i.e., the
probed effective region grows, for increasing $q_{T_s}$,
suggesting that long-range correlation could be present, in
association with Tsallis statistics. The study also showed that a
very small deviation from the Boltzmann statistics, corresponding
to $q_{T_s}=1.015$, could lead to clear differences in the
correlation function under some particular choices of the
combination $T/(\gamma v)$, as shown in the top right plot of Fig.
21. We see that, due the strong dependence on the combination
$T/(\gamma v)$, the search for such a deviation from the Boltzmann
statistics, as suggested by the Tsallis extensive statistics,
would be favored at lower energies. Nevertheless, it was shown
that the experimental data on event-by-event pion
transverse-momentum fluctuations, that could not be explained by a
model based on standard extensive statistics, was compatible with
a small deviation, i.e., for a value of the non-extensive
parameter  $q_{T_s}=1.015$, which inspired our analysis. Also,
NA35 Collaboration data on pion transverse momentum distribution
from $S+S$ collisions at SPS showed a better agreement with fit
based on non-extensive statistics, with $q_{T_s}=1.038$\cite{ALQ},
mainly for the tail of the distribution, which is of power-law
type.

The above approach to the problem, however, is very simple, since
the connection of $q_{T_s}$ to the space-time variables was
considered only through the non-decoupled phase-space present in
the adopted model\cite{spratt}. Nevertheless, the possibility is
not excluded that a more general form for the Wigner function, as
the one derived in Eq.(\ref{wignergen}) of Section II.6.2, would
be more appropriate. Maybe long-range correlations, as suggested
by Tsallis statistics, would not allow for reducing the Wigner
function to its conventional form, as in Eq.(\ref{wigner}).

\subsection{II.8~ PHASE-SPACE DENSITY}

Victor Vizcarra-Ruiz\cite{vvr}, another student working with me,
investigated a suggestion made by George
Bertsch\cite{bertschphsp}, according to which the average
phase-space density could be estimated by means of the two-pion
correlation function and the single-particle spectrum. In his
Master dissertation, a generalization of Bertsch's suggestion is
proposed, by applying the wave-packet formalism proposed in
Ref.\cite{pgg}.

Bertsch's proposition could be briefly summarized as follows. He
considered that, in ultra-relativistic heavy-ion collisions, the
local phase-space density in the final state is frozen and gives a
measure of the dynamics in the priory interacting region. He
starts by converting the source function to an equivalent one at a
common instant, $t_0$, times the phase-space density at that time,
i.e.,
\begin{equation}
g(x,\vec K)\rightarrow \delta(t-t_0) f(\vec r, \vec K)/(2\pi)^3
.\label{bertschwigner}\end{equation} Using the Wigner formulation
for the spectrum $ \frac{d^3 N}{d^3k}=\int d^2x g(x,\vec K) $ and
assuming the approximate validity of the relation $d^3N/d^3k_1
\approx d^3N/d^3k_2 \approx d^3N/d^3K$, independently of the
momentum difference $\vec q$, he wrote the average phase-space
density as
%\begin{eqnarray}
\begin{equation}
<f>_{\vec K}=\frac{\int d^3r f^2({\vec r},{\vec K})}
{\int d^3r f({\vec r},{\vec K})}
\approx \frac{d^3 N}{d^3k} \int d^3 q [C({\vec q},{\vec K}) - 1]
. \label{averagef}\end{equation}

For instance, in case the experimental single-inclusive
distribution could be well reproduced by the expression
\begin{equation}
\frac{d^3 N}{d^3k}=\frac{1}{E_K}\frac{dN_y}{dy}\frac{e^{-K_T/T}}{2\pi T^2}
\label{expspec}\end{equation}
and the correlation function by Eq.(\ref{bertsch}), the
phase-space density would be written as
\begin{equation}
<f>_{\vec K}=\frac{\lambda\sqrt{\pi}}{2 E_K T^2}\frac{dN_y}{dy}
\frac{e^{-K_T/T}}{R_OR_SR_L}
\,\label{phsp}\end{equation}
since the right-hand-side of Eq.(\ref{averagef}) results in
$\frac{\lambda \pi^{3/2}}{R_OR_SR_L}$.

The generalization proposed in \cite{vvr} was derived using the
powerful formulation based on the Wigner formalism\cite{pgg} and
summarized in Section II.2.2, applying to the case of two-boson
interferometry. In this first approach, we tried to keep our
derivation as close as possible to the one proposed by Bertsch, so
that the differences could be clearly seen. With this in mind, we
maintained expression (\ref{bertschwigner}) for the
non-relativistic source function as suggested in
\cite{bertschphsp}.

The single-particle distribution (spectrum) is then written as
\begin{eqnarray}
\left[E_{k_i}\frac{d^3N}{d^3k_i}\right]_\Delta\!\!\!\!\!\!
&=&\!\!\!\frac{1}{(2\pi)^3(2\pi\Delta p)^{\frac{3}{2}}}
\int \!\!d^3r d^3\!p \;f(\vec r,\vec p)\; e^{-(\vec p - \vec k_i)^2/
2\Delta p^2}\nonumber \\
&\approx&\frac{1}{(2\pi)^3} \int d^3r d^3p \;f(\vec r,\vec p,\vec K;\Delta p)
,\!\!\!\!\!\!\label{genspectrum}\end{eqnarray}
where we have defined
\begin{equation}
f(\vec r,\vec p,\vec K;\Delta p)=
\frac{1}{(2\pi\Delta p)^{\frac{3}{2}}} \; f(\vec r,\vec p)\;
e^{-(\vec p - \vec K)^2}
.\label{fgen}\end{equation}

The average phase-space density is then defined as
\begin{eqnarray}
&&<f(\vec K;\Delta p,\Delta x)>=\nonumber\\
&& \frac{\int d^3r d^3p
\;f(\vec r,\vec p,\vec K;\Delta p,\Delta x)\;f(\vec r,\vec p,\vec K;\Delta
p)}{\int d^3r d^3p\;f(\vec r,\vec p,\vec K;\Delta p)}
.\label{averf}\end{eqnarray}

Then, the reformulation of the average phase space density in terms
of the generalized Wigner formalism was derived as
\begin{equation}
<f(\vec K;\Delta p,\Delta x)> \approx\!\!
\left[\frac{d^3N}{d^3K}\right]_\Delta \int d^3q [C_\Delta(\vec
q,\vec K) -1] .\label{averfgen}\end{equation}
\begin{figure}[!hbt22]
\hskip2mm\resizebox{13pc}{!}{\hskip1.1cm\includegraphics{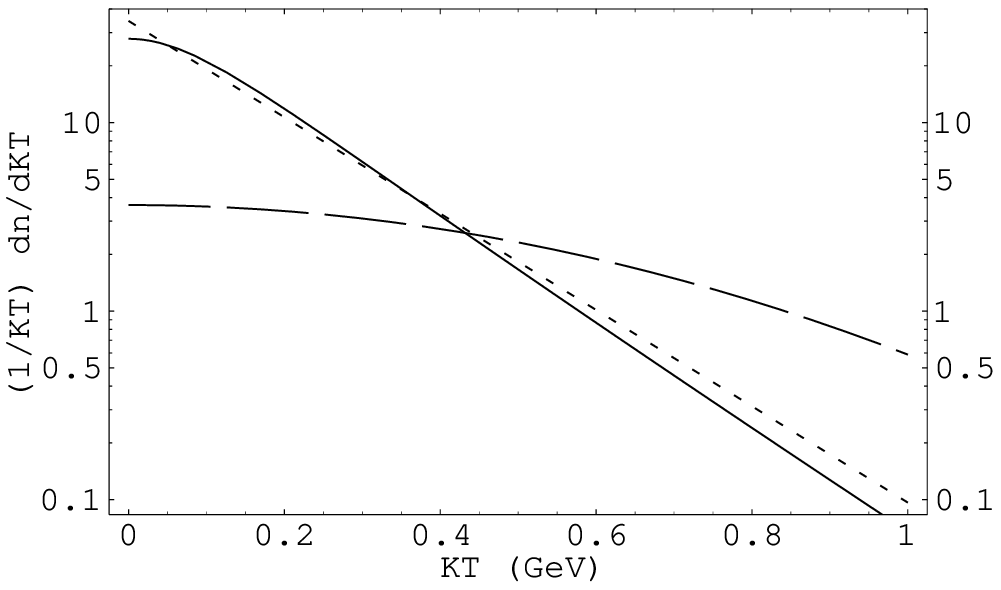}}\vskip-2.2cm
  \resizebox{13pc}{!}{\includegraphics{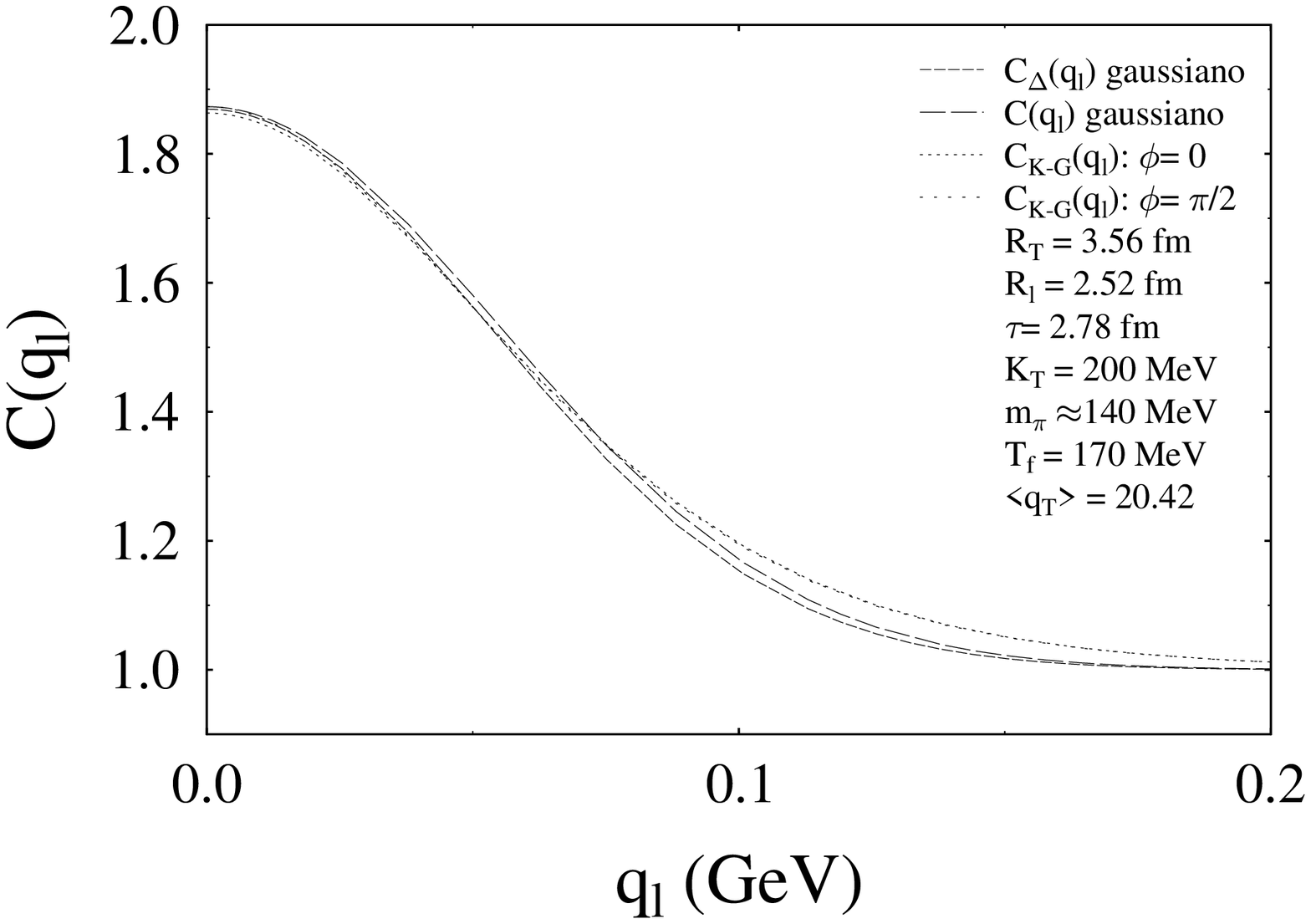}}\vskip-2.6cm
  \resizebox{13pc}{!}{\includegraphics{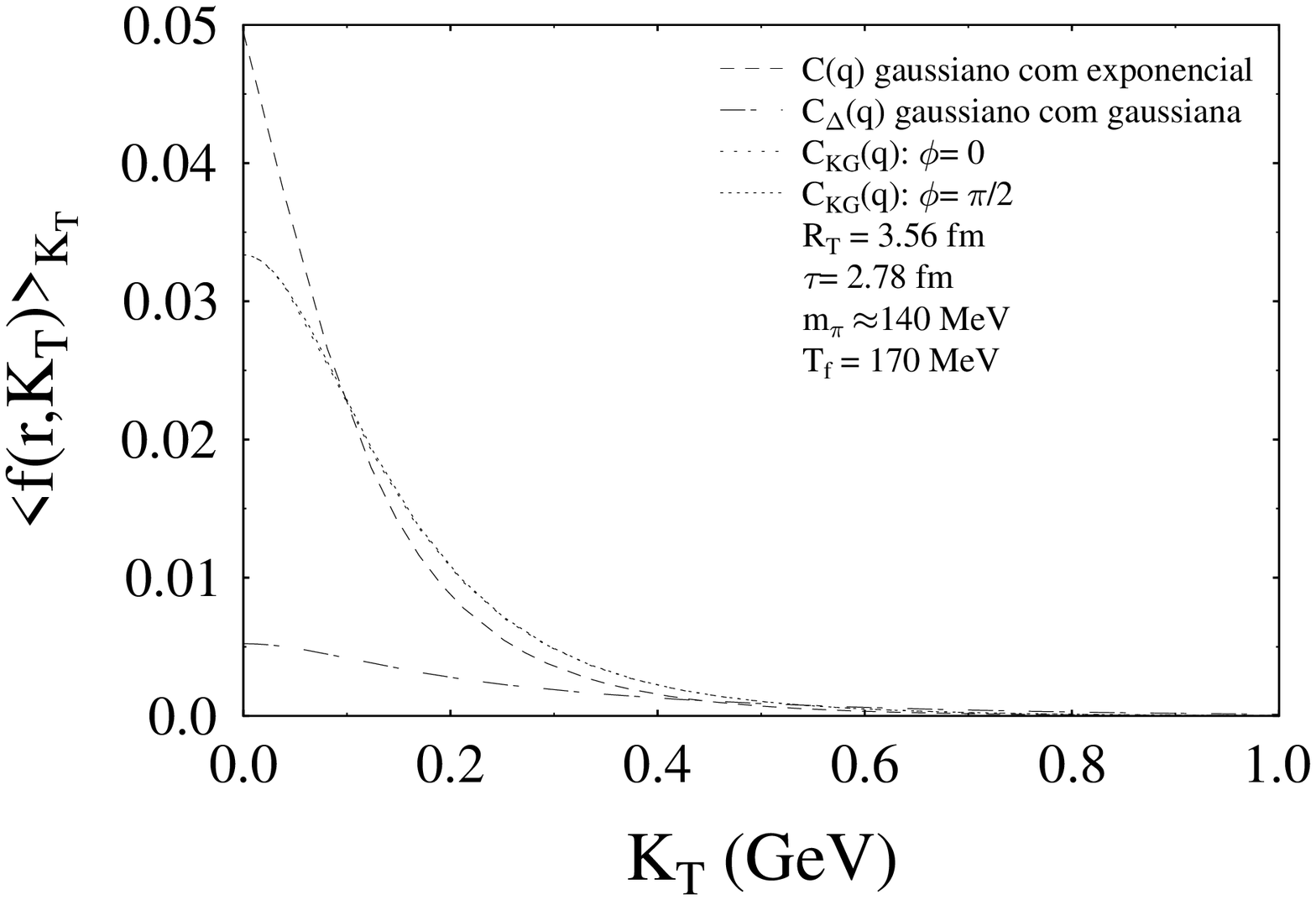}}
\caption{The top plot shows the pion spectral distribution. The
dashed line represents the transverse distribution in Eq.
(\ref{expspec}), the continuous one, the Kolehmainen-Gyulassy
type, and the long-dashed curve, the Gaussian spectral
distribution. The plot in the middle shows the correlation
function in terms of $q_L$, for all the parameterizations used:
Gaussian without and with the inclusion of the wave-packets (two
bottom ones), as well as the one on top corresponding to minimal
packets (or Kolehmainen-Gyulassy). The plot on the bottom shows
the estimate for the average phase-space (divided by the integral
in rapidity), for all the models: the highest curve corresponds to
the simple Gaussian parametrization for the source with
exponential spectrum, the lowest one, to the Gaussian source with
Gaussian spectrum and minimal wave-packets, and the middle curve,
to the Kolehmainen-Gyulassy type.}
\end{figure}

With the definition in Eq.(\ref{fgen}), we see from Eq.
(\ref{averfgen}) that the form of expression for the average phase
density $<f>$ does not change significantly with the introduction
of the wave-packets coming from the generalization of the Wigner
formulation for interferometry,. What actually changes is the way
to estimate the spectrum and the two-pion correlation function,
where the wave packets are implicitly present.

For illustrating an application of Eq.(\ref{averfgen}), we can look
into the simplest example, the Gaussian emission function, similarly
to the one proposed by Bertsch, but without splitting the transverse
variables, i.e.,
\begin{equation}
C(k_1,k_2) = 1 + \; e^{-q^2_{\small T} R^2_T} e^{-q^2_{\small L} R^2{\small L}}
\;,\label{corrwig}\end{equation}
whereas the correspondent expression in the generalized Wigner
formulation of Ref.\cite{pgg} is
\begin{equation}
C(k_1,k_2) = 1 +  \; e^{-q^2_{\small T} R^2_{\Delta,{\small T}}}
e^{-q^2_{\small L} R^2_{\Delta,{\small L}}}
\;,\label{corrgenwig}\end{equation} from where we see that the
intercept parameter,$\lambda$, is implicitly being fixed to unity.
In Eq. (\ref{corrgenwig}), we have
$R^2_{\Delta,i}=R^2_i(K)+1/(4m_\pi T_f)$. Thus, we see that the
results for the radial parameters in the standard Wigner
formulation would correspond to the limit $\Delta x\sim1/\Delta p
\rightarrow 0$. The parameters used in this example are:
$R_T=3.56$ fm, $R_L=2.52$ fm, as suggested by experimental results
from the BNL/AGS\cite{e859}, $T_f=140$ MeV, and minimum packets
were considered, i.e., $\Delta x \Delta p=1/2$, being $\Delta p
\approx \sqrt{m_\pi T_f}$. Relative to these values of radii, the
factor $1/\Delta p$ is very small ($0.5$ fm, or $\approx 4\%
$).%, so the wave packet correction is not significant.
Consequently, the corrections to the correlation function and to
$<f>$ are expected to be small, which is indeed what is observed.
Nevertheless, we should notice that the time dependence was
supposed to be a delta function, similar to Eq.
(\ref{bertschwigner}), to keep the analysis simple and closer to
Bertsch's proposition, and for emphasizing the influence of the
wave-packets. However, we know that the duration of the emission
has an enormous influence into the correlation function and the
expectations are that more pronounced differences would be seen if
a broader time interval was introduced, for instance, by Gaussian
time spread instead of the delta one used.

The proposed extension of the average phase-space estimate was
applied to some examples. The first was the static Gaussian case,
with and without the inclusion of wave packets. Another one was
the case corresponding to the Bjorken picture, IOC, which, for
minimal packets in the generalized Wigner formalism, is reduced to
the Kolehmainen-Gyulassy solution, where the spectrum is close to
an exponential. In the generalized Wigner formalism, however, it
is possible to choose other possibilities for parameterizing the
spectrum and, for a test, a Gaussian version of the momenta
distribution is tried, although it is well known that the
experimental results favor an exponential type of parametrization.

In the estimate of the average phase-space density, a strong
sensitivity to the adopted spectral shape was observed. The study
compared cases of static Gaussian source distribution, having
exponential and Gaussian corresponding to minimum packets, with
the Bjorken type from the Kolehmainen-Gyulassy approach. This is
shown in Fig. 22.

\subsection{II.9~ THE RHIC PUZZLE}

The new millennium started together with the first run of the
Relativistic Heavy Ion Collider, at BNL. The main preliminary
surprising result released was, coincidentally, in HBT. Since The
conference Quark Matter 2001, in January of that year, it has been
known as the RHIC Puzzle, first seen by the STAR
collaboration\cite{star} and later confirmed by PHENIX
Collaboration\cite{phenix}. The result was related to the ratio of
two of the radii parameters, $R_{out}/R_{sid}$ ($R_{out} \equiv
R_O$ is the transverse radius in the direction of the transverse
average momentum, ${\bf K_T}$, whereas $R_{sid}\equiv R_S$ is its
component orthogonal to ${\bf K_{T}}$). The predictions based on
microscopic models, such as UrQMD, suggested that the ratio would
either grow indefinitely (considering  hadronic re-scattering
only) or increase up to $K_{T}>0.15$ GeV/c and then decrease,
apparently saturating towards unity at large $K_{T}$ (when
combining hydrodynamics and cascading). In any case, the ratio was
above unity. The data, however, indicated an opposite trend, i.e.,
the data points for that ratio appeared to be steadily decreasing,
reaching values around 0.8 at $K_{T}\approx 0.4$ GeV/c.
Hydrodynamical models, such as proposed by U. Heinz and P.
Kolb\cite{upkolb}, failed in general to describe the experimental
results for the individual radii, approaching data only if they
considered that the system would immediately freeze-out at the
hadronization point. Even in that case, however, they did not
succeed in describing the ratio.

\subsection{II.9.1~ {\bf\sl HOT TAMALE} MODEL}

Challenged as well by this puzzle, Larry MacLerran and myself
tried and built a model assuming that the QGP, initially formed at
RHIC, constituted an opaque source. The opacity was implemented by
considering a model where pions are emitted from the surface of
the system, at fixed radius, all along its lifetime, from its
formation at $\tau_0$, up to the freeze-out. We neglected
transverse flow for simplicity, considering the ideal Bjorken
picture of 1+1 longitudinal expansion. The phase transition
started at $\tau_c$ (at temperature $T_c=175$ MeV, to be
consistent with lattice Monte-Carlo data), ending at $\tau_h$. The
system, pions only, further expanded until reaching $\tau_f$ (at
$T_f=150$ MeV, to be consistent with the energy per particle at
RHIC and to fit the $p_T$ distribution of pions), when it broke
up, in a volumetric emission.

Many of the features of the model we propose are embodied in the
hydrodynamic computations of Heinz and Kolb\cite{upkolb}, the
essential difference being the treatment of surface emission. The
energy was emitted already from the partonic phase.  This quark
and gluon matter was assumed to be directly converted into a flux
of pions with the same energy and a blackbody distribution at the
temperature of emission.  This energy conservation condition
allowed us to directly take a flux of gluons and quarks and
convert it into a spectrum of purely pions.  No complex mechanism
for the QGP hadronization was considered in detail, although
hadronization must take place. In other words, in first
approximation, we considered the evaporation of ``gluons'' and
``quarks'' (as hadronized pions) from the external surface of the
system in the same way as emission of pions, except for the
different number of degrees of freedom.

We estimated the emitted energy and the total entropy at each
stage. In the initial phase, lasting from $\tau_0$ to $\tau_c$,
the emitted energy as a function of time was estimated considering
the emission by an expanding cylinder of transverse radius $R_T$
and length $h$, in the time interval between $\tau$ and
$\tau+d\tau$, as  \vskip-5mm
\begin{equation}
dE_{in} = - \kappa \sigma T^4 \; 2 \pi \; R_T \; h \; d\tau -
\frac{1}{3} \; \sigma T^4\;  \pi \; R_T^2  dh
\; \; , \label{denergy}\end{equation}
%\vskip-3mm
where the first term comes from the black-body type of energy
radiated from the surface of the cylinder, and the second term
results from the mechanical work due to its expansion. The
$\kappa$ factor was introduced to take into account that the
system has some opacity to surface emission.  The constant
$\sigma$ is proportional to the number of degrees  of freedom in
the system. Integrating this equation we get for the energy
density (i.e., $\epsilon = E/V$)
\begin{equation}
\epsilon_{in} = \epsilon_0 (\frac{\tau_0}{\tau})^\frac{4}{3} \;
e^{\!-\frac{2 \kappa}{R_T}(\tau-\tau_0)} \; \; .
\label{energyin}\end{equation}
From Eq.(\ref{energyin}) we see that an extra factor, $e^{-\frac{2
\kappa}{R_T}(\tau-\tau_0)}$, is obtained in additional to that
coming from the Bjorken picture. Consequently, the temperature in
this model changes more rapidly, according to
\begin{equation}
T(\tau) = T_0 (\frac{\tau_0}{\tau})^\frac{1}{3}
e^{-\frac{\kappa}{2 R_T}(\tau-\tau_0)} \; \; (T_0 < T < T_c) \; \;
. \label{temp}\end{equation} The initial temperature was estimated
by equating the initial entropy to the number of produced pions
(pion yield) at the end, i.e.,
$%\begin{equation}
S_0 = \Gamma {\cal N} = \left[ (g_{\small g} + g_{\small q}) \times
(\frac{4}{3}) \frac{\pi^2}{30} T_0^3 \right] \pi R_T^2 \tau_0
,$%\; \; , \label{N}\end{equation}
where ${\cal N} \sim 1000$ is the average produced pion
multiplicity per unit of rapidity at RHIC, and
$\Gamma=S_\pi/N_\pi\approx 3.6$. Then, we get $T_0=411$ MeV. The
degeneracy factors, $g$, are given by the gluon degrees of
freedom, $g_{\small g} = 2 (spin) \times 8 (color) $, and the
quark/anti-quark degrees of freedom, $g_{\small q} = \frac{7}{8}[2
(spin) \times 2 (q+\bar{q}) \times 3 (color) \times N_f
(flavor)]$, which add up to $g_{\tiny qgp} = g_{\small g} +
g_{\small q}$. In the case of pions, the degeneracy factor is
$g_{\small \pi} = 3$.

The
initial time is estimated by means of the uncertainty principle,
i.e., $E_0 \tau_0 \sim 1$ ($\hbar c =1$) and $E_0 \approx 3T_0$.

\smallskip
\begin{center}
{\bf TABLE 4: Parameters of the {\sl Tamale} Model} \\
%\vskip0.5cm
\begin{table}[htb]
\label{table:1}
\newcommand{\m}{\hphantom{$-$}}
\newcommand{\cc}[1]{\multicolumn{1}{c}{#1}}
%\renewcommand{\tabcolsep}{1.2pc} % enlarge column spacing
%\renewcommand{\arraystretch}{1.2} % enlarge line spacing
%\begin{tabular}{@{}lllllll}
\begin{tabular}{@{}|c|c|c|c|c|c|c|}
\hline
$\kappa$& $\tau_0$&$\tau_c$&$\tau_h$&$\tau_f$&
${\cal S}/{\cal N}_{tot}$ & {\large ${\cal V}$}$/{\cal N}_{tot}$ \\
&($\frac{fm}{c}$)&($\frac{fm}{c}$)&($\frac{fm}{c}$)&($\frac{fm}{c}$)&
($\tau_0 \le \tau \le \tau_f$)&(at $\tau_f)$\\
\hline\hline
1&0.160&1.54&5.73&6.97&0.844&0.156\\
0.5&0.160&1.75&8.37&10.5&0.758&0.242\\
\hline
\end{tabular}\\[2pt]
%%The experimental values are given in ref. \cite{Eato75}.
\end{table}
\end{center}

During the mixed phase (i.e., $\tau_c < \tau < \tau_h$), the
temperature remains constant and the total entropy is conserved.
If the fraction of the fluid in the QGP phase is $f$, the fraction
in the pionic phase is ($1-f$), so that the entropy density is
written as $\tilde{s}=f\tilde{s}_{QGP}+(1-f)\tilde{s}_h$, where
\begin{equation}
f = \left( \frac{\tau_h \; e^{-\frac{2 \kappa}{R_T}(\tau-\tau_c)}-
\tau \; e^{-\frac{2 \kappa}{R_T}(\tau_h-\tau_c)}} {\tau_h - \tau_c
\; e^{-\frac{2 \kappa}{R_T}(\tau_h-\tau_c)}} \right)
\frac{\tau_c}{\tau} \; . \label{fractionf}\end{equation}
%\begin{equation}
%(1 - f) = \left( \frac{\tau- \tau_c \; e^{-\frac{2 \kappa}{R_T}(\tau-\tau_c)}}
%{\tau_h - \tau_c \; e^{-\frac{2 \kappa}{R_T}(\tau_h-\tau_c)}} \right)
%\frac{\tau_h}{\tau}
%\; \; . \label{fraction1-f}\end{equation}
%
Finally, the instants corresponding to the initial of the
first-order phase transition, $\tau_c$, its end, $\tau_h$, and the
breakup of the system, $\tau_f$, were estimated as
\begin{equation}
\tau_c \; e^{\frac{3 \kappa}{2 R_T} \tau_c} =
\left( \frac{T_0}{T_c} \right)^3 \; \tau_0 \;
e^{\frac{3 \kappa}{2 R_T} \tau_0}
\; \; , \label{tauc}\end{equation}
\begin{equation}
\tau_h \; e^{\frac{2 \kappa}{R_T} \tau_h} =
\left( \frac{g_g+g_q}{g_\pi} \right) \tau_c \;
e^{\frac{2 \kappa}{R_T} \tau_c}
\; \; , \label{tauh}\end{equation}
\begin{equation}
\tau_f \; e^{\frac{3 \kappa}{2 R_T} \tau_f} =
\left( \frac{T_c}{T_f} \right)^3 \; \tau_h \;
e^{\frac{3 \kappa}{2 R_T} \tau_h}
\; \; . \label{tauf}\end{equation}
\begin{figure}[!hbt23]
\includegraphics*[angle=-90, width=6cm]{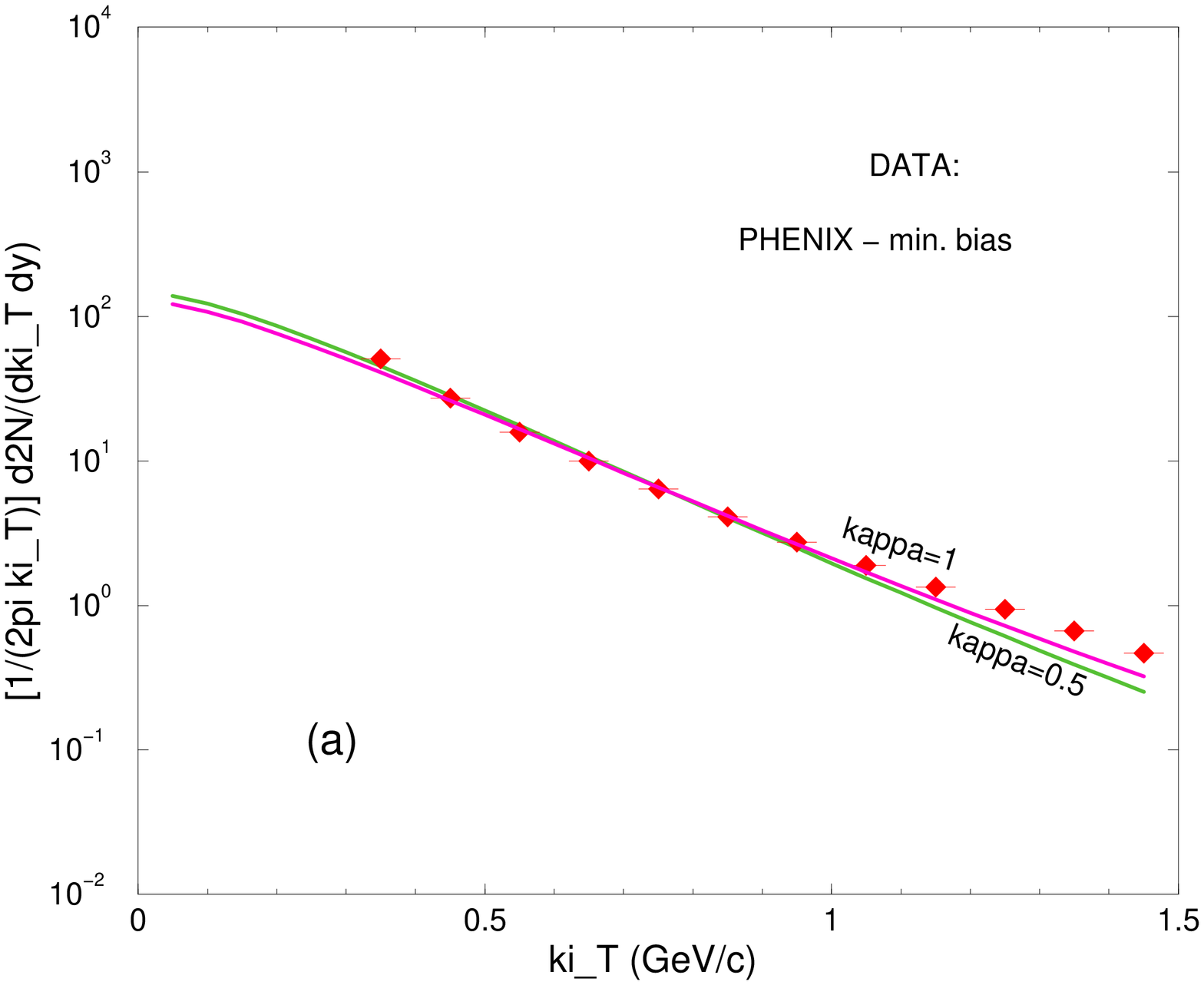}
\includegraphics*[angle=-90, width=6cm]{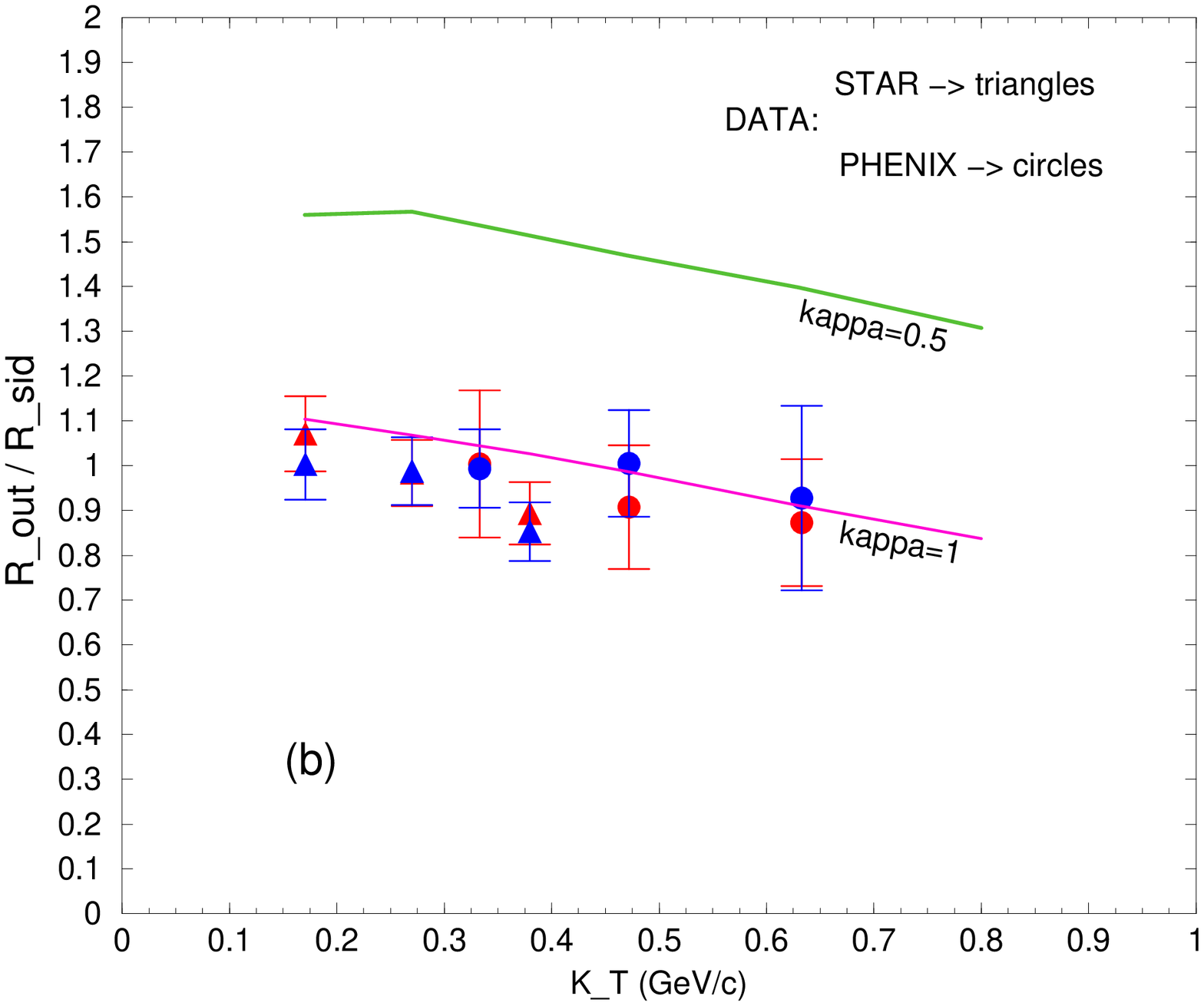}
\caption{Part (a) shows the prediction for the spectrum based on
the Tamale model. The points are from the minimum bias data from
PHENIX Collaboration. The emissivity $\kappa$ is indicated in each
curve. Part (b) shows results for the ratio $R_{out}/R_{sid}$ for
our model and for two values of the emissivity. The ratio
corresponding to $\kappa=1$ agrees very well with the PHENIX data
points, within the experimental error bars but the one with 50\%
emissivity is completely excluded by data. The plots were
extracted from Ref.\cite{sspqm02}. }
\end{figure}
Table 4 illustrates the time variables for two different
assumptions on the emissivity, $\kappa$. I also write the
estimated fraction of the particles emitted from the surface
during the period $\tau_0 \le \tau \le \tau_f$, ${\cal S}/{\cal
N}_{tot}$, relative to the total number of produced particles,
${\cal N}_{tot}$, as well as the remnant portion at freeze-out,
{\large ${\cal V}$}$/{\cal N}_{tot}$, then emitted from the entire
volume.
%
%\vskip-0.5cm
With the above parameters, we estimated the spectra and compared
with experimental data from RHIC/Phenix
Collaboration\cite{phenix}, restricted to the central rapidity
region. This is shown in Fig. 23.

%\bigskip\bigskip\bigskip
With this model we managed to roughly explain the trend of data
for $R_{out}$ but failed for $R_{sid}$, since it was independent
on $K_T$. This, however, was expected, since the flow was
neglected. The ratio, however, as can be seen from Fig. 23, came
up just right, through the middle of the data
points\cite{sspqm02}! Nevertheless, only much afterwards we did
realize our emissivity parameter, $\kappa$, which controlled the
surface emission, differed from the conventional one associated to
a black-body. In fact, only by then we realized that the system
would have to irradiate about four times more than an usual
black-body for matching the data points. We then concluded that it
was another strong evidence that flow should be included in the
estimates, which is still under investigation.

\subsection{II.9.2~  IC, SPheRio, CEM \& HBT }

In Ref.\cite{sghk}, Ot\'avio, Fr\'ed\'erique, Yogiro and Takeshi
analyze the effects of fluctuations in the initial conditions
(IC), and of the continuous emission (CE) model, in two-pion
interferometry within the hydrodynamical description of the
SPheRio code\cite{spherio}, comparing the results with the
RHIC/STAR data.

The IC are, for simplicity, usually chosen as highly symmetric and
smooth distributions of velocities and thermodynamical quantities.
These IC correspond to mean distributions of hydrodynamical
variables, averaged over several events. Nevertheless, for the
typical finite systems formed in high energy collisions, large
fluctuations are expected, varying from event to event. Moreover,
the IC on the event-by-event basis often show small high-density
spots\cite{sghk} in the energy distribution. Being so, it would be
expected that such spots would manifest themselves in the particle
emission, contributing to decrease the HBT radii. In order to
produce event-by-event fluctuating IC, in Ref.\cite{sghk} they
used the NeXus event generator\cite{nexus}: once the incident
nuclei and incident energy are given, it produces the
energy-momentum tensor distribution at the proper time
$\sqrt{t^2-z^2}$, in an event-by-event basis. This, together with
the baryon-number density distributions constitute the fluctuating
IC. No strangeness was introduced in the calculation.

On the other hand, the CE model we already discussed in Sec. II.5,
is a alternative picture for describing the final state of the
high energy collisions, i.e., the particle emission. According to
the CE model, this emission occurs not only from the sharply
defined freeze-out surface but also happens continuously, from the
whole expanding volume at different temperatures and times. The
main ingredients of the model are discussed in more detail by
Fr\'ed\'erique Grassi in her review, in this volume. The basic
formulation for CE interferometry was discussed in Sec. II.5.

In between the assumptions in the IC and the final stages
described by the CE picture, there is the system evolution. This
is considered in Ref.\cite{sghk} by means of hydrodynamics,
developed in the code SPheRio, which was based on the {\sl
Smoothed Particle Hydrodynamics} (SPH), first used in astrophysics
and more recently adapted to nuclear collisions\cite{spherio}. SPH
uses discrete Lagrangian coordinates attached to small volumes
(``particles") with some conserved quantities, which they take as
the entropy and baryon number. For details about SPheRio and
hydrodynamics, see the review by Yogiro Hama and Takeshi Kodama,
in this volume.

Some of the results shown in Ref.\cite{sghk} are summarized in Fig.
24.
\begin{figure}[!hbt24]
\hskip-1cm  \resizebox{12.5pc}{!}
{\includegraphics{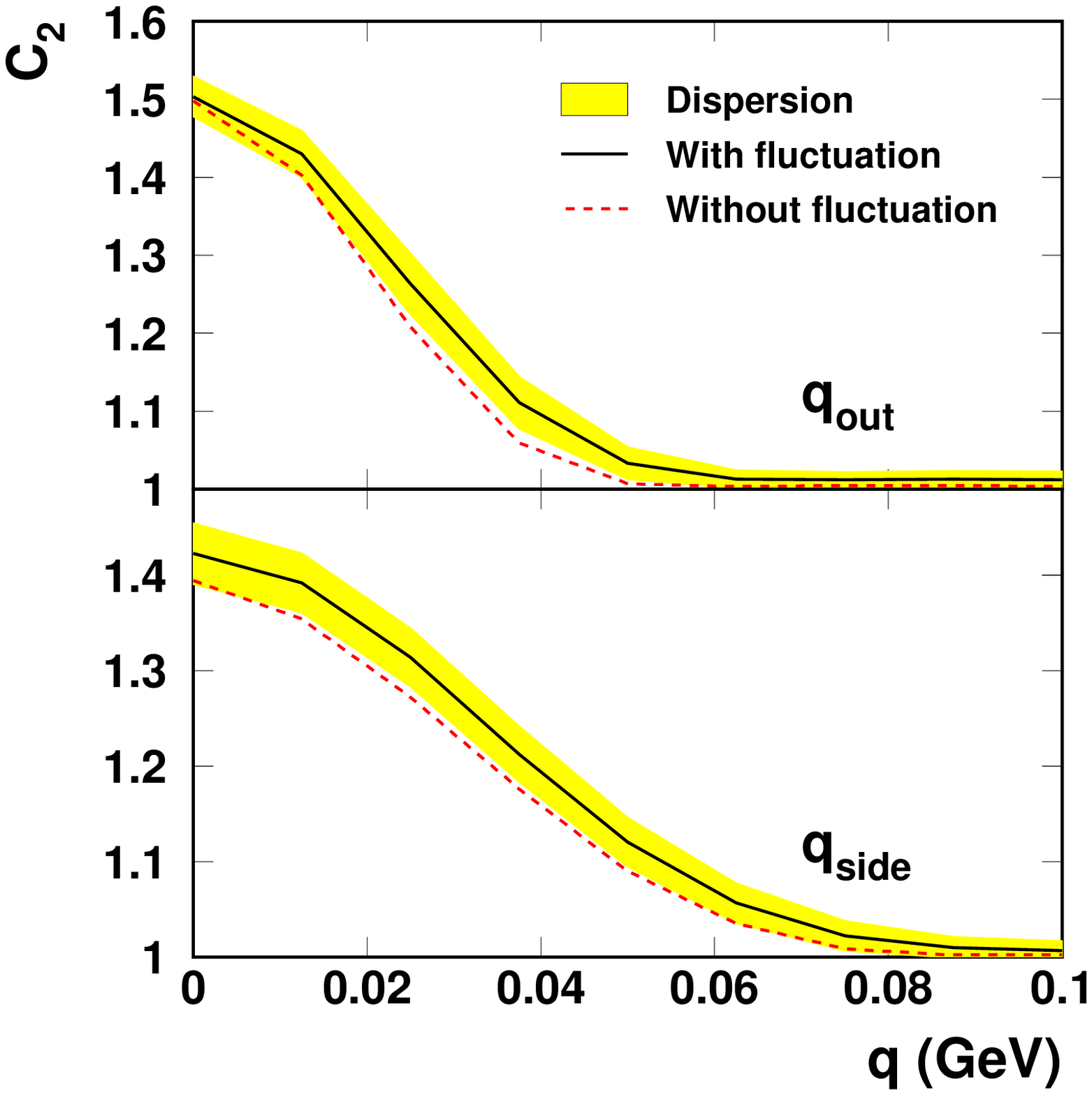}}\hskip-0.7cm\resizebox{12pc}{!}{\includegraphics{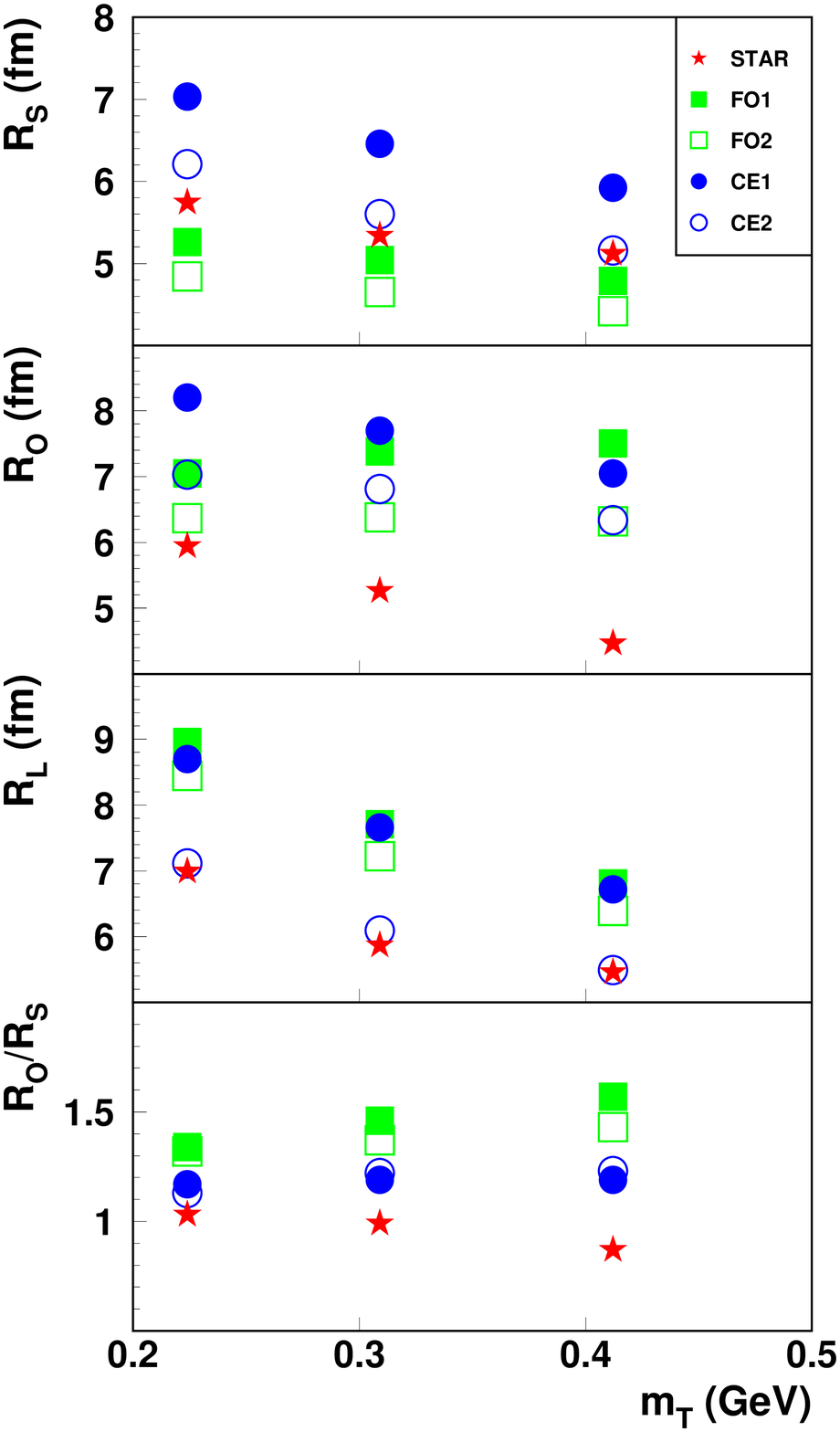}}
 \caption{The plot on the left shows correlation functions
 assuming fluctuating IC and average IC, with sudden freeze-out. The
 $q_{O,S,L}$ that are not displayed in the horizontal axes were
 integrated in the interval $0\le q_{O,S,L}\le35$ MeV. The plot on
 the right shows the radii $R_O, R_S, R_L$ together with the ratio
 $R_O/R_S$, as a function of $m_T$. The complement 1 stands for
 average IC and 2, for fluctuating IC. The plots were extracted from
 Ref.\cite{sghk}. }
\end{figure}
To test the effect of fluctuations in the IC, a sudden freeze-out
is initially assumed, at $T_{f.o.}=128$ MeV. In the left part of
Fig. 24 the correlation function $C_2$, averaged over 15
fluctuating events, is compared  with the corresponding $C_2$ with
average IC (i.e., no fluctuations). The IC fluctuations are
reflected in the fluctuations seen in the correlation functions.
Averaging the correlations over the fluctuating range, we see that
the resulting $C_2$ is broader than the correspondent ones
computed with average IC. So, the effect of the IC fluctuations is
to reduce the effective radii extracted from the correlation
functions. Besides, the shapes of the resulting correlations with
fluctuating IC and with smoothed average IC, are different. On the
right-hand-side of Fig. 24, it is shown the $m_T$ dependence of
the HBT radii. These were estimated by fitting Gaussians to $C_2$.
Results based on the CE model, as well as RHIC data from STAR
Collaboration\cite{star} are also included in that plot.

From Fig. 24, we see that joint effects of the smooth IC with
sudden freeze-out make the $m_T$ dependence of the outwards radii,
$R_O$, flat or slightly increasing. The fluctuating IC push the
effective radii towards smaller values and flat $m_T$ dependence.
Adding the CE hypothesis to the analysis, we also see from Fig. 24
that all effective radii, $R_O$, $R_S$, and $R_L$ decrease with
increasing $m_T$. This is expected, since in the CE picture,
high-$k_T$ particles have a bigger chance of being emitted at
early times, when the system is hot, mostly from its surface. On
the contrary, low-$k_T$ particles are emitted later, when the
system has already cooled down and has larger size due to the
expansion. The role of the fluctuating IC in the CE case is mostly
of reducing the values of the effective radii, without
significantly changing its $m_T$ dependence. These two effects
together improve the description of the data points, as seen in
Fig. 24. Nevertheless, the ratio $R_O/R_S$ is still above unity
and does not describe the data points yet. Another ingredient may
still be lacking.

\subsection{II.10~ BACK-TO-BACK CORRELATIONS}

The Back-to-Back Correlations (BBC) have a much more recent
history. Under the historical perspective we are pursuing in this
review, it is worthy to briefly revise its origin. In 1991,
Andreev, Pl\"umber and Weiner\cite{bbcapw} wrote a paper in which
they pointed out the surprise existence of a new quantum
statistical correlation, among particle-antiparticle pairs. Thus,
$\pi^+\pi^-$ correlations of this type would be similar to
$\pi^0\pi^0$ (since $\pi^0$ is its own antiparticle) but entirely
different from the Bose-Einstein correlations we discussed in the
previous sections. They were related to the expectation value of
the annihilation (creator) operator, more specifically,
$<\hat{a}^{(\dagger)}(k_1) \hat{a}^{(\dagger)}(k_2)>\ne0$,
analogous to what is observed in two-particle squeezing in optics,
where the averages are estimated using a density matrix that
contains squeezed states. Although not all they discussed was
correct, they pointed out that $C(\pi^+\pi^-)>1$ and
$C(\pi^0\pi^0)>1$, reflecting particle-antiparticle quantum
statistical effects. Later, Sinyukov\cite{YS94}, discussed a
similar effect for $\pi^+\pi^-$ and $\pi^0\pi^0$ pairs, claiming
that the effects would be due to inhomogeneities in the system, as
opposite homogeneity regions in HBT, which comes from a
hydrodynamical description of the system evolution. He used Wick's
theorem for expanding the two-particle inclusive distribution in
terms of bilinear forms.

We can understand the origin of this effect in a simpler way in terms
of creation and annihilation operators, in the $\pi^0$ case. For instance,
the single-inclusive distribution is written as
\begin{equation}
N_1({\mathbf k}_i) =  \omega_{{\mathbf k}_i}\frac{d^3N} {d{\mathbf
k}_i}
 =   \omega_{{\mathbf k}_i}  \langle \hat{a}^{\dagger}_{{\mathbf k}_i}
 \hat{a}_{{\mathbf k}_i} \rangle
\;, \label{spec1}\end{equation}
and, after the decomposition that follows from Wick's theorem, the
two-particle distribution is written as
\begin{eqnarray}
&& N_2({\mathbf k}_1,{\mathbf k}_2)  =
\omega_{{\mathbf k}_1} \omega_{{\mathbf k}_2}
\langle \hat{a}^\dagger_{{\mathbf k}_1} \hat{a}^\dagger_{{\mathbf k}_2}
\hat{a}_{{\mathbf k}_2} \hat{a}_{{\mathbf k}_1} \rangle
\nonumber\\
&=&
\omega_{{\mathbf k}_1} \omega_{{\mathbf k}_1}
\{ \langle \hat{a}^{\dagger}_{{\mathbf k}_1} \hat{a}_{{\mathbf k}_1} \rangle
\langle \hat{a}^{\dagger}_{{\mathbf k}_2} \hat{a}_{{\mathbf k}_2} \rangle
+ \langle \hat{a}^{\dagger}_{{\mathbf k}_1} \hat{a}_{{\mathbf k}_2} \rangle
\langle \hat{a}^{\dagger}_{{\mathbf k}_2} \hat{a}_{{\mathbf k}_1} \rangle
\nonumber\\
&& \hskip.2cm + \hskip.2cm \langle \hat{a}^{\dagger}_{{\mathbf k}_1}
\hat{a}^{\dagger}_{{\mathbf k}_2} \rangle
\langle \hat{a}_{{\mathbf k}_1} \hat{a}_{{\mathbf k}_2} \rangle \}
\;. \label{spec2}\end{eqnarray}

In Eq. (\ref{spec2}), the first term corresponds to the product of
the two single-inclusive distributions, the second one gives rise
to the Bose-Einstein identical particle correlation, reflecting
the last position, just after particle emission. The third term,
absent in the $\pi^\pm \pi^\pm$ case, is the responsible for the
particle-antiparticle correlation. Under analogous conditions as
HBT, this last term usually vanishes but is non zero if the
Hamiltonian of the system is of the type $H=H_0+H_1$, where $H_0$
is the free part in the vacuum (final particles) and $H_1$
represents the interaction of quasi-particles, resulting in an
effective shift of their masses. Alternatively, as in the pioneer
work in Ref. \cite{bbcapw}, this could be similar to having a {\sl
chaotic superposition of coherent states} and the density matrix
containing {\sl squeezed states}.

Later, Andreev and Weiner\cite{andwei} continued the discussion.
They considered that, in high energy collisions, a pion blob
(strongly interacting pion system, seen as a liquid) is formed and
later undergoes a sudden freeze-out, where the pionic system
(having background and excited states) is rapidly converted into
free pions. They postulate that, at the moment of transition, they
can relate in-medium creation and annihilation operators
($\hat{b}^{\dagger},\hat{b}$) to the corresponding free ones
($\hat{a}^{\dagger},\hat{a}$), by means of a squeezing
transformation, with a real squeezing parameter
[$r=\frac{1}{2}\ln(E_{fr}/E_{in})$]. In the same year, Asakawa and
Cs\"org\H o proposed a similar structure to this previous
approach, but relating in-medium operators  to free ones by means
of a Bogoliubov transformation. They suggested that, while the
in-medium Hamiltonian, $H_1$,  is diagonalized in terms of
$\hat{b}^{\dagger},\hat{b}$, the free one is diagonalized by
$\hat{a}^{\dagger},\hat{a}$. They also proposed to observe hadron
mass modification in hot medium by means of {\sl Back-to-Back
Correlations}. However, the theory was not yet completely correct.

There were a few more tentative works by the two groups but the
correct approach was finally written in 1999, by Asakawa,
Cs\"org\H o, and Gyulassy\cite{acg}.

The formalism developed for the case of an infinite medium can be
summarized as follows. The in-medium Hamiltonian is written as
%\begin{eqnarray}
${H}  =   H_0 - \frac{1}{2} \int d {\mathbf x}~ d {\mathbf y}
\phi({\mathbf x})
\delta M^2({\mathbf x}-{\mathbf y}) \phi({\mathbf y}),
\;\;$%\label{ham} \\
where
$H_0  =  \frac{1}{2} \int d {\mathbf x} \left(
\dot{\phi}^2+ |\nabla \phi|^2 + m^2 \phi^2  \right),
\;$%\end{eqnarray}
is the asymptotic (i.e., free) Hamiltonian, in the rest frame
of matter.
The scalar field $\phi({\mathbf x})$ in the Hamiltonian $H$
corresponds to
quasi - particles that propagate with a momentum-dependent
medium-modified effective mass,
which is related to the vacuum  mass, $m$,  via
$$ m_*^2({|{\mathbf k}|}) =  m^2 - \delta M^2({|{\mathbf k}|}).$$
The mass-shift is assumed to be limited to long wavelength
collective modes:
$\delta M^2({|{\mathbf k}|}) \ll m^2$ if $|{\mathbf k}| >
\Lambda_s$.
Given such a mass shift, the dispersion relation is modified to
$\Omega_{\mathbf k}^2 =\omega^2_{\mathbf k}-\delta M^2
(|{\mathbf k}|)$,
where $\Omega_{\mathbf k}$ is the frequency of the in-medium mode
with momentum ${\mathbf k}$.

The in-medium, thermalized annihilation (creation) operator is denoted by
$b_{\mathbf k}$ ($b^\dagger_{\mathbf k}$), whereas the
corresponding asymptotic operator for the observed quantum with four-momentum
$k^{\mu}\, = \, (\omega_{\mathbf k},{\mathbf k})$,
$\omega_{\mathbf k}^2= m^2 + {\mathbf k}^2$ ($\omega_{\mathbf k}
> 0$) is denoted by $a_{\mathbf k}$ ($a^\dagger_{\mathbf k}$).
These operators are related by the Bogoliubov transformation,
i.e., $ a_{{\mathbf k}_1} = c_{{\mathbf k}_1} b_{{\mathbf k}_1} +
s^*_{-{\mathbf k}_1} b^\dagger_{-{\mathbf k}_1} $, which is
equivalent to a squeezing operation. For this reason, $r_{\mathbf
k}$ is called mode-dependent {\bf squeezing parameter}. Note that
the relative and the average pair momentum coordinates are
$q^0_{1,2}= \omega_1-\omega_2,$  ${\mathbf q}_{1,2}={\mathbf k}_1-
{\mathbf k}_2$, $E_{i,j}=\frac{1}{2}(\omega_i+\omega_j)$,  and
${\mathbf K}_{1,2}=(1/2)({\mathbf k}_1+ {\mathbf k}_2)$. For
shortening the notation, we wrote the squeezed functions as
$c_{i,j} = \cosh[r({i,j},x)]$ and $s_{i,j} = \sinh[r({i,j},x)]$,
where
\begin{eqnarray}
r(i,j,x) &=& \frac{1}{2}\log \left[( K^{\mu}_{i,j} u_\mu (x))/
(K^{* \nu}_{i,j}(x) u_\nu(x) ) \right]
\nonumber\\
&=& \frac{1}{2} \log
\left[\frac{\omega_{k_i}(x)+\omega_{k_j}(x)}{\Omega_{k_i}(x)+
\Omega_{k_j}(x)} \right].
\end{eqnarray}
Also, $n_{i,j}$ is the density distribution, which is taken as the
Boltzmann limit of the Bose-Einstein distribution, i.e., $
n^{(*)}_{i,j} (x) \approx \exp{\{ - [K^{(*)^\mu}_{i,j} u_\mu(x) -
\mu (x)]/T(x)\}} $, where the symbol $^*$ implies the use of
in-medium mass, whereas it should be dropped where there is no
mass-shift.

In cases where the boson is its own anti-particle, as for $\pi^0
\pi^0 $ or $\phi \phi$ correlations, the full correlation function
consists of a HBT part (related to the chaotic amplitude,
$G_c(1,2)$) together with a BBC portion (related to the squeezed
amplitude, $G_s(1,2)$), as shown below
\begin{eqnarray}
&&C_2({\mathbf k}_1,{\mathbf k}_2)  =
 \frac{N_2({\mathbf k}_1,{\mathbf k}_2)}
 {N_1({\mathbf k}_1) N_1({\mathbf k}_2)}
\nonumber \\
&&=
1 + \frac{| G_c(1,2) |^2}{G_c(1,1) G_c(2,2) }
+ \frac{| G_s(1,2) |^2}{G_c(1,1) G_c(2,2) },
\label{fullcorr}
\end{eqnarray}
being the invariant
single-particle and two-particle momentum distributions given by
\begin{eqnarray}
G_c(1,2)&=&
\sqrt{\omega_{{\mathbf k}_1} \omega_{{\mathbf k}_2} }
\langle \hat{a}^\dagger_{{\mathbf k}_1} \hat{a}_{{\mathbf k}_2}\rangle,
\nonumber \\
G_s(1,2) &=&
\sqrt{\omega_{{\mathbf k}_1} \omega_{{\mathbf k}_2} }\langle \hat{a}_{
{\mathbf k}_1} \hat{a}_{{\mathbf k}_2} \rangle , \nonumber\\
G_c(i,i) &=& G_c(k_i,k_i) = N_1({\mathbf k}_i)
.\label{rand}\end{eqnarray}

So far, the derivation considered an infinite system. The effects
of finite size on BBC will be considered afterwards in the text.
In Ref.\cite{acg}, they discussed the influence of finite emission
times, observing that the BBC in this case was suppressed when
compared to the instant emission. Including the finite emission
time as $\theta(t-t_0)\Gamma \exp[-\Gamma(t-t_0)]$, $C_2({\mathbf
k}_1,{\mathbf k}_2) - 1$ acquires a multiplicative factor
$[1+(\omega_{\mathbf k}-\omega_{-\mathbf k})^2/\Gamma^2]$. They
illustrated the maximum BBC, corresponding to $C_2({\mathbf
k},-{\mathbf k})$ for the $\phi$-meson, obtaining a considerable
magnitude for the BBC correlation function, in spite of
considering the time suppression. We will see this more explicitly
later.

\bigskip
As we discussed in the beginning of this section, the BBC is a
different type of correlation, discovered for boson-antiboson
pairs. More recently, Tam\'as Cs\"org\H o, Yogiro Hama, Gast\~ao
Krein, Prafulla K. Panda, and myself demonstrated that a similar
correlation existed between fermion-antifermion pairs, if the mass
of the fermions were modified in a thermalized medium. We already
saw in the Introduction that, regarding the HBT effect, identical
bosons have an opposite behavior as compared to identical
fermions, as illustrated in Fig. 4, where we see that quantum
statistics suppresses the probability of observing pairs of
identical fermions with nearby momenta, while it enhances such a
probability in the case of bosons. With respect to BBC, however,
we found out a very different situation: fermionic BBC are
positive and similar in strength to bosonic BBC! And, contrary to
the HBT correlations, the BBC are unlimited.

In the fermion BBC case, there are expressions similar to
Eq.(\ref{spec1}) and (\ref{spec2}),
\vskip-.5cm\begin{equation}
N_1({\mathbf k}_i) =  \omega_{{\mathbf k}_i}  \langle a^{\dagger}_{{\mathbf k}_i}
 a_{{\mathbf k}_i} \rangle \; \; ; \; \;
 \tilde{N}_1({\mathbf k}_i) = \omega_{{\mathbf k}_i}
 \langle \tilde{a}^{\dagger}_{{\mathbf k}_i}
 \tilde{a}_{{\mathbf k}_i} \rangle \;, \label{spec1f}\end{equation}
\vskip-.8cm\begin{equation} N_2({\mathbf k}_1,{\mathbf k}_2)  =
\omega_{{\mathbf k}_1} \omega_{{\mathbf k}_2} \langle
a^\dagger_{{\mathbf k}_1} \tilde{a}^\dagger_{{\mathbf k}_2}
\tilde{a}_{{\mathbf k}_2} a_{{\mathbf k}_1} \rangle \; .
\label{spec2f}\end{equation} \vskip-.2cm In the above
expressions,$<\hat{O}>$ denotes the expectation value of the
operator $\hat{O}$ in the thermalized medium and $a^\dagger, a,
\tilde{a}^\dagger, \tilde{a}$ are, respectively, creation and
annihilation operators of the free baryons and antibaryons of mass
$M$ and $\omega_k=\sqrt{M^2+|\vec k^2|}$, which are defined
through the expansion of the baryon field operator as $\Psi(\vec
x) = \frac{1}{V} \sum_{\lambda,\lambda',\vec k} (u_{\lambda,\vec
k} a_{\lambda,\vec k}+v_{\lambda',-\vec k}
a^\dagger_{\lambda',-\vec k}) e^{i \vec k.\vec x}$; $V$ is the
volume of the system, $u_{\lambda,\vec k}$ and $v_{\lambda',-\vec
k}$ are the Dirac spinors, where the spin projections are
$\lambda,\lambda'=1/2,-1/2$. The in-medium creation and
annihilation operators are denoted by $b^\dagger, b,
\tilde{b}^\dagger, \tilde{b}$. While the $a$-quanta are observed
as asymptotic states, the $b$-quanta are the ones thermalized in
the medium. They are related by a fermionic Bogoliubov-Valatin
transformation,
\begin{equation}
\left(
\begin{array}{c}
a_{\lambda,{\mathbf k}} \\
\tilde{a}^{\dagger}_{\lambda',-{\mathbf k}} \end{array}
\right) =
\left(
\begin{array}{cc}
c_{\mathbf k} &
\frac{f_{\mathbf k}}{|f_{\mathbf k}|} \,s_{\mathbf k} \,A  \\
-\frac{f^*_{\mathbf k}}{|f_{\mathbf k}|}\, s^*_{\mathbf k}\, A^{\dagger} &
c^*_{\mathbf k}
\end{array}
\right)
\left(
\begin{array}{c}
b_{\lambda,{\mathbf k}} \\
\tilde{b}^{\dagger}_{\lambda',-{\mathbf k}}
\end{array}      \right ),
\label{BVtransf}
\end{equation}
here $c_1=\cos r_1$, $s_1=\sin r_1 $,
and
\begin{equation}
\tan(2 r_1) = - \frac{|{\bf k}_1| \Delta M({\bf k}_1) }
{\omega({\bf k}_1)^2 - M \Delta M({\bf k}_1) }
\label{fk}
\end{equation}
is the fermionic squeezing parameter. Note that in the fermionic
case, the squeezing parameter is the coefficient of sine and
cosine functions, differently than the bosonic cases in which
appeared their hyperbolic counterparts. In Eq.~(\ref{BVtransf})
$A$ is a $2~\times~2$ matrix with elements
$A_{\lambda_1\lambda_2}=
\chi^{\dagger}_{\lambda_1}\sigma\cdot{\hat{\bf k}}_1\tilde
\chi^{\phantom\dagger}_{\lambda_2}$, where $\hat{\bf k}_1 = {\bf
k}_1/|{\bf k}_1|$,  $\chi$ is a Pauli spinor and $\tilde \chi=
-i\sigma^2 \chi$. Since $r$ is real in the present case, we drop
the complex-conjugate notation in what follows.

In order to evaluate the thermal averages above, the system is
modelled as a globally thermalized gas o quasi-particles
(quasi-baryons). In this description, the medium effects are taken
into account through a self-energy function, which, for a
spin-$\frac{1}{2}$ particle (we will focus on proton and
anti-proton pairs), under the influence of mean fields in a
many-body system, can be written as $\Sigma = \Sigma^s + \gamma^0
\Sigma^0 + \gamma^i \Sigma^i$. In this expression, $\Sigma^0$ is a
weakly momentum-dependent function which, for locally thermalized
systems that we are considering, has the role of shifting the
chemical potential, i.e., $\mu_* = \mu - \Sigma^0$. The vector
part is very small and is neglected. The scalar part can be
written as $\Sigma^s = \Delta M({\bf k})$. With these
approximations we describe the system with a momentum-dependent
in-medium mass, $M_*({\bf k}) = M - \Delta M(|{\bf k}|)$.

We are mainly interested here in the study of the squeezed
correlation function, which corresponds to considering only the
joint contribution of the first and third terms on the rhs of Eq.
(\ref{fullcorr}). In the fermionic case and for an infinite,
homogeneous thermalized medium, the correlation BBC part of the
correlation function is written as
\begin{eqnarray}
&&\!\!\!\!\!\!\!\!\!\!\!\!C_2^{(+-)}({\bf k}_1,-{\bf k}_1) = 1 + \;
[1+(2\Delta t \; \omega_{\bf k})^2]^{-1} \; \times
\nonumber\\
&&\!\!\!\!\!\!\!\!\!\!\!\!\{\frac{(1 - n_{\bf k} - \tilde{n}_{\bf k})^2
    (c_{\bf k} s_{\bf k})^2}
{ \left[ c_{\bf k}^2 n_{\bf k} + s_{\bf k}^2 (1-\tilde{n}_{\bf k})
    \right]
 \left[ c_{\bf k}^2 \tilde{n}_{\bf k} + s_{\bf k}^2  (1-n_{\bf k})
    \right] }\},\; \label{BBCFa}
\end{eqnarray}
where $n_{\bf k} =  \frac{1}{\exp
\left[(\Omega_{\bf k} - \mu_*)/T\right] +1} \; ; \;
\tilde{n}_{\bf k}  =  \frac{1}{\exp
\left[(\Omega_{\bf k} + \mu_*)/T\right] +1}$ in terms of which the
net baryonic density is written as
$\rho_B = (g/V)\sum_{\bf k}~\bigl( n_{\bf k} - n_{-{\bf k}}\bigr)$.
In Eq. (\ref{BBCFa}) we have included a more gradual freeze-out by
means of a finite emission interval,
similarly to what was done in Ref\cite{acg}, which has the effect of
suppressing the BBC signal.

For our numerical study of the fermionic back-to-back correlations
(fBBC), we considered, for simplicity, momentum independent
in-medium masses, i.e., $M_* = M - \Delta M$. There is no
difficulty in considering momentum dependent self-energies,
however, this requires the commitment to a specific model and we
preferred to leave it for a future investigation. In Fig. 25 we
show fBBC for $\bar{p}p$ pairs as a function of the in-medium mass
$M_*$, for three values of the net baryonic density $\rho_B$: for
the normal nuclear matter, one tenth of this value and for the
baryon free region, i.e., $\rho_B=0$. We show in the same plot
results for the bosonic case, bBBC, corresponding to $\phi$ meson
pair, whose mass is close to the proton mass and was the example
used in Ref.\cite{acg}.
\smallskip
\begin{figure}[!hbt25]
%%\resizebox{10pc}{!}{\includegraphics{fbbcfig1}}resizebox{5pc}{!}{\includegraphics{fbbcfig2}}
\hskip-0.45cm\includegraphics*[angle=0,
width=4cm]{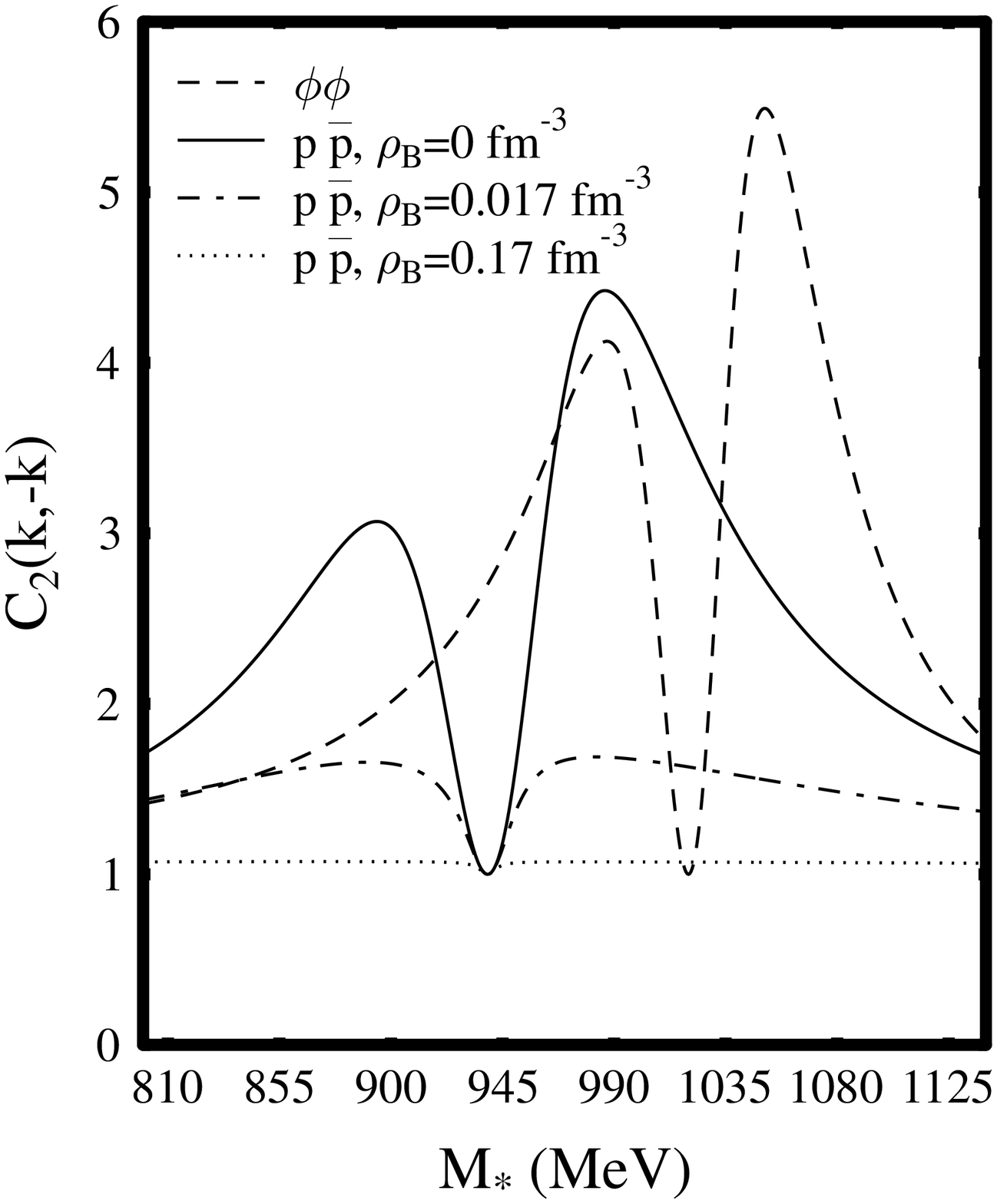}\includegraphics*[angle=0,
width=5cm]{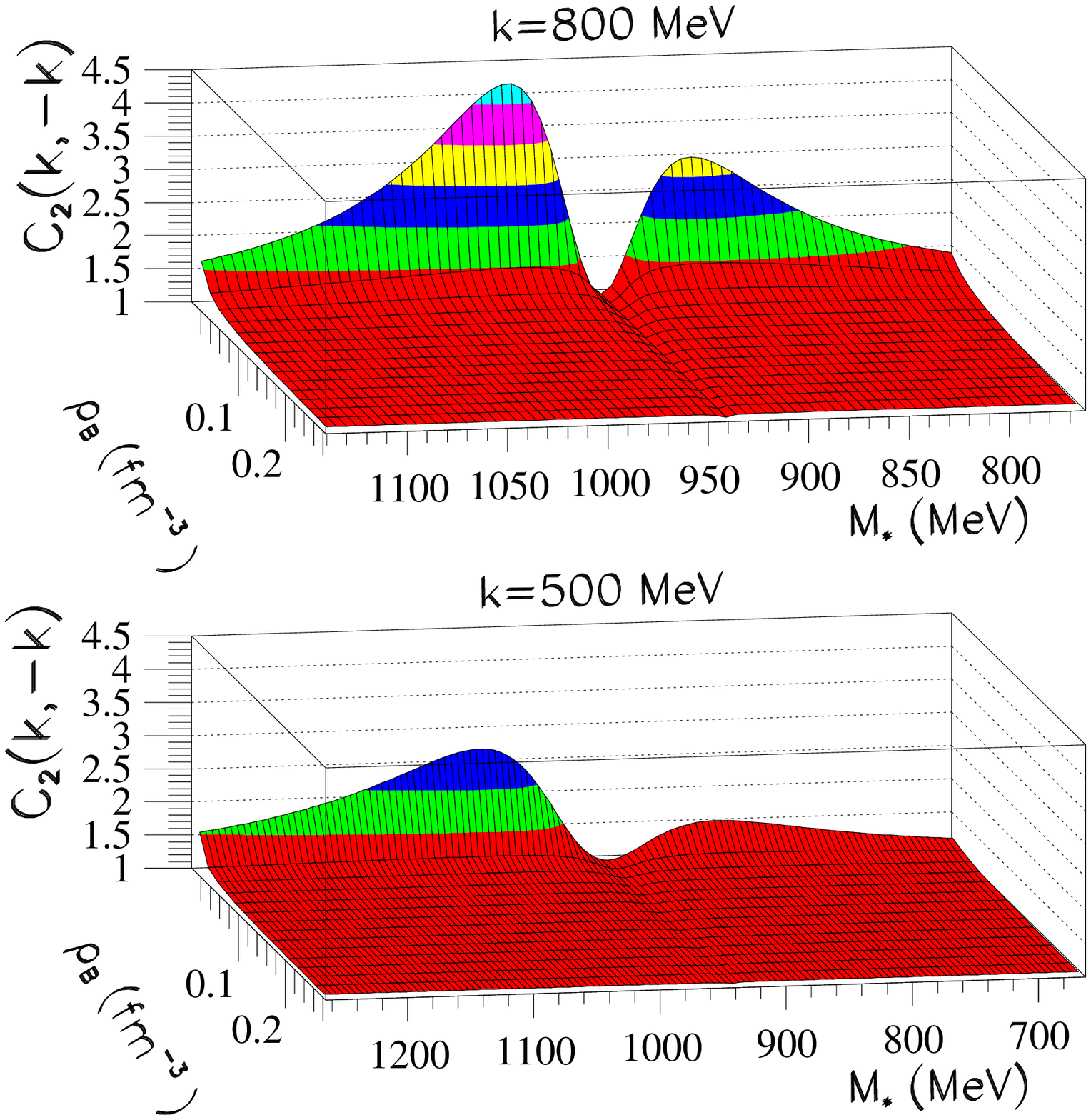} \caption{The plot on the left shows
back-to-back correlations of $\bar{p}p$ (fBBC) and of $\phi$-meson
pairs, for $|\vec k|=800$ MeV/c. The plot on the right shows the
dependence of the fBBC on the in-medium modified proton mass,
$M_*$, and on the net baryon density, $\rho_B$ and two typical
values of $|\vec k|$. In both cases $T=140$ MeV and $\Delta t=2$
fm/c. The plots were extracted from Ref.\cite{pchkp}. }
\label{bbc1}\end{figure}

We see from Fig. 25 that fBBC and bBBC are, indeed, both positive
correlations, with similar shape, and of the same order of
magnitude. We also observe that fBBC is strongly enhanced for
decreasing net baryonic density, being maximal for $\rho_B \approx
0$, i.e., for approximately equal baryon and anti-baryon
densities. Fig. 25 also shows that fBBC increases with increasing
momentum. Besides, we see that the shape of fBBC is very sensitive
to the shape of the freeze-out distribution and, in the limit of a
very long freeze-out, both fBBC and bBBC would vanish.

\bigskip
So far, we considered an infinite and homogeneous medium, although
we know that the systems produced in high energy collisions,
including the ones at RHIC, have finite sizes. Thus, we need to
know if the BBC signal would survive when more realistic spacial
and dynamical considerations were made. For pursuing this purpose,
we studied the effects on the squeezing parameter and on the
back-to-back correlation of a finite size medium moving with
collective velocity. For this, we considered a hydrodynamical
ensemble, in which the amplitudes $G_c$ and $G_s$ in Eq.
(\ref{fullcorr}) and (\ref{rand}) are extended to the special form
derived by Makhlin and Sinyukov \cite{sm}.
\begin{eqnarray}
G_c(1,2)&=&\frac{1}{(2 \pi)^3 }
\int d^4\sigma_{\mu}(x) K_{1,2}^{\mu} e^{i q_{1,2} \cdot x}
\{|c_{1,2}|^2 n_{1,2}
\nonumber\\
&& \hskip1cm + \;\;\; | s_{-1,-2}|^2  (n_{-1,-2} + 1) \},
\label{e:gc}
\end{eqnarray}
\begin{eqnarray}
G_s(1,2)\!\!&=&\!\!\frac{1}{(2 \pi)^3 }
\int d^4\sigma_{\mu}(x) K_{1,2}^{\mu} e^{2 i  K_{1,2} \cdot x}
\{ s^*_{-1,2} c_{2,-1} n_{-1,2}
\nonumber\\
&& \hskip1cm + \;\;\; c_{1,-2} s^*_{-2,1} (n_{1,-2} + 1) \}.
\label{e:gd}
\end{eqnarray}

\smallskip\smallskip
\noindent In Eq.(\ref{e:gc}) and (\ref{e:gd}) $d^4\sigma^{\mu}(x)
= d^3\Sigma^{\mu}(x;\tau_f)\, F(\tau_f) d\tau_f $ is the product
of the normal-oriented volume element depending parametrically on
$\tau_f$ (the freeze-out hyper-surface parameter) and on the
invariant distribution of that parameter $F(\tau_f)$. We consider
two possibilities: i) an instant freeze-out, corresponding to
$F(\tau)=\delta(\tau-\tau_0)$; ii) an extended freeze-out, with a
finite emission interval, with
$F(\tau)=[\theta(\tau-\tau_0)/\Delta t] e^{-(\tau-\tau_0)/\Delta
t}$. These cases lead, after performing the integration in $d
\tau$ in Eq. (\ref{e:gc}) and (\ref{e:gd}) with weight ($E_{i,j}
\;e^{-i 2 E_{i,j} \tau}$), respectively to: i)
$(\omega_i+\omega_j) \; e^{- i (\omega_i+\omega_j) \tau_0}$; ii)
$(\omega_i+\omega_j)[1+[(\omega_i+\omega_j)\Delta t]^{-2}$.

According to the hydrodynamical solution, we can express
the chemical potential  as
$\frac{\mu(x)}{T(x)} = \frac{\mu_0}{T} - \frac{{\mathbf r}^2}{2R^2}$,
being $R$ the radius of the system, $T=T(x)$ the temperature of
the system in each space-time point $x$, and $\mu_0$ a constant.
We assume that the system expands with four-dimensional flow
velocity $u^\mu = \gamma (1,{\mathbf v})$, where
${\mathbf v}=<\!\!u\!\!>\! \frac{\mathbf r}{R}$. In the non-relativistic
limit, we can write $\gamma = (1+{\mathbf v})^{-1/2}
\approx 1+\frac{1}{2}{\mathbf v}^2$, thus taking into account all
terms up to ${\cal O}(mv^2)$. We estimate the geometrical and
dynamical effects on the BBC in the bosonic case, considering the
in-medium changes for the $\phi$-meson.
\begin{figure}[!hbt26]
\resizebox{11pc}{!}{\includegraphics{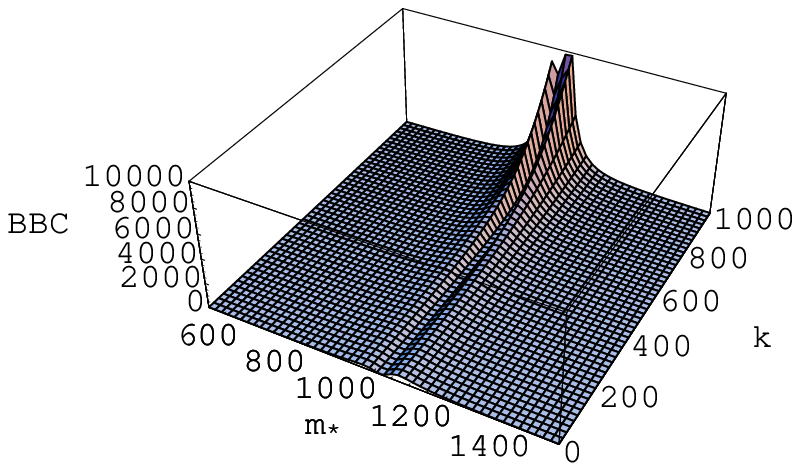}}\resizebox{11pc}{!}{\includegraphics{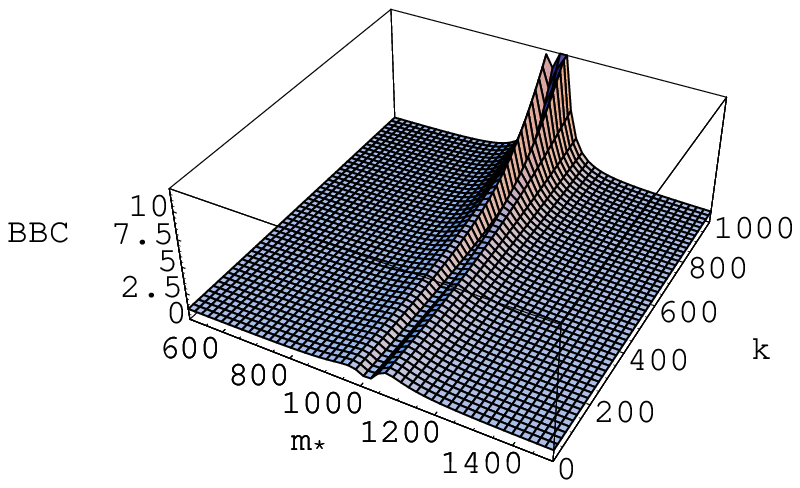}}
\resizebox{11pc}{!}{\includegraphics{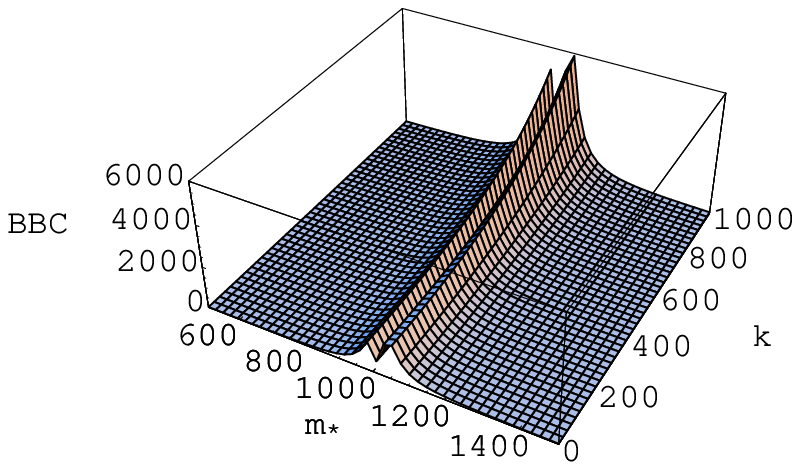}}\resizebox{11pc}{!}{\includegraphics{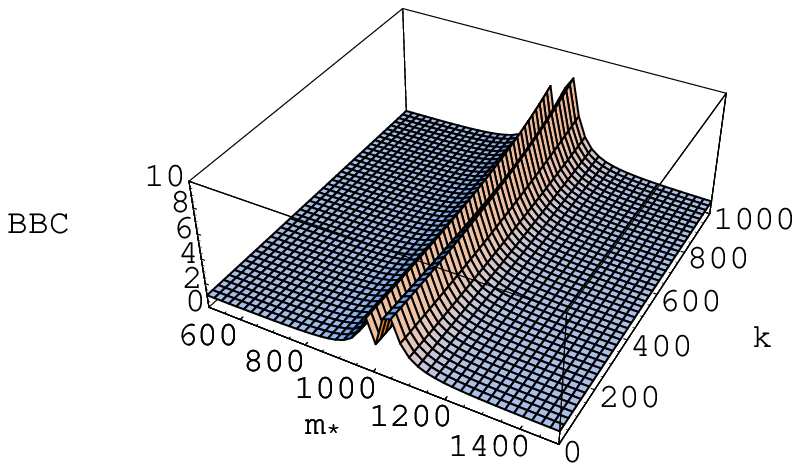}}
%\vskip.5cm
\caption{The Back-to-Back Correlation is shown as a function of
the pair average momentum, for ${\mathbf k}_1=-{\mathbf
k}_2={\mathbf k}$, when considering the values of the shifted mass
around the vacuum mass, $m_\phi=1020$ MeV). The squeezed mass was
considered to change in the entire Gaussian volume, with rms $R=6$
fm. In the top plots no flow was included, i.e., ${\mathbf
v}=<\!\!u\!\!>\! \frac{\mathbf r}{R} = 0$, but radial flow with
$<\!\!u\!\!>\! = 0.5$ was included in the bottom cases. Plots on
the left side correspond to the case of particles emitted
instantly, and those on the right, to the case where particles are
emitted during a finite time interval, $\Delta t$.}
\label{bbc2}\end{figure}

For small mass shifts, i.e. $\frac{(m-m_*)}{m} \ll 1$, the flow
effects on the squeezing parameter are of fourth order, i.e.,
${\cal O} \left( \frac{Kin. \; energy}{m} \right) (\frac{\delta
m^2}{m^2})$. As a consequence, the flow effects on $r_{i,j}$ can
be neglected, and the factor $c_{i,j}$ and $s_{i,j}$ become flow
independent, although they could still depend on the coordinate
$r$ through the shifted mass, $m_*$ (e.g., through $T(x)$, as in
hydrodynamics), which is not considered here.

For the sake of simplicity, trying to keep the results as much
analytical as possible (for details, see \cite{chkppflow}), we
make the hypothesis that the mass-shift is independent on the
position within the fireball. We further assume that this last one
has a sharp surface, i.e., $\delta m=0$ on the surface, and also
the density vanishes outside the system volume. The spatial
integration in Eq. (\ref{e:gc}) and (\ref{e:gd}) extends over the
region where the mass-shift is non-vanishing,  which is {\it not}
infinitely large. For instance,I in relativistic heavy ion
collisions is a finite region $V \approx R^3 \approx (5-10)^3$
fm$^3$. We should keep in mind that the vacuum term in the
integrand vanishes outside the mass-shift region, since it is
proportional to $s_{i,j}$, which vanishes in that region. On the
other hand, the terms proportional to $n^{(*)}_{i,j} (x)$ are
finite.
%due to the finiteness of the hydrodynamical solution.
Being so, we can extend the integration in Eq. (\ref{e:gc}) and
(\ref{e:gd}) to infinity and, without much loss of generality, we
can choose for $V$ a Gaussian profile, $\exp[-{\mathbf
r}^2/(2R^2)]$.

We studied the flow effects in two cases. The first one considers
the mass-shift as occurring in the entire volume of the system,
for which the parameter $R$, in the exponential, represents the
cross-sectional area of the Gaussian profile with rms $R$ . The
other case corresponds to considering the shift of mass in a
smaller volume, associated to a certain $R_s < R$. %We estimated
%the corresponding numerical results for $R=6$ fm and $R_s=5$ fm.
We show in Fig. \ref{bbc2} the results for the first case only,
where we adopted $R=6$ fm. The BBC signals for the mass shift in
the partial volume, although not shown, are very similar to the
ones in Fig. 26, except for the maximum strength of the bBBC
signal, which is smaller than seen in the plots of Fig. 26,
emphasizing that the effect is directly proportional to the size
of the region where the mass-shift occurs. By comparing the top
panels (no flow) with the bottom ones ($<\!\!u\!\!>\! = 0.5$) in
Fig. 26, we see that the flow has a suppression effect on the bBBC
strength in the high momentum region. However, the presence of
flow surprisingly enhances the BBC signal in the low momentum
region (i.e., for $|{\mathbf k}|\lesssim 1000$ MeV), as compared
to the no-flow case on top panel, suggesting this as the region to
focus the search for the BBC effect. In any case, we conclude with
enthusiasm that the BBC signal survives more realistic conditions,
like finite sizes, time spread, and flow. This strongly encourages
us to analyze even more realistic systems than the ones discussed
here, and to optimize the way we should look for this interesting,
although not yet observed, effect.

\section{BRIEF CONCLUDING REMARKS}

I would like to finalize by emphasizing that, although a lot was
written above about the evolution and contributions related to
HBT, since its discovery about half a century ago, the present
text merely covers a tiny fraction of what has been produced on
this subject. An increasing number of people over these five
decades have given countelss contributions to the field.
Unfortunately, the lack of space would not allow me to mention and
discuss them all. Finally, I would like to dedicate this review to
the memory of my father, who passed away this year, as well as to
Robert Hanbury-Brown and Richard Q. Twiss, on celebrating the
50$^{th}$ anniversary of their first publication on this fabulous
method and discovery.
%the memory of Robert Hanbury-Brown, who passed away in January,
%2002, and to Richard Q. Twiss, on celebrating the 50$^{th}$
%anniversary of the first publication about their fabulous method
%and discovery.

%\bigskip

%\acknowledgments

I would like to express my gratitude to Yogiro Hama and Takeshi
Kodama for devising, 15 years ago, and organizing ever since, this
active group working on hadronic and nuclear interactions at high
energies. %, already 15 years ago.
%, and for continuously making the
%effort of keeping the group alive and growing in a productive way.

This work was partially supported by CNPq (Proc. Proc. N$^{\b{o}}
200410/82-2$) and Funda\c c\~ao de Amparo \`a Pesquisa do Estado
de S\~ao Paulo (FAPESP), {\sl Projetos Tem\'aticos} $90/4074-5,
93/2463-2, 95/4635-0, 98/2249-4$ and $00/04422-7$.

\end{document}